\documentclass[useAMS,usenatbib]{mn2e}
\usepackage{graphicx}
\usepackage{natbib}
\usepackage{array,booktabs,tabularx}
\usepackage{rotating}
\usepackage{lscape}
\usepackage{epstopdf}
\usepackage{amssymb}
\usepackage[caption=false]{subfig}

\def\mch{M$\rm^{c}$Hardy\,}
\def\Msol{$M_{\odot}\ $}
\def\mdote{\.m$\mathrm{_{Edd}}$}
\def\mdotes{\.m$\mathrm{_{Edd}}~$}
\def\mdot{\.m}
\def\mcrit{\.m$\mathrm{_{crit}}$}

\begin{document}
\title[Long-term X-Ray Spectral Variability in AGN from the Palomar sample observed by {\it Swift}]{Long-term X-Ray Spectral Variability in AGN from the Palomar sample observed by {\it Swift}}

\author[S. D. Connolly, I.M.  \mch, C. J. Skipper and D. Emmanoulopoulos]{S. D. Connolly$^{1}$\thanks{E-mail: sdc1g08@soton.ac.uk}, I.M.
M\parbox[b][3.4mm][t]{2.0mm}{c}Hardy$^{1}$, C. J. Skipper$^{1,2}$, D. Emmanoulopoulos$^{1}$ \\ \\
$^{1}$Physics \& Astronomy, University of Southampton, Highfield, Southampton, SO17 1BJ, UK\\
$^{2}$Jodrell Bank Centre for Astrophysics, Alan Turin Building, The University of Manchester, Manchester, M13 9PL, UK}

\pagerange{\pageref{firstpage}-\pageref{lastpage}} \pubyear{2012}

\maketitle

\label{firstpage}


\begin{abstract}

We present X-ray spectral variability of 24 local active galactic nuclei (AGN) from the Palomar sample of nearby galaxies, as observed mainly by Swift. 
From hardness ratio measurements, we find that 18 AGN with low accretion rates show hardening with increasing count rate, converse to the 
softer-when-brighter behaviour normally observed in AGN with higher accretion rates. Two AGN show softening with increasing count rate, two 
show more complex behaviour, and two do not show any simple relationship. 

Sufficient data were available for the spectra of 13 AGN to be summed in flux-bins. In 9 of these sources, correlated luminosity-dependent changes in 
the photon index ($\Gamma$) of a power-law component are found to be the main cause of hardness variability. For 6 objects, with a low accretion 
rate as a fraction of the Eddington rate (\mdote), $\Gamma$ is anticorrelated with \mdote, i.e. `harder-when-brighter' behaviour is observed. The 3 
higher-\mdote-rate objects show a positive correlation between $\Gamma$ and \mdote. This transition from harder-when-brighter at low \mdotes to 
softer-when-brighter at high \mdotes can be explained by a change in the dominant source of seed-photons for 
X-ray emission from cyclo-synchrotron emission from the Comptonising corona itself to thermal seed-photons from the accretion disc. 
This transition is also seen in the `hard state' of black hole X-ray binaries (BHXRBs). The results support the idea that LINERs are 
analogues of BHXRBs in the hard state and that Seyferts are analogues of BHXRBs in either the high-accretion-rate end of the hard state or in the hard-intermediate state.

\end{abstract}


\section{Introduction}

In many studies of black holes with relatively high accretion rates as a fraction of their Eddington accretion rates (\mdote), 
including both single-epoch observations of samples of objects 
and multi-epoch observations of individual objects, the X-ray spectrum becomes softer as the luminosity increases. This 
softer-when-brighter behaviour is seen in both black hole X-ray binary systems (BHXRBs) \citep[e.g.][]{Gu2009} and in AGN 
\citep[e.g.][]{McHardy1998, Sobolewska2009,Shemmer2006}. This behaviour is usually attributed to a real 
increase in the photon index, $\Gamma$, of the power law assumed to describe the  shape of the intrinsic emission spectrum.

This `softer-when-brighter' correlation has not been found to hold, however, for lower accretion rate systems.
In a sample of single-epoch observations of low-luminosity AGN (LLAGN), \citet{Gu2009} found
$\Gamma$ to be {\it anticorrelated} with the accretion rate (i.e. `{\it harder}-when-brighter'). 
A similar conclusion was reached by both \citet{Younes2011} and \citet{Hernandez-Garcia2014} 
in samples of low-ionisation nuclear emission region galaxies (LINERs). 
Harder-when-brighter behaviour has also been observed in multi-epoch observations of individual BHXRBs 
by e.g. \citet{Kalemci2004} and \citet{Yuan2007}. This behaviour has, however, 
only been seen once in multi-epoch observations of in an individual AGN, the LLAGN NGC 7213, by \citet{Emmanoulopoulos2012},
and was once again thought to be caused by an anticorrelation between $\Gamma$ and the accretion rate.

In a study of a sample of BHXRBs by \citet{Wu2008}, harder-when-brighter behaviour was observed below a `critical' accretion rate (\mcrit) 
of $\mathrm{\sim 0.01}\ $ \mdote, above which softer-when-brighter behaviour was seen. 
The same change in behaviour above a certain critical accretion rate has also been seen into multi-epoch observations of the XRB Cyg X-1 \citep{Skipper2013,Axelsson2008}. 
Single-epoch observations of large samples of AGN have shown that they follow the same behaviour globally 
\citep{Constantin2009,Gu2009,Younes2011}, supporting the idea that AGN behave in a similar way to BHXRBs. Note that the change from harder-when-brighter to softer-when-brighter 
at \mcrit$\ $ occurs within the hard state and is not considered to be a state transition. 
The accretion rate below which BHXRBs are never seen in the soft state ($\mathrm{~2\%}$ \mdote; \citet{Maccarone2003}) occurs factors of 2-3 above \mcrit.

At extremely low accretion rate $\ $($\mathrm{< \sim10^{-6}}$ \mdote), the behaviour has been found to change once again, 
with the spectral hardness no longer changing, which is thought to be due to $\Gamma$ saturating at a 
relatively constant, and high, level as the accretion rate decreases \citep{Yang2015}. This saturation is once again seen in observations of both 
AGN and BHXRBs, using a small number of observations from large samples of sources \citep{Yang2015, Gu2009, Hernandez-Garcia2013}. 

Due to a number of similarities, including a low accretion rate and radio loudness, it has been suggested that LLAGN are the analogues of the `hard state' of BHXRBs, whilst more 
luminous Seyferts and quasars are the analogues of 'soft-state' BHXRBs. In this paper we will re-examine these analogies.

The change in the relationship between the $\Gamma$ and \mdotes in these systems
at a particular critical \mdotes from a 
negative correlation to a positive correlation as \mdotes increases implies that the seed photon source to the X-ray corona is changing  \citep[e.g.][]{Emmanoulopoulos2012, Skipper2013}. It has been suggested that the reason 
for this change may be that the accretion disc is transitioning from an advection dominated accretion flow (ADAF) to a standard thin disc as \mdotes increases
\citep{Esin1997,Narayan1994,Wu2008,Constantin2009}. In ADAF models, an increase in \mdotes leads to an increasing optical depth and hence to a higher Compton parameter in the hot accretion flow, 
`hardening' the X-ray spectrum and producing an anticorrelation between $\Gamma$ and \mdote, and therefore harder-when-brighter behaviour.

The `hard state', with which ADAFs are associated in BHXRBs, is usually also associated with the presence of a jet \citep{Russell2010}.
The X-ray emission mechanism in BHXRBs is not known unambiguously, but self-Comptonisation of synchrotron emission from the corona, 
or possibly from a jet at higher energies, is a possibility. Synchrotron self-Compton emission dominates blazar jet emission, with luminosity-increases 
associated with an injection of hard-spectrum electrons, e.g. from a shock \citep[e.g.][]{Ghisellini2009}.
The harder-when-brighter spectral relationship is commonly observed in blazars for this reason \citep[e.g.][]{Krawczynski2004,Gliozzi2006,Zhang2006} and
could therefore also explain the same behaviour in LLAGN \citep{Emmanoulopoulos2012}.

In an ADAF, the inner edge of the optically thick part of the disc is truncated a long way from the black hole.
Further possible explanations for harder-when-brighter behaviour which do not require a truncated disc also exist, for example by invoking a outflowing 
hot thermal corona - for details see \citet{Sobolewska2011} and references therein.

In this study we present results on changes is hardness with luminosity of a sample of 24 AGN from the Palomar sample \citep{Ho1997} which have been observed by {\it Swift}. 
For those sources for which there is enough data, spectral fitting is performed with a variety of models, and changes in $\Gamma$ with luminosity are measured.

The objects cover a range of masses between $\mathrm{5 \times 10^{4}}$ \Msol and $\mathrm{8 \times 10^{9}}$ \Msol
and a range of \mdotes between $\mathrm{10^{-6}}$ and $\mathrm{0.5}$ (see Table \ref{radio}). Approximately half of these
objects possess accretion rates which lie in the regime in which softer-when-brighter behaviour is expected (\mdotes $\mathrm{\geq 10^{-3}}$).
The rest occupy the accretion rate range (\mdotes $\mathrm{~\sim 10^{-6} - 10^{-3}}$) in which individual XRBs \citep[e.g.][]{Axelsson2008,Skipper2013} and samples of AGN show  harder-when-brighter behaviour. 
We also include NGC 7213 in our discussion, using the data from \citet{Emmanoulopoulos2012}. 

\section{Observations \& Data Reduction}

\subsection{The AGN Sample}

The sample used in this study consists of all known AGN in the Palomar sample of galaxies for which there is sufficient data for meaningful analysis. 
The Palomar sample contains 163 such
objects (Seyfert galaxies and LINERs), 70 of which have been observed by {\it Swift}. 
So that sufficient data exist for at least a rudimentary spectral analysis, 
only objects for which there are a total of $\mathrm{>350}$ total photon counts and 4 or more separate observations were included in the analysis,
leaving a final sample of 24 AGN. Table \ref{obsTable} gives a summary of the {\it Swift} observations
of each of the selected AGN. The Palomar sample contains almost every galaxy in the northern sky with a magnitude 
of $\mathrm{B_T > 12.5}$; the AGN within this sample are therefore all relatively nearby and
a large fraction of them are low-luminosity - 85\% lie below $\mathrm{L(H\alpha) = 10^{40} ergs ~s^{-1}}$ \citep{Ho1997,Ho1997a}. 
As our sample contains only a relatively small subset of the Palomar AGN sample it is not statistically
complete. However, whilst some galaxies were observed because they contained known AGN, others were observed for different reasons, 
e.g. because they contained a supernova, thus the subset is not particularly biased.

\begin{table*}
\footnotesize
\begin{tabular}{l c c c c c c}
\hline
Object & Observations & GTIs & Start \& end date & Total exposure (s) & Total counts & No. summer spectra \\ 
\hline	
NGC 315  		& 5 	& 42	& 2007-05-31 - 2009-03-03 & 26054 	&  644		& 1 \\
NGC 1052 		& 21 	& 126 	& 2007-01-19 - 2011-03-06 & 81041 	&  3043	& 3	\\
NGC 1068 (M77) 		& 13 	& 44	& 2007-06-19 - 2015-02-05 & 32681 	&  15819	& 5 \\
NGC 2655 		& 33 	& 120 	& 2011-01-11 - 2014-12-31 & 60351 	&  1056	& 1	\\
NGC 3031 (M81) 		& 605 	& 1033 	& 2005-04-21 - 2014-06-22 & 771530	&  205159 & 28	\\
NGC 3147 		& 28 	& 63	& 2009-01-11 - 2014-07-13 & 52929 	&  2417		& 3 \\
NGC 3226 		& 26 	& 63	& 2008-10-28 - 2015-07-06 & 52841 	&  699		& 1 \\
NGC 3227 		& 26 	& 63	& 2008-10-28 - 2015-05-23 & 19269 	&  38816	& 10 \\	 
NGC 3628 		& 6 	& 32	& 2006-11-12 - 2014-07-21 & 26518 	&  379		& 1\\
NGC 3998 		& 2 	& 20	& 2007-04-20 - 2007-04-29 & 27285 	&  6362		& 4 \\
NGC 3998$^\dagger$	& 376 	& 376	& 2010-12-31 - 2011-12-26 & 287568	&  7628314	& 25 \\
NGC 4051 		& 55 	& 100 	& 2009-02-14 - 2013-10-10 & 58002 	&  30821& 10 	\\	
NGC 4151 		& 5 	& 33	& 2005-12-26 - 2012-11-12 & 27387 	&  17808	& 7 \\	
NGC 4258 (M106)		& 12 	& 8 	& 2008-03-01 - 2014-06-22 & 32238 	&  1270		& 1 \\
NGC 4321 (M100)		& 26 	& 88	& 2005-11-06 - 2011-04-15 & 27426 	&  282		& 1 \\
NGC 4388 		& 4 	& 34	& 2005-12-27 - 2008-05-09 & 10594 	&  1219		& 4 \\
NGC 4395 		& 225 	& 412 	& 2005-12-31 - 2013-05-10 & 302125	&  34277	& 22 \\	
NGC 4486 (M87) 		& 19 	& 46	& 2006-12-23 - 2015-05-18 & 29601	&  4264		& 4 \\	
NGC 4472 (M49) 		& 6 	& 19	& 2007-11-13 - 2010-03-30 & 14298 	& 1268		& 1 \\
NGC 4579 (M58) 		& 27 	& 41	& 2007-05-15 - 2007-05-18 & 20463	& 3555		& 3 \\
NGC 4736 (M94)		& 18 	& 39	& 2013-04-15 - 2015-05-12 & 23700 	&  104 		& 1 \\
NGC 5194 (M51)		& 112 	& 261 	& 2005-06-30 - 2014-09-24 & 233095	&  5599	& 1 	\\
NGC 5548 		& 656 	& 878	& 2005-04-08 - 2014-06-21 & 574670	&  206387	 & 42 \\	
NGC 5806 		& 49  	& 147	& 2012-01-14 - 2014-09-17 & 141994	&  381		& 1 \\
NGC 7331 		& 34  	& 90	& 2007-02-12 - 2015-05-19 & 77403 	&  616		& 1 \\
\hline

\end{tabular}

\caption{ Details of the observations of the 24 AGN considered in this study.
All data are from {\it Swift} except for the observations of NGC 3998 denoted by $^\dagger$, from {\it RXTE}.
The number of summed spectra is also given, denoting the number of flux bins into which the observations
for each source were divided. Where the number is one, all the observations for that source were summed
into a single spectrum (i.e. not flux-binned).}

\label{obsTable}

\end{table*}

The data for all Palomar AGN except NGC 3998 were obtained by {\it Swift} and, in almost all cases, are composed of observations from a number of different programmes covering a number
of years. The data for NGC 3998 are from a single campaign by the 
{\it Rossi X-ray Timing Explorer} ({\it RXTE}), a preliminary analysis of which is described in  \citet{SkipperThesis}; whilst there are some {\it Swift} data available for NGC 3998,
the {\it RXTE} data are of better quality. 

In all {\it Swift} pointings, the observations were performed using the {\it Swift} XRT in `photon counting mode'. Exposure times of individual good time interval (GTI) observations ranged from less than
$10$ seconds to over $10^4$ seconds.  In each case, GTIs were excluded if they had a low signal-to-noise ratio (S/N $\mathrm{< 3}$), very short exposure time 
($\mathrm{T_{exp} < 50\ s}$) or the source was in close proximity to bad pixels. The raw data
for all {\it Swift XRT} observations were downloaded from the HEASARC archive$^1$\footnotetext[1]{http://heasarc.gsfc.nasa.gov/cgi-bin/W3Browse/swift.pl}.

In addition to the data from this work, we also include data from NGC 7213 \citep{Emmanoulopoulos2012} in the discussion,
to allow comparison with what was, until now, the only confirmed detection of harder-when-brighter behaviour in an individual AGN. 
NGC 7213 is not part of the Palomar sample, and has not been observed extensively by {\it Swift}. However, it has been observed extensively 
(882 times) by {\it RXTE} and is known to possess a relatively simple power law spectrum \citep{Emmanoulopoulos2013b}. We therefore do not 
include NGC 7213 within our discussion of {\it Swift} spectral fitting (Section \ref{spectralFitting}) or of the properties of the Palomar sample, 
but take the results of the {\it RXTE} spectral fitting from  \citet{Emmanoulopoulos2012}.

\subsection{Data Reduction}

The XRT data were reduced using our automatic pipeline, described in e.g \citet{Cameron2012}, \citet{Connolly2014}. The most recent version of the standard {\it Swift} 
XRTPIPELINE software (versions 0.12.4 - 0.12.6) was used in each case. Spectra and light curves were extracted using the XSELECT
tool, using flux-dependent source and background extraction regions chosen such that background contamination at faint fluxes was minimised, and to account for
the effects of pile-up at high fluxes. The effects of vignetting and the presence of bad pixels and columns on the CCD were removed by using the {\it Swift} XRTEXPOMAP
and XRTMKARF tools to create an exposure map and an ancillary response file (ARF) for each visit. The relevant redistribution matrix file (RMF) from the {\it
Swift} calibration database was also used in each case. The local X-ray background was estimated and subtracted from the instrumental count rates, using the
area-scaled count rate measured in a background annulus region. The observed XRT count rates were corrected to take into account the fraction of counts lost
due to bad pixels and columns, vignetting effects, and the finite extraction aperture (including regions excised in order to mitigate pileup effects).

\textit{RXTE} observed NGC 3998 376 times under our own proposal 96398, from 2010 Dec 31 until 2011 Dec 26, and with exposure times of
between 542 seconds and 6,770 seconds. The earlier observations were each separated by a period of roughly two days, but after approximately 2011 Oct 25 the frequency of
observations increased to roughly once every six hours. All the data reduction was performed only upon the standard-2 data, with a time-resolution of $\mathrm{16\ s}$.

For each observation a GTI was generated based upon the spacecraft elevation angle being greater than 10 degrees, the pointing offset being less than
0.02 degrees, and the time since the last south Atlantic anomaly (SAA) being at least 30 minutes. Synthetic background data were generated using the script
\textsc{runpcabackest} with the faint background model (pca\_bkgd\_cmfaintl7\_eMv20051128.mdl), and suitable response files were created using \textsc{pcarsp}.

\begin{landscape}
\begin{figure}
	\centering
	\includegraphics[width = 24cm,clip = true, trim = 0 0 0 0]{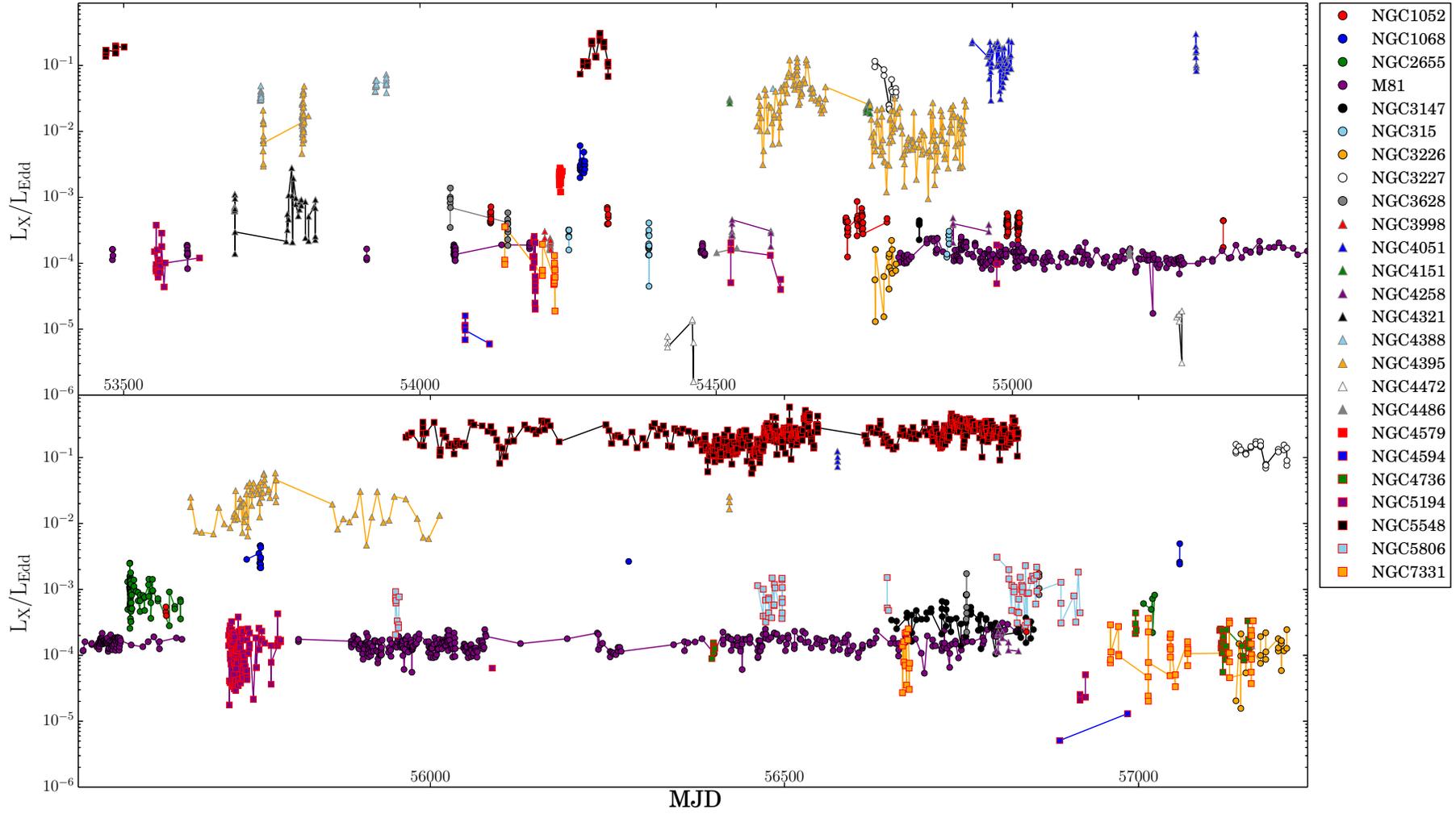}
		
	\caption{Light curve showing the ratio of the X-ray Luminosity to the Eddington Luminosity of each of the sources over time,
			showing the large range of Eddington rates present in the sample. The light curve is split into two sections for clarity.}
	\label{allLC}
\end{figure}
\end{landscape}

\begin{landscape}
\begin{figure}
	\includegraphics[width = 24cm,clip = true, trim = 0 0 0 0]{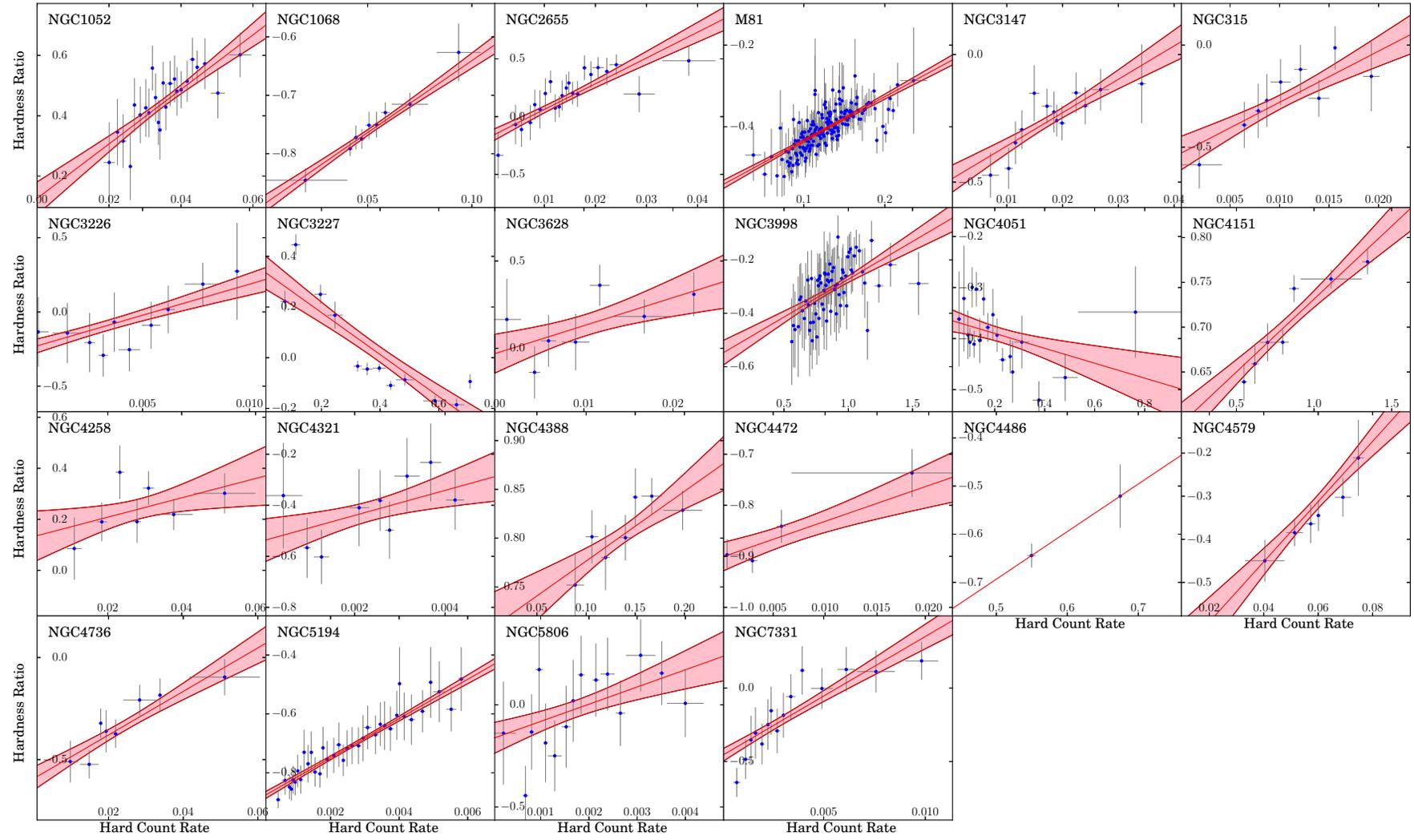}
		
	\caption{The hardness ratio of the spectrum of each {\it Swift} object using the 0.5-2 and 2-10 keV bands, plotted against its 2-10 keV count rate, and the hardness 
	ratio of NGC 3998 using the 3.3-7 and 7-20 keV bands, plotted against its 7-20 keV count rate. The data are binned into bins of 10 data points. 
	Error bars show the one-sigma confidence levels. In each case, the red line shows the
	best-fitting linear model and the pink region bounded by red indicates the one-sigma error region of the model parameters. 
	A clear correlation between the two parameters, either positive or negative, is present in all
	objects except NGC 4321 and NGC 4468. NGC 4486 possessed too little data for meaningful binning with more than ten bins,
so errors are not given. }
	\label{allhardness}
\end{figure}
\end{landscape}

Spectra were extracted from the standard-2 data and the background data for all layers of PCU 2 using \textsc{saextract}, and subsequently rebinned using \textsc{rbnpha}
into 14 channels. All the analysis described in this paper uses channels 2-13, which roughly corresponds to the energy range 3.3 to 20 keV. The observations covered a large range of exposure times, with many being too short for useful spectral analysis.
In many cases, time-binning would not have produced spectra of sufficient
quality for meaningful spectral analysis without using broad time bins, which would itself have lead to spectra consisting of data taken across a large range of fluxes. 
Whilst there is some scatter in the hardness of a given source at a given flux, the scatter is not large (see Section \ref{hardness}), indicating that spectral shape 
is a function mainly of luminosity rather than of time. Thus, flux-binning is legitimate. To improve the S/N, the data were therefore binned in flux.  

Where there were sufficient data, the {\it Swift} spectra were binned such that each resultant summed spectrum had a minimum of 1000 total counts and a maximum width 
in count rate of $\mathrm{5 \times 10^{-3} counts\ s^{-1}}$, with the exception of NGC 1052, NGC 3147 and NGC4579 which, due to fewer data, were binned with a minimum of 
900, 600 and 900 total counts per spectrum, respectively. For objects for which there were insufficient data to obtain at least 3 flux-binned spectra with a minimum of 350 total counts,
a total summed spectrum was produced. The {\it RXTE} spectra, having a considerably higher count rate but lower signal-to-noise ratio, 
were binned such that each resultant summed spectrum had a minimum of 8000 total counts per spectrum. The number of summed spectra produced for
each source is shown in Table \ref{obsTable}.

The spectra were combined using the {\it HEADAS} tool `addspec'. As the amount of data varied between sources, the number of spectra produced in each case also varied. 
The energy channels of each of the summed spectra were then grouped, using the {\it HEADAS} tool `grppha', such that each group contained a minimum of 15 counts. 

The light curves of the ratio of the X-ray luminosity to the Eddington luminosity of each of the 24 AGN are shown in Fig. \ref{allLC}.

\section{Spectral Hardness}
\label{hardness}

\begin{figure}
	\includegraphics[width = 8.5cm,clip = true, trim = 0 0 0 0]{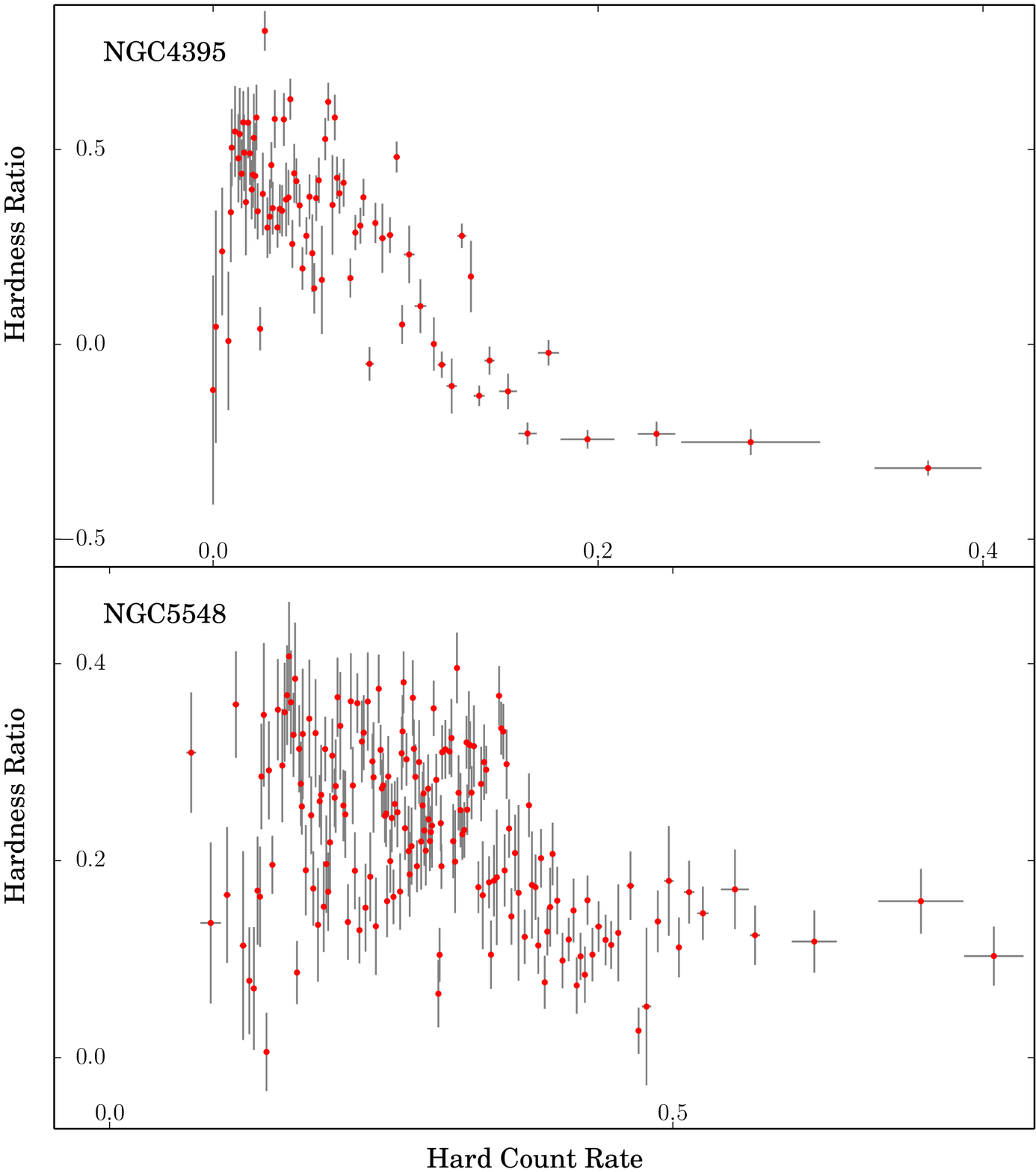}
		
	\caption{The hardness ratio of the spectra of NGC 4395 and NGC 5548 using the 0.5-2 and 2-10 keV bands, plotted against the 2-10 keV count rate. 
	The data are binned into bins of 10 data points. Error bars show the one-sigma confidence levels. Both objects show
	more complex behaviour in their hardness ratios with increasing count rate when compared to the other objects. Both
	also show a general softer-when-brighter trend at intermediate count rates.}
	\label{swbhardness}
	\vspace{-0.5cm}
\end{figure}

For the objects observed by {\it Swift}, hard emission is defined as $2.0 - 10.0$ keV and soft emission as $0.5-2.0$ keV; for NGC 3998 (observed by 
{\it RXTE}), hard emission is defined as $7.0-20.0$ keV and soft emission as $3.3 - 7.0$ keV.
In all cases, the hardness ratio is defined as: 

\begin{equation}
\centering
\mathrm{Hardness\ Ratio} = \frac{\mathrm{H} - \mathrm{S}}{\mathrm{H} + \mathrm{S}}
\label{hardnessEqn}
\end{equation}
 
where `H' is the hard count rate and (`S') is the soft count rate.
Plots of the hardness ratio against the hard count rate of each object are shown in Figs.  \ref{swbhardness} and \ref{allhardness}.
These plots show, in a model-independent way, the way in which the spectrum of
a given source is changing with increasing luminosity; if the spectral hardness increases with increasing luminosity,
the source is harder-when-brighter, if the opposite is true the source is softer-when-brighter.
For 22 of the 24 AGN, the hardness ratio changed with luminosity in an approximately linear way. The hardness ratios vs. luminosity of these objects were
fitted with a linear model in order to determine numerically whether they were harder- or softer-when-brighter.
In each case, errors on the best-fitting parameters of the linear fits
were calculated using 10000 Monte Carlo simulations in which the data were varied within their error distributions
(assuming Gaussian errors) and refitted. The reduced $\mathrm{\chi^2}$ ($\chi^2_R = \chi^2 / DoF$) and best-fitting parameters from these fits are
shown in Table \ref{hardnessTable}; the best-fitting lines and the errors on the parameters of the lines are also
shown in the hardness ratio vs. luminosity plots in Fig. \ref{allhardness}.

NGC 4395 and NGC 5548 showed more complex variations in hardness ratio with increasing
luminosity and were not well fit by a linear model. Both objects show evidence of being very soft at low luminosity
and hardening rapidly with increasing luminosity, then changing to becoming softer-when-brighter at higher luminosity (see Fig. \ref{swbhardness}).
This behaviour is also very similar to that observed in the Seyfert galaxies NGC 1365 and Mkn 335 \citep{Connolly2014, Connolly2015}.

The linear fits to the hardness ratio vs. count rate data showed that 18 of the AGN possess harder-when-brighter behaviour.
In addition to NGC 4395 and NGC 5548, two other sources showed softer-when-brighter behaviour. Two of the sources
(NGC 4486 and NGC 4321) could not be constrained as being either harder- or softer-when-brighter, as  errors in the
gradients of the linear fits to their hardness ratio vs. count rate data do not constrain the gradients to be positive or negative. 
A high fraction (72\%) of the sample therefore show harder-when-brighter behaviour.

\begin{table*}
\footnotesize
\begin{tabular}{l c  c c  c}
\hline
Object & $\mathrm{\chi^2 / DoF}$ & Gradient & Intercept  & Behaviour when brighter \\ 
\hline
NGC 315  		& 0.70 & $26.26^{+11.76}_{-9.39}$ 	& $-0.53^{+0.11}_{-0.11}$ & harder \\ 
NGC 1052 		& 0.54 & $8.95^{+2.33}_{-1.90}$		& $0.14^{+0.07}_{-0.10}$  & harder  \\ 
NGC 1068 		& 0.40 & $2.44^{+0.96}_{-0.56}$ 	& $-0.88^{+0.03}_{-0.06}$ & harder \\
NGC 2655 		& 1.04 & $21.82^{+4.45}_{-4.67}$	& $-0.15^{+0.07}_{-0.07}$ & harder \\ 
NGC 3031 		& 0.80 & $1.10^{+0.13}_{-0.10}$ 	& $-0.54^{+0.01}_{-0.01}$ & harder \\ 
NGC 3147 		& 0.54 & $13.11^{+3.21}_{-3.56}$ 	& $-0.49^{+0.07}_{-0.06}$ & harder \\ 
NGC 3226 		& 0.49 & $40.52^{+17.39}_{-14.40}$	& $-0.24^{+0.08}_{-0.07}$ & harder \\ 
NGC 3227 		& 13.44& $-0.83^{+0.06}_{-0.05}$ 	& $0.34^{+0.03}_{-0.03}$  & softer \\ 
NGC 3628 		& 1.11 & $14.32^{+9.07}_{-7.97}$ 	& $-0.01^{+0.13}_{-0.14}$ & harder \\ 
NGC 3998$^\dagger$	& 2.43 & $1.74^{+1.50}_{-0.67}$ 	& $-0.53^{+0.07}_{-0.11}$ & harder \\ 
NGC 4051 		& 2.22 & $-0.19^{+0.16}_{-0.18}$ 	& $-0.34^{+0.04}_{-0.05}$ & softer \\ 
NGC 4151 		& 1.64 & $0.16^{+0.03}_{-0.03}$ 	& $0.57^{+0.02}_{-0.02}$  & harder \\ 
NGC 4258 		& 1.13 & $3.69^{+3.70}_{-2.67}$ 	& $0.12^{+0.11}_{-0.10}$  & harder \\ 
NGC 4321 		& 2.69 & $25.69^{+30.60}_{-25.80}$	& $-0.46^{+0.05}_{-0.05}$ & undetermined\\ 
NGC 4388 		& 0.76 & $0.73^{+0.27}_{-0.27}$ 	& $0.71^{+0.03}_{-0.04}$  & harder \\ 
NGC 4395 		& 14.27& $-1.75^{+0.18}_{-0.17}$ 	& $0.40^{+0.02}_{-0.02}$  & softer \\ 
NGC 4472 		& 1.78 & $7.81^{+7.95}_{-5.17}$ 	& $-0.91^{+0.05}_{-0.05}$ & harder \\
NGC 4486 		& 0.68 & $0.04$ 			& $-0.68$ 		  & undetermined\\ 
NGC 4579 		& 0.19 & $6.38^{+3.12}_{-2.05}$ 	& $-0.71^{+0.12}_{-0.18}$ & harder \\ 
NGC 4736 		& 0.80 & $10.39^{+4.99}_{-2.88}$ 	& $-0.57^{+0.08}_{-0.11}$ & harder \\ 
NGC 5194 		& 1.28 & $35.59^{+11.66}_{-9.96}$ 	& $-0.80^{+0.04}_{-0.03}$ & harder \\ 
NGC 5548 		& 5.03 & $-0.32^{+0.03}_{-0.02}$ 	& $0.32^{+0.01}_{-0.01}$  & softer \\ 
NGC 5806 		& 0.58 & $105.89^{+25.23}_{-30.11}$ 	& $-0.22^{+0.06}_{-0.04}$ & harder \\ 
NGC 7331 		& 0.86 & $80.45^{+15.10}_{-15.01}$ 	& $-0.46^{+0.06}_{-0.06}$ & harder \\ 
\hline 
\end{tabular}

\caption{ The reduced $\mathrm{\chi^2}$ and best-fit parameters (with one-sigma errors) for
linear fits to the hardness ratio vs. $\mathrm{2 - 10 keV}$ count rate plots of each object (see Fig. \ref{allhardness}).
The behaviour of each source with increasing count rate is always shown,
 determined by whether the slope of the linear fit is positive (softer-when-brighter) or negative (harder-when-brighter).
The behaviour of objects with increasing luminosity is `undetermined' if the errors on the
gradient of the fit cross zero. NGC 4486 possessed too little data for meaningful binning with more than ten bins,
so errors are not given.}

\label{hardnessTable}

\end{table*}

\section{Resolving a Constant Spectral Component}

Whilst the hardness ratio vs. flux plots indicate that there is a division in the behaviour of spectral variation in the AGN sample,
the causes of the spectral variation cannot be discerned from hardness ratio changes alone. If the spectrum is 
a simple absorbed power law, with no or constant absorption, correlated hardness changes imply changes in
the photon index which are correlated with luminosity. However, a hardness ratio correlated with luminosity 
would also be seen if an additional constant soft component is present in the spectrum, or could be produced by variable absorption.
In both of these cases, the photon index of the underlying power law would not be required to vary.

A constant soft spectral component can be produced by background contamination from hot gas due to circumnuclear star formation,
which is seen in some AGN spectra \citep{Nemmen2006,Wang2009,Perez-Olea1996}. 
As the source brightens, the contribution of such a component would become smaller, increasing
the hardness of the overall spectrum without the need for any change in the photon index. 
This soft component must, however, be luminous enough to produce the observed hardness changes.

\begin{figure}
	\includegraphics[width = 8.5cm,clip = true, trim = 0 0 0 0]{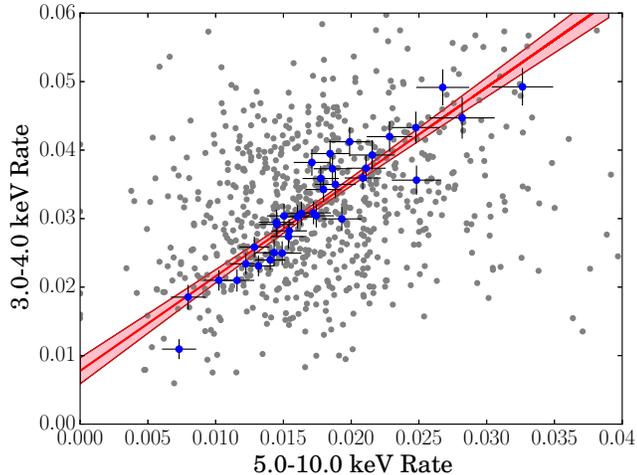}
		
	\caption{An example of a flux-flux diagram for M81, of the 3-4 v. 5-10 keV bands (binned). Blue points are binned data, with a minimum of
	20 points per bin. Grey points shown behind are unbinned data. Error bars show the one-sigma confidence levels. The red line shows the
	best-fitting model, assuming a linear trend, and the pink region bounded by red indicates the one-sigma error region of the model parameters.}
	\label{colour}
\end{figure}

A model-independent method to determine whether a constant spectral component is present, and if so what form it takes, 
is to plot the flux in different X-ray bands against one another \citep[see e.g.][]{Taylor2008}.
If one assumes a linear relationship between two bands, a linear model can be fitted to the data and extrapolated 
to estimate the flux in one band when the other is at zero flux (see Fig. \ref{colour}). 
If it is assumed that any constant component present becomes negligibly small at
high energies, as is reasonable for models of e.g. emission from hot plasma due to star formation, 
a high-energy reference band can be used to find the flux of the lower-energy bands when the variable component is zero.
In this way, the shape of the underlying constant spectral component can be estimated. 
If the spectral shape of any constant component is known, it is then possible
to determine in a model-independent way whether this component is the sole cause of hardness changes with luminosity, 
or whether intrinsic variations in the spectrum are required.

The only harder-when-brighter source for which this method could produce reasonably well-constrained results was M81.
The X-ray count rates in 7 soft energy bands (0.3-0.5, 0.5-1.0, 1.0-1.5, 1.5-2.0, 2-3, 3.-4 and 4-5 keV) were plotted against the
5-10 keV hard energy band. These data were then fitted with a linear model in the same manner as were the hardness ratio vs. flux plots
in Section \ref{hardness}. By assuming that there is no underlying component in the 5-10 keV band, the intercepts
of these linear fits can be taken as the count rates of any soft component in each band.
The bands all showed a weak positive correlation, with an average Pearson correlation coefficient of 0.32.
Fig. \ref{colour} shows an example of a plot of the count rate in two bands in the spectrum of M81, the best-fitting
linear model and the errors in its parameters.

\begin{figure}
	\includegraphics[width = 8.5cm,clip = true, trim = 0 0 0 0]{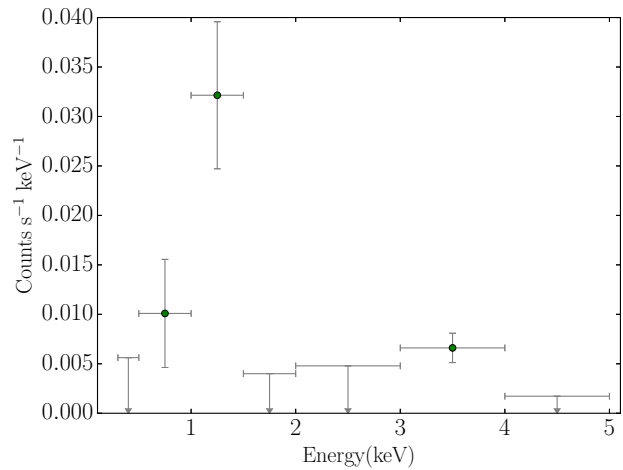}
		
	\caption{The constant, soft component of the spectrum of M81, assuming a linear relationship between the flux in
		 different bands and zero flux in the 5-10 keV band. The data are folded through a typical response
		 for {\it Swift}.}
	\label{softComp}
\end{figure}

The constant spectral component is measured in count space from the instrumental data and must
therefore be deconvolved using the instrumental response for {\it Swift} in the same way as for
any other X-ray spectrum. A typical instrumental response matrix for {\it Swift} was therefore applied
to the derived constant spectrum  during spectral modelling. A significant excess was only found in
3 of the 7 bands, two of these were in the $\mathrm{0.5-1.5 ~keV}$ range; Fig. \ref{softComp} shows
the extrapolated count rate in each band.

This spectrum was fitted with three models - a power law, a hot gas model (mekal) and combination of
the two. None of the models fitted the spectrum well ($\mathrm{ \chi^2_R}$ of 3.31, 3.67 and 4.04, for 6, 6 and 4 degrees of freedom, respectively),
however the power law was the best-fitting model. 
Whilst this model may not be a perfect physical approximation of the constant component, it may still be used to test 
the possibility that such a component is causing the observed spectral variability. The lack of a good spectral fit, 
combined with the lack of a significant excess rate in 4 of the 7 bands may indicate that a constant component is not present
at all, or extremely weak; despite this, the model fitted to the extrapolated spectrum was used in subsequent spectral modelling of M81 in order to
test whether its contribution would be large enough to cause the observed hardness changes, under the assumption that
it may be real.

\section{Spectral Modelling}
\label{spectralFitting}

Enough data existed for spectral fitting of 14 of the 24 AGN at multiple flux levels.
A set of models, in increasing degrees of complexity from a simple power law to a partially covered power law (see Table \ref{models}), 
were successively fitted to the spectra of each source until a good fit was found, using the \textsc{xspec} 12.7 analysis package \citep{Arnaud1996}.
Models were fitted to data in the $\mathrm{0.5-10 keV}$ energy range in all cases except that of the NGC 3998 {\it RXTE} data, for which
the $\mathrm{3-20 keV}$ energy range was used.
In all models, a neutral absorber whose column was fixed at the Galactic column in the direction of the source was included.

Table \ref{models} contains descriptions of all of the spectral models fitted to the spectra. In all cases, the model was fitted simultaneously
to all of the flux-binned spectra for a given object - for some models, some parameters were `tied', meaning they were allow to vary,
but constrained to be the same value for all spectra. In each case, F-tests were used to determine whether the
addition of complexity to a model improved the fit sufficiently to better describe the spectra.

The only models for which only the power law normalisation was allowed to vary were the two-component models, i.e. those containing a power law plus a constant
(hot gas) component or the partial covering model. Only for these models can simple changes in the normalisation of the power law change the hardness ratio. 

The simplest model fitted was a power law with the photon index and normalisation allowed to vary with luminosity (model 1 in Table \ref{models}).
The next simplest set of models consisted of an absorbed power law, including models in which the absorbing column or the photon index were tied between spectra (models 2-4).
The next were a pair of models consisting of an absorbed power law plus a hot gas model, for which the normalisation of the power law was free to vary
and the photon index was either free to vary or tied between spectra;
all parameters of the hot gas model (i.e. temperature and normalisation) were tied, as it was assumed to be constant (models 56). 
Also fitted was a set of models consisting of a power law absorbed by an ionised absorber (absori), in which the normalisation
of the power law and either the photon index, ionisation state, absorbing column or all three were free to vary between spectra,
with the remaining parameters tied (models 8-11). These spectral models were fitted to all 14 objects with flux-binned spectra; the
results of these fits are in Tables \ref{modelfits} and Table \ref{absmodelfits}.

Objects whose spectra were not well fitted by any of these models were also fitted with a
set of models consisting of a power law and a partially-covering absorber (pcfabs) in which, in addition to the normalisation of
the power law, either the photon index, absorbing column or covering fraction, the absorbing column {\it and} the covering fraction,
or all three were free to vary (models 12-17). These objects were also fitted with a set of similar models with the addition
of a hot gas component, whose parameters were tied between spectra (models 18-23). The
results of these fits are in Tables \ref{pcmodelfits} and Table \ref{complexmodelfits}.

In addition, NGC 1068 was fitted with a model consisting of
a power law with a neutral absorber plus a Gaussian, to account for excess emission at $\mathrm{\sim 6.7 keV}$
which is likely to be due to an iron fluorescence line (model 7) (see Section \ref{ngc1068}).
The results of these fits are also in Table \ref{complexmodelfits}.

The data from the remaining 11 AGN, for which there were not enough data to fit multiple flux-binned spectra,
were instead combined to produce a single, total spectrum for each object. These total spectra were
fitted in the same fashion as the flux-binned spectra.
As only one spectrum was fitted in each case, all parameters were free to vary in all models (except the galactic absorbing column and redshift). 
Table \ref{singleFits} shows the results of these fits.

Table \ref{bestFits} summarises the best-fitting spectral model for each of the objects for which
flux-binned spectra could be produced.

\begin{table*}	
  \centering
  \footnotesize 
  \begin{tabular}{l l c c c c c c c}

  \hline 
  N	&	Description						&	Xspec	Description		&	Free Parameters							 \\
  \hline 
  1	&	Simple power law 					&	pow				& $\mathrm{\Gamma}$				\\
  2	&	Power law with neutral absorber 			&	wabs*pow			& $\mathrm{\Gamma}$				\\
  3	&	Power law with neutral absorber  			&	wabs*pow			& $\mathrm{N_H}$			\\
  4	&	Power law with neutral absorber 			&	wabs*pow			& $\mathrm{N_H}$, $\mathrm{\Gamma}$	\\
  5	&	Power law with neutral absorber and hot gas 		&	mekal + (wabs*pow)		& -					\\
  6	&	Power law with neutral absorber and hot gas 		&	mekal + (wabs*pow)		& $\mathrm{\Gamma}$				\\
  7	&	Power law with neutral absorber, Gaussian and hot gas   &	mekal + (wabs*(pow + gauss))	& $\mathrm{\Gamma}$				\\
  8	&	Power law with ionised absorber 			&	absori*pow			& $\mathrm{\Gamma}$	\\
  9	&	Power law with ionised absorber 			&	absori*pow			& $\mathrm{N_H}$	\\
  10	&	Power law with ionised absorber 			&	absori*pow			& $\mathrm{\xi}$	\\
  11	&	Power law with ionised absorber 			&	absori*pow			& $\mathrm{N_H}$, $\mathrm{\xi}$, $\mathrm{\Gamma}$	\\
  12	&	Partially covered power law 				&	pcfabs*pow			& -				\\
  13	&	Partially covered power law 				&	pcfabs*pow			& $\mathrm{\Gamma}$				\\
  14	&	Partially covered power law 				&	pcfabs*pow			& $\mathrm{N_H}$			\\
  15	&	Partially covered power law 				&	pcfabs*pow			& $\mathrm{CF}$	\\
  16	&	Partially covered power law 				&	pcfabs*pow			& $\mathrm{N_H}$, $\mathrm{CF}$	\\
  17	&	Partially covered power law 				&	pcfabs*pow			& $\mathrm{N_H}$, $\mathrm{CF}$, $\mathrm{\Gamma}$ 	\\
  18	&	Partially covered power law and hot gas			&	mekal + pcfabs*pow		& -	\\
  19	&	Partially covered power law and hot gas			&	mekal + pcfabs*pow		& $\mathrm{\Gamma}$	\\
  20	&	Partially covered power law and hot gas			&	mekal + pcfabs*pow		& $\mathrm{N_H}$	\\
  21	&	Partially covered power law and hot gas			&	mekal + pcfabs*pow		& $\mathrm{CF}$	\\
  22	&	Partially covered power law and hot gas			&	mekal + pcfabs*pow		& $\mathrm{N_H}$, $\mathrm{CF}$	\\
  23	&	Partially covered power law and hot gas			&	mekal + pcfabs*pow		& $\mathrm{N_H}$, $\mathrm{CF}$, $\mathrm{\Gamma}$	\\

  \hline 
\end{tabular}
		
	\caption{A description of the series of models applied successively to the spectra of each AGN. All models also included the `wabs' model of neutral absorption set to the Galactic absorption column
		  in the direction of each AGN. The free parameters, which can vary between spectra, are shown as symbols: $\mathrm{\Gamma}$ (photon index), $\mathrm{N_H}$ (absorbing column), 
		  $\mathrm{\xi}$ (ionisation parameter) and CF (covering fraction). In all cases the normalisations of the power law was free to vary.  In models involving a mekal component, its normalisation 
		  was always tied to be the same value in all spectra of a given source.
		  }
	\label{models}
\end{table*}

\begin{table*}	

   \centering
  \footnotesize 
  \begin{tabular}{l | c c | c c c c c c | c c c c  | c c }

    Xspec Model	& \multicolumn{2}{c |}{pow}	&		 \multicolumn{6}{c| }{	wabs*pow }						& \multicolumn{4}{c |}{	wabs*pow +mekal	}			& \multicolumn{2}{c }{ wabs*(pow + gauss) }	\\
		& \multicolumn{2}{c |}{}	&		 \multicolumn{6}{c| }{	 }							& \multicolumn{4}{c |}{			}			& \multicolumn{2}{c }{ + mekal }	\\
  \hline 
   Model No.	&	\multicolumn{2}{c |}{1}	& 	\multicolumn{2}{c }{2}	&\multicolumn{2}{c }{3}		&\multicolumn{2}{c |}{4}	&\multicolumn{2}{c }{5}		&\multicolumn{2}{c |}{6}		&	\multicolumn{2}{c }{7}\\	
Free Parameters &\multicolumn{2}{c |}{$\mathrm{\Gamma}$}&\multicolumn{2}{c }{$\mathrm{\Gamma}$}&\multicolumn{2}{c }{$\mathrm{N_H}$}&\multicolumn{2}{c |}{$\mathrm{\Gamma,N_H}$}&\multicolumn{2}{c }{$\mathrm{-}$}&\multicolumn{2}{c |}{$\mathrm{\Gamma}$}&\multicolumn{2}{c }{$\mathrm{\Gamma}$}\\					
		&$\chi^2_R$	&	DoF	&$\chi^2_R$	&	DoF	&$\chi^2_R$	&	DoF	&$\chi^2_R$	&	DoF	&$\chi^2_R$	&	DoF	&$\chi^2_R$	&	DoF	&	$\chi^2_R$	&	DoF	\\
    \hline 
  NGC 1052	&	2.09	&	162	&	2.11	&	161	&	2.17	&	161	&	2.13	&	159	&	1.42	&	161	&	1.41	&	159	&			&		\\
  NGC 1068	&	2.69	&	436	&	2.58	&	435	&	2.58	&	435	&	2.59	&	431	&	1.62	&	437	&	1.61	&	433	&{\bf 1.0743}	&{\bf 430}	\\
  NGC 3031	&	1.01	&	5902	&	1.01	&	5901	&	1.03	&	5901	&	1.01	&	5874	&	1.01	&	5927	&{\bf 0.98}&{\bf 5899}	&			&		\\
  NGC 3147	&	0.73	&	125	&{\bf 0.70}	&{\bf 124}	&	0.72	&	124	&	0.70	&	122	&	0.73	&	124	&	0.71	&	122	&			&		\\
  NGC 3227	&	1.17	&	1094	&	1.10	&	1093	&	1.57	&	1093	&	1.09	&	1082	&	1.52	&	1102	&	1.08	&	1091	&			&		\\
  NGC 3998	&{\bf1.40}	&{\bf 225}	&	1.39	&	224	&	1.69	&	224	&	1.42	&	200	&	1.63	&	271	&      1.42	&	247	&			&		\\
  NGC 4051	&	2.04	&	784	&	2.05	&	783	&	2.07	&	783	&	2.11	&	774	&	1.25	&	790	&{\bf 1.10}&{\bf 781}	&			&		\\
  NGC 4151	&	2.11	&	703	&	2.11	&	702	&	2.62	&	702	&	2.13	&	696	&	1.29	&	707	&	1.31	&	695	&			&		\\
  NGC 4388	&	1.92	&	260	&	1.93	&	259	&	1.92	&	259	&	1.94	&	256	&{\bf 1.17}&{\bf 261}	&	1.17	&	258	&			&		\\
  NGC 4395	&	1.15	&	1373	&	1.15	&	1372	&	1.30	&	1372	&	1.17	&	1351	&	1.31	&	1392	&	1.15	&	1371	&			&		\\
  NGC 4486	&	3.01	&	261	&	1.83	&	260	&	1.82	&	260	&	1.83	&	257	&{\bf 1.02}&{\bf 261}	&	1.02	&	258	&			&		\\
  NGC 4579  	&	1.09	&	201	&	1.10	&	200	&	1.09	&	200	&	1.11	&	198	&{\bf 1.00}&{\bf 200}	&	1.00	&	198	&			&		\\
  NGC 5548	&	1.66	&	8659	&	1.46	&	8658	&	1.43	&	8658	&	1.43	&	8617	&	1.15	&	8697	&	1.11	&	8656	&			&		\\
  
  \hline 
\end{tabular}
	
	\caption{ The reduced $\chi^2$ values, $\chi^2_R$, and number of degrees of freedom (DoF) of the best fit with models 1-7 to each of the sources.
			In each case, the free parameters are indicated by $\mathrm{\Gamma}$ (photon index) and $\mathrm{N_H}$ (absorbing column). 
			The best fitting model for each source is highlighted in bold.}
	\label{modelfits}

\end{table*} 

\begin{table*}	

   \centering
  \footnotesize 
  \begin{tabular}{l | c c c c c c c c c}

    Xspec Model	& \multicolumn{8}{c }{absori*pow}	&\\
  \hline 
   Model No.	&	\multicolumn{2}{c }{8}	& 	\multicolumn{2}{c }{9}	&\multicolumn{2}{c }{10}		&\multicolumn{2}{c }{11}	\\	
Free Parameters &\multicolumn{2}{c }{$\mathrm{\Gamma}$}&\multicolumn{2}{c }{$\mathrm{\xi}$}&\multicolumn{2}{c }{$\mathrm{N_H}$}&\multicolumn{2}{c }{$\mathrm{\Gamma,\xi,N_H}$}\\					
		&$\chi^2_R$	&	DoF	&$\chi^2_R$	&	DoF	&$\chi^2_R$	&	DoF	&$\chi^2_R$	&	DoF	\\
    \hline 
  NGC 1052	&	1.42	&	159	&	1.51	&	159	&	1.85	&	159	&	1.48	&	155	\\
  NGC 1068	&	2.37	&	433	&	2.59	&	433	&	2.59	&	433	&	2.56	&	425	\\
  NGC 3147	&	0.71 	&	122	&	0.72 &	122	&	0.73 &	122	&	0.72	&	118 \\
  NGC 3031	&	1.01	&	5900	&	1.03	&	5900	&	1.03	&	5900	&	1.02	&	5846	\\
  NGC 3227	&{\bf 1.05}&{\bf 1091}&	1.27	&	1091	&	1.31	&	1091&	1.17 &	1069	\\
  NGC 3998	&	1.47	&	247	&	1.64	&	247	&	1.82	&	247	&	1.78	&	199	\\
  NGC 4051	&	1.15	&	781	&	1.14	&	781	&	1.13	&	781	&	1.11	&	763	\\
  NGC 4151	&	2.02	&	701	&	1.78	&	701	&	1.96	&	701	&	1.16	&	689	\\
  NGC 4388	&	2.38	&	258	&	1.92	&	258	&	1.93	&	258	&	1.66	&	252	\\
  NGC 4395	&{\bf 1.03}&{\bf 1370}&	1.08	&	1370	&	1.06	&	1370	&	1.01	&	1328	\\
  NGC 4486	&	1.82	&	259	&	1.81	&	259	&	1.84	&	259	&	1.86	&	253	\\
  NGC 4579  	&	1.09	&	198	&	1.18	&	198	&	1.08	&	198	&	1.08	&	194	\\
  NGC 5548	&	0.96	&	8656	&	0.95	&	8656	&	0.96	&	8656	&{\bf 0.94}&{\bf 8574}	\\
  
  \hline 
\end{tabular}
	
	\caption{The reduced $\chi^2$ values, $\chi^2_R$, and number of degrees of freedom (DoF) of the best fit with models 8-11 to each of the sources.
			In each case, the free parameters are indicated by $\mathrm{\Gamma}$ (photon index), $\mathrm{N_H}$ (absorbing column) and $\mathrm{\xi}$  (ionisation state).
			The best fitting model for each source is highlighted in bold. }
	\label{absmodelfits}

\end{table*} 

\begin{table*}	

   \centering
  \footnotesize 
  \begin{tabular}{l |  p{0.3cm} p{0.45cm} p{0.3cm} p{0.45cm}p{0.3cm} p{0.45cm}p{0.3cm} p{0.45cm}p{0.3cm} p{0.45cm}p{0.3cm} p{0.45cm} }

    Xspec Model	&  	\multicolumn{12}{c }{pcfabs*pow}																			\\  
  \hline 
   Model No.	&\multicolumn{2}{c }{12}	&\multicolumn{2}{c }{13}	&\multicolumn{2}{c }{14}	&\multicolumn{2}{c }{15}	&\multicolumn{2}{c}{16}		&\multicolumn{2}{c}{17}	\\	
Free Parameters &\multicolumn{2}{c }{$\mathrm{-}$}&\multicolumn{2}{c }{$\mathrm{\Gamma}$}&\multicolumn{2}{c }{$\mathrm{N_H}$}&\multicolumn{2}{c }{$\mathrm{CF}$}&\multicolumn{2}{c }{$\mathrm{N_H, CF}$}&\multicolumn{2}{c }{$\mathrm{\Gamma, N_H, CF}$}\\					
		&$\chi^2_R$	&	DoF	&$\chi^2_R$	&	DoF	&$\chi^2_R$	&	DoF	&$\chi^2_R$	&	DoF	&$\chi^2_R$&	DoF	&$\chi^2_R$	&	DoF	\\
    \hline 
  NGC 1052	&	1.11	&	162	&{\bf 1.01}	&{\bf 160}	&	1.08	&	160	&	1.03	&	160	&	1.00	&	158	&	0.99	&	156	\\
 NGC 1068	&	2.24	&	438	&	2.24	&	434	&	2.25	&	434	&	2.26	&	434	&	2.27	&	430	&	2.28	&	426	\\
 NGC 4151	&	1.23	&	708	&	1.20	&	702	&	1.22	&	702	&	1.16	&	702	&	1.15	&	696	&	1.13	&	690	\\

  \hline 
\end{tabular}
	
	\caption{The reduced $\chi^2$ values, $\chi^2_R$, and number of degrees of freedom (DoF) of the best fit with models 12-17 to the sources not well fit by models 1-11.
			In each case, the free parameters are indicated by $\mathrm{\Gamma}$ (photon index), $\mathrm{N_H}$ (absorbing column) and $\mathrm{CF}$  (covering fraction). 
			The best fitting model for each source is highlighted in bold.}
	\label{pcmodelfits}

\end{table*} 

\begin{table*}	

   \centering
  \footnotesize 
  \begin{tabular}{l |  p{0.3cm} p{0.45cm} p{0.3cm} p{0.45cm}p{0.3cm} p{0.45cm}p{0.3cm} p{0.45cm}p{0.3cm} p{0.45cm}p{0.3cm} p{0.45cm}   }

    Xspec Model	&	\multicolumn{12}{c |}{pcfabs*pow + mekal} 																			\\  
  \hline 
   Model No.	&\multicolumn{2}{c }{18}	&\multicolumn{2}{c }{19}	&\multicolumn{2}{c }{20}	&\multicolumn{2}{c }{21}	&\multicolumn{2}{c}{22}		&\multicolumn{2}{c}{23}	\\	
Free Parameters &\multicolumn{2}{c }{$\mathrm{-}$}&\multicolumn{2}{c }{$\mathrm{\Gamma}$}&\multicolumn{2}{c }{$\mathrm{N_H}$}&\multicolumn{2}{c }{$\mathrm{CF}$}&\multicolumn{2}{c }{$\mathrm{N_H, CF}$}&\multicolumn{2}{c }{$\mathrm{\Gamma, N_H, CF}$}\\					
		&$\chi^2_R$	&	DoF	&$\chi^2_R$	&	DoF	&$\chi^2_R$	&	DoF	&$\chi^2_R$	&	DoF	&$\chi^2_R$	&	DoF	&$\chi^2_R$	&	DoF		\\
    \hline 
  NGC 1068	&	1.12	&	436	&	1.10	&	432	&	1.11	&	432	&	1.12	&	432	&	1.12	&	428	&	1.12	&	424		\\
  NGC 4151	&	1.11	&	706	&	1.11	&	700	&	1.12	&	700	&	1.07	&	700	&	1.06	&	694	&{\bf 1.03}	&{\bf 688}			\\
 
  \hline 
\end{tabular}
	
	\caption{The reduced $\chi^2$ values, $\chi^2_R$, and number of degrees of freedom (DoF) of the best fit with models 18-23 to the sources not well fit by models 1-17.
			In each case, the free parameters are indicated by $\mathrm{\Gamma}$ (photon index), $\mathrm{N_H}$ (absorbing column) and $\mathrm{CF}$  (covering fraction).
			The best fitting model for each source is highlighted in bold.}
	\label{complexmodelfits}

\end{table*} 

\subsection{M81 (NGC 3031)}

\begin{figure}
	\includegraphics[width = 8.5cm,clip = true, trim = 0 0 0 0]{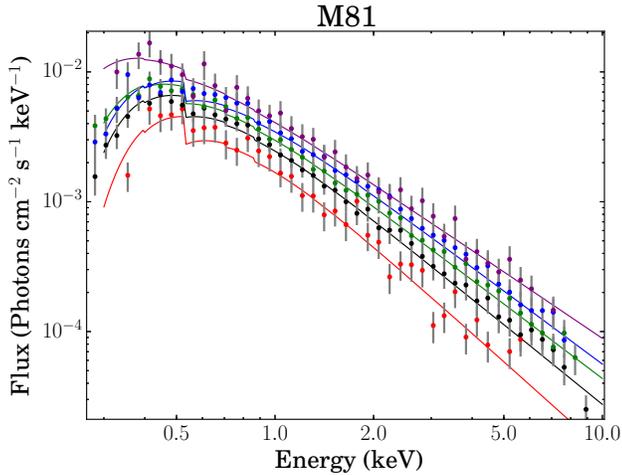}
	\caption{A sample of the flux-binned spectra of M81, fitted with the best-fitting spectral model (see Table \ref{modelfits}). Error bars show the one-sigma confidence levels.}
	\label{m81spectra}
\end{figure}

As the most extensively observed object to show harder-when-brighter behaviour, the spectra of M81 were studied in the greatest
detail. The soft component of M81 in particular, extrapolated from its colour-colour plots (see section \ref{colour}), was derived in order to be able to
draw conclusions about similar sources in the sample. 
Fig. \ref{m81spectra} shows a sample of the spectra from M81, covering the full observed flux range. 

The X-ray spectrum of M81 is well described by a simple power law absorbed by the Galactic absorbing column of $\mathrm{7.2 \times 10^{21} cm^{-2}}$. This is consistent
with the many previous shorter-term X-ray studies  of M81 \citep[e.g.][]{Elvis1982,Swartz2003,Markoff2008}; past studies discovered narrow features
attributed to high-ionisation emission lines including Fe lines \citep[e.g.][]{Ishisaki1996,Pellegrini2000,Page2003,Page2004}, but these are not detectable in the {\it
Swift} data.

The variability in the spectral data is well described by changes in $\Gamma$ alone; no extra spectral components other than the Galactic absorption
in the direction of M81 are required.  If the spectrum is assumed to be absorbed by an additional neutral absorber, the variability in the spectral data is equally well described by changes in $\Gamma$ 
with the absorbing column fixed or by changes in the absorbing column with a fixed $\Gamma$. As these models require an extra component and do not provide a better fit,
it is, however, unlikely to be a better description of the data. 

If $\Gamma$ is held constant, we require that the absorbing column decreases as the luminosity rises. In the case of NGC 1365 \citep{Connolly2014}, 
which has complex luminosity-dependent absorption, 
greatly in excess of the Galactic column in its direction, such a model is readily explained in terms of a luminosity-dependent wind. However, for M81 there is no evidence in the 
total summed spectrum for any absorption in excess of the Galactic column. No previous spectral studies have found  evidence for such an excess column. 
As the model with variable absorption requires an additional variable absorbing column in addition to the fixed Galactic column, we consider it less likely than the 
model of variable $\Gamma$, which has one parameter fewer.

Spectral models including the constant component extrapolated from the colour-colour plots (see section \ref{colour}) were also tested; the component
was found not to make a significant enough contribution to the spectrum to account for the variability. The inclusion of this constant component in 
a model consisting of a power law and a neutral absorber, with a fixed photon index, gave significantly worse fit ($\mathrm{\chi^2_R = 1.15}$).

In addition to the constant component extrapolated from the colour-colour plots, we also separately test a model consisting of a constant hot gas (mekal) component 
together with a power law with varying normalisation but fixed photon index and fixed absorption. In this model, the parameters of the hot gas component are tied to 
be the same between spectra, but allowed to vary, in order to test whether any significant constant soft component can be found in this way.
This model fits the data approximately as well as the model of an absorbed power law with varying photon index. As the photon index is free to vary between flux-binned spectra in the 
model without the hot gas component, it has more free parameters (i.e. is more complex). 
However an F-test shows that the simple power law with variable $\Gamma$ is preferred to the constant-$\Gamma$ model with a hot component
(f = 2.26, p = $\mathrm{3.30 \times 10^{-4}}$). 
Similarly, if the model with a hot gas component in which the photon
index is constant is compared to the same model but with the photon index allowed to vary between flux-binned spectra, 
an f-test shows that the  model in which $\Gamma$ is variable is preferred (f = 7.78, p = $\mathrm{1.27 \times 10^{-29}}$)
and the varying-photon index model is therefore overwhelmingly preferred; this model is also statistically preferable to the simple absorbed power law, despite
its higher level of complexity (f = 76.0, p = $\mathrm{2.48 \times 10^{-33}}$). The integrated flux of the extrapolated constant component is within a factor of 2 of 
that of the hot gas component in the best-fitting model of the full spectrum, and therefore likely to be the same component, though perhaps not well recovered.
The data therefore require a constant hot gas component, but it does not contribute to the spectrum enough to explain the overall spectral variability. 
It is therefore most likely that the observed spectral variability is driven by changes to $\Gamma$.

The top-left panel in Fig. \ref{allGamma} shows how $\Gamma$ changes with increasing flux for the best-fitting model to M81 (model 6). 
$\Gamma$ shows a clear anticorrelation with the luminosity of the source, well fitted by a linear model (see Table \ref{gammaFit}). Significantly, the same anticorrelation is seen 
in all models in which $\Gamma$ is free to vary, regardless of the presence of a hot gas component and whether or not the absorbing column was also left free to vary,
strengthening still more the evidence that intrinsic variability of the photon index is required to explain the spectral variability of M81.
This anticorrelation between $\Gamma$ and luminosity is therefore found to be the cause of the observed harder-when-brighter behaviour in M81.

\subsection{The Palomar {\it \bf Swift} Sample of AGN}

Details on the spectral fitting of individual AGN in the sample are given in Appendix \ref{individual},
along with examples of the spectra (Figs. \ref{totalSpectra}, \ref{fluxSpectra}).

\subsubsection{Flux-Binned Spectral Modelling}

A simple power law, either not absorbed, absorbed by a neutral absorber, or absorbed by an ionised absorber, fitted the spectra of 4 of the 13
flux-binned sources well; in all 4 cases, a varying photon index was required to account for spectral variability.
Two of the sources were, however, better fit when the absorbing column and ionisation state of the absorber were
also allowed to vary.

The addition of a constant hot gas component to the absorbed power law was required in 6 of the 13 flux-binned sources (including M81). 
The spectral variability  of only 1 of these 6 sourced can only be explained by variability in the normalisation of a power law with constant $\Gamma$ 
together with this constant hot gas component. The other 5 sources required $\Gamma$ to also be free to vary. 
Allowing the absorber to vary (i.e. absorbing column and ionisation state where the absorber was ionised)
did not improve the fit in any of these sources.

A more complex model was required for a good fit to the remaining 3 of the 13 flux-binned objects -  NGC 1052, NGC 4151 and NGC 4388. 
NGC 1052 and NGC 4151 both required a partial
covering absorber; Allowing the photon index to vary did not improve the fit in this model for NGC 1052, whilst
in NGC 4151 a varying photon index was required, but a varying absorbing column and covering fraction were also
needed. NGC 4388 required a two-component model consisting of two power laws, one of which is absorbed by
a neutral absorber. The normalisations of the two power laws were correlated in this model, implying
both originate from the central source. In this model, allowing the photon index to vary did not improve the fit.

The results of the fits to the spectra of NGC 1068, NGC 4151, NGC 4388 were not included in subsequent analysis, as
they were deemed unphysical (see Appendix \ref{individual}).

Fig. \ref{allGamma} shows plots of the value of $\Gamma$ as a function of flux of the 9 objects observed by {\it Swift} whose best-fitting models required
variations in  $\Gamma$ (including M81), plus NGC 7213 for comparison. The data are fitted with linear models in the same manner as were the hardness data described in Section \ref{hardness}.
The results of these fits are shown in Table \ref{gammaFit}.
Of these 10 objects, 9 show a clear trend of changing $\Gamma$ with luminosity.
6 of the objects showed a negative correlation, 3 showed a positive correlation and 1 shows no clear correlation. 
The best linear fits to NGC 3998 and NGC 5548 (harder-when-brighter), NGC 3227, NGC 4051 and NGC 4395 (softer-when-brighter) all have high $\mathrm{\chi^2_R}$ values,
due to the larger degree of scatter in the measured values of $\Gamma$ for these systems. The Pearson correlation coefficients
for these systems still show them to be fairly strongly correlated, however.
All of the objects' behaviours are consistent with that implied by their hardness ratio vs. flux plots, except for NGC 5548;
however the behaviour of its hardness ratio with changing flux is not as simple as the others (as described above) and spectral modelling shows that it possesses complex, variable absorption \citep{Kaastra2014}. 

The data therefore imply that a significant fraction of objects which displayed harder-when-brighter behaviour 
do show hardening of their intrinsic spectra, i.e. a decrease in photon index, $\Gamma$, with increasing luminosity. All of the objects showing softer-when brighter
showed softening of their intrinsic spectra (i.e. an increase in $\Gamma$) with increasing luminosity.
One can therefore conclude that the behaviour of the hardness ratio with changing luminosity is a reasonable predictor of the behaviour of the spectral slope of the underlying power law,
and that it is likely that a similar fraction of the remaining objects which showed harder-when-brighter behaviour in their hardness ratio vs. count rate plots,,
whose data did not allow spectral fitting, would also possess intrinsic hardening.

\subsubsection{Total Spectral Modelling}

\begin{table*}
\footnotesize
	\begin{tabular}{l | c c c c c c c c c c c c c c c}
\hline
Xspec Model	&	\multicolumn{2}{c }{pow} 	& \multicolumn{2}{c }{wabs*pow}	&  \multicolumn{2}{c }{absori*pow}&  \multicolumn{2}{c }{wabs*pow}& \multicolumn{2}{c }{pcfabs*pow}& \multicolumn{2}{c }{pcfabs*pow} & \multicolumn{2}{c }{wabs*(pow+gauss)} \\	
		 &	\multicolumn{2}{c }{} 		& \multicolumn{2}{c }{}	&  \multicolumn{2}{c }{}&  \multicolumn{2}{c }{ + mekal}& \multicolumn{2}{c }{}& \multicolumn{2}{c }{+ mekal} & \multicolumn{2}{c }{ + mekal} \\	
\hline
		&	 $\chi^2_R$	&	DoF	& $\chi^2_R$	& DoF	&  $\chi^2_R$	& DoF	&  $\chi^2_R$	& DoF	& $\chi^2_R$	& DoF 	& $\chi^2_R$	& DoF & $\chi^2_R$	& DoF 	\\	
\hline
NGC 315		&	 3.80		&	26	 &	3.28	&  26	 & 	4.92	& 24	&  {\bf 1.02}	& {\bf 24}	& 		&       &		&    	& 		&	\\
NGC 2655	&	 10.37		&	47	 &	9.76	&  46	 & 	4.65	& 44	& 2.51		& 44	& 3.36		& 45 	&{\bf 1.15}		& {\bf 43 } 	&		&	\\		
NGC 3226	&	 2.64		&	31	 &	{\bf 0.90}	&  {\bf 30}	 & 	0.97	& 28	&  		& 	& 		&       &		&    	& 		&	\\	
NGC 3628	&	 2.66		&	16	 &	{\bf 0.88}	&  {\bf 15}	 & 	1.02	& 13	&  		& 	& 		&       &		&    	& 		&	\\
NGC 4258	&	 6.77		&	56	 &	6.90	&  55	 & 	3.27	& 52	&  2.70		& 53	& 2.91		& 54	& {\bf 1.03}		& {\bf 52}  	&		&	\\
NGC 4321	&	 3.71		&	11	 &	2.56	&  10	 & 	5.04	& 8	&  {\bf 1.62}		& {\bf 7}	& 		&       &		&    	& 		&	\\			
NGC 4472	&	13.36		&	46	&	4.84	& 45	& 	6.37	& 45	&  {\bf 1.03}		& {\bf 43}    &		&	&		&	&		& 	\\ 
NGC 4736	& 	1.69		&	82	&	1.55	& 81	& 	1.75	& 79	& {\bf 0.97	}	& {\bf 79}	& 		&	& 		&	&		&	\\
NGC 5194	&	 12.25		&	134 	&	7.13	&  133   & 	12.27	& 131	&  1.80		& 131	& 9.75		& 132   & 1.46		& 131 	&  {\bf 1.18 }   	& {\bf 127}	\\
NGC 5806	&	 1.27		&	16	 &	{\bf 1.16}	&  {\bf 15}	 & 	1.45	& 13	&  		& 	& 		&       &		&    	& 		&	\\
NGC 7331	&	 2.63		&	26	 &	2.69	&  25	 & 	2.79	& 23	&  {\bf 0.83}		& {\bf 23}	& 		&       &		&    	& 		&	\\	
\hline
	\end{tabular}
		
 	\caption{Results of spectral fitting to the total spectra of those objects with too little data for flux binning. 
 	All models included a neutral absorber fixed at Galactic absorbing column in that direction.}

	\label{singleFits}
\end{table*}

Spectral fitting was also carried out on the total spectra of the 11 objects for which there was sufficient data to produce 
plots of hardness ratio vs. luminosity, but not sufficient data for flux-binned spectral fitting. 
All of the available good-quality data on each of these objects was summed and fitted in the
same manner as for the flux-binned spectra. The results of this spectral analysis are presented in Table \ref{singleFits}.
Of these 11 objects, 7 possessed evidence for either complex absorption, i.e. variable partial-covering absorbers, or strong contamination, e.g. from hot gas. 
The remaining 4 objects possessed spectra which can be described by simple power laws, with evidence
of absorption in 3 of the objects. These four objects also show harder-when-brighter behaviour in the 
plots of hardness ratio vs. luminosity (Fig. \ref{allhardness}). The spectral modelling therefore shows the harder-when-brighter
behaviour of these objects due to changes in their photon indices and/or absorbing columns which are correlated with luminosity.

The harder-when-brighter behaviour seen in 6 of the 7 objects with more complex spectra could
be caused by variations in parameters other than $\gamma$, i.e. the absorption, ionisation state and/or covering fraction. NGC 4321, whose hardness ratio behaviour with increasing
luminosity did not follow any simple pattern, has a complex spectrum which could not be well fitted by any of the models.
As we have limited spectral data in all of these cases, we cannot be sure of the cause of their behaviour without further data to allow multi-epoch or flux-binned spectral modelling.

\begin{table*}	

   \centering
  \footnotesize 
  \begin{tabular}{l c c c c c c }

  Source	& Best-Fitting   & \multicolumn{5}{c }{Best-Fitting Parameters}										\\
  		&Spectral Model&$\mathrm{N_H (10^{22} cm^{-2})}$	& 	$\mathrm{\Gamma}$ 	& 	kT (keV)	& 	CF		 &	$\Xi$			\\
  \hline 
  		
NGC 315		&	1	&	0.78			&	1.94			&	0.56		&	-		&	-			\\
NGC 1052	&	13	&	8.60			&	$1.37-1.70$	&	-		&	0.91		&	-			\\
NGC 1068	&	7	&	55.7			&	$2.90-3.06$	&	0.67		&	0.95		&	-			\\
NGC 2655	&	23	&	20.5			&	1.35			&	0.57		&	0.95		&	-			\\
NGC 3031	&	6	&	0			&	1.69-2.05		&	0.55		&	-		&	-			\\
NGC 3147	&	2	&	-			&	$1.35-1.58$	&	-		&	-		&	-			\\
NGC 3226	&	4	&	0.54			&	2.57			&	-		&	-		&	-			\\
NGC 3227	&	8	&	0.13			&	$1.43-1.65$	&	-		&	-		&	0.045		\\
NGC 3628	&	4	&	0.35			&	1.36			&	-		&	-		&	-			\\
NGC 3998	&	1	&	-			&	$1.87-2.50$	&	-		&	-		&	-			\\
NGC 4051	&	6	&	0			&	$1.35-1.97$	&	0.20		&	-		&	-			\\	
NGC 4151	&	23	&	$3.01-5.07$	&	$0.76-1.20$	&	0.23		&$0.76-0.95$	&	-			\\
NGC 4258	&	23	&	9.89			&	1.66			&	0.60		&	0.95		&	-			\\
NGC 4321	&	6	&	1.51			&	4.46			&	0.43		&	-		&	-			\\
NGC 4388	&	5	&	26.3			&	$1.05-1.24$	&	0.62		&	-		&	-			\\
NGC 4395	&	8	&	2.17			&	$0.91-1.60$	&	-		&	-		&	215.16		\\
NGC 4472	&	6	&	0.81			&	3.83			&	0.74		&	-		&	-			\\
NGC 4486	&	5	&	0.45			&	2.68			&	1.52		&	-		&	-			\\
NGC 4579	&	5	&	0.01			&	1.67			&	0.52		&	-		&	-			\\
NGC 4736	&	6	&	0.055		&	1.48			&	0.35		&	-		&	-			\\
NGC 5194	&	7	&	0			&	0.7			&	0.60		&	-		&	-			\\
NGC 5806	&	4	&	0.057		&	1.65			&	-		&	-		&	-			\\
NGC 5548	&	11	&	$0.64-1.43$ 	&	$1.09-1.40$	&	-		&	-		&	$7.46-35.88$	\\
NGC 7331	&	6	&	0.075		&	1.37			&	0.37		&	-		&	-			\\
  \hline 
\end{tabular}
\caption{The best-fitting spectral model for each of the sources,
	  and the values of the parameters which were free to vary (excluding normalisations, which were always free but are not included).
	  See Table \ref{models} for a description of each model.}
\label{bestFits}
\end{table*}

\begin{figure*}
	\includegraphics[width = 18cm,clip = true, trim = 0 0 0 0]{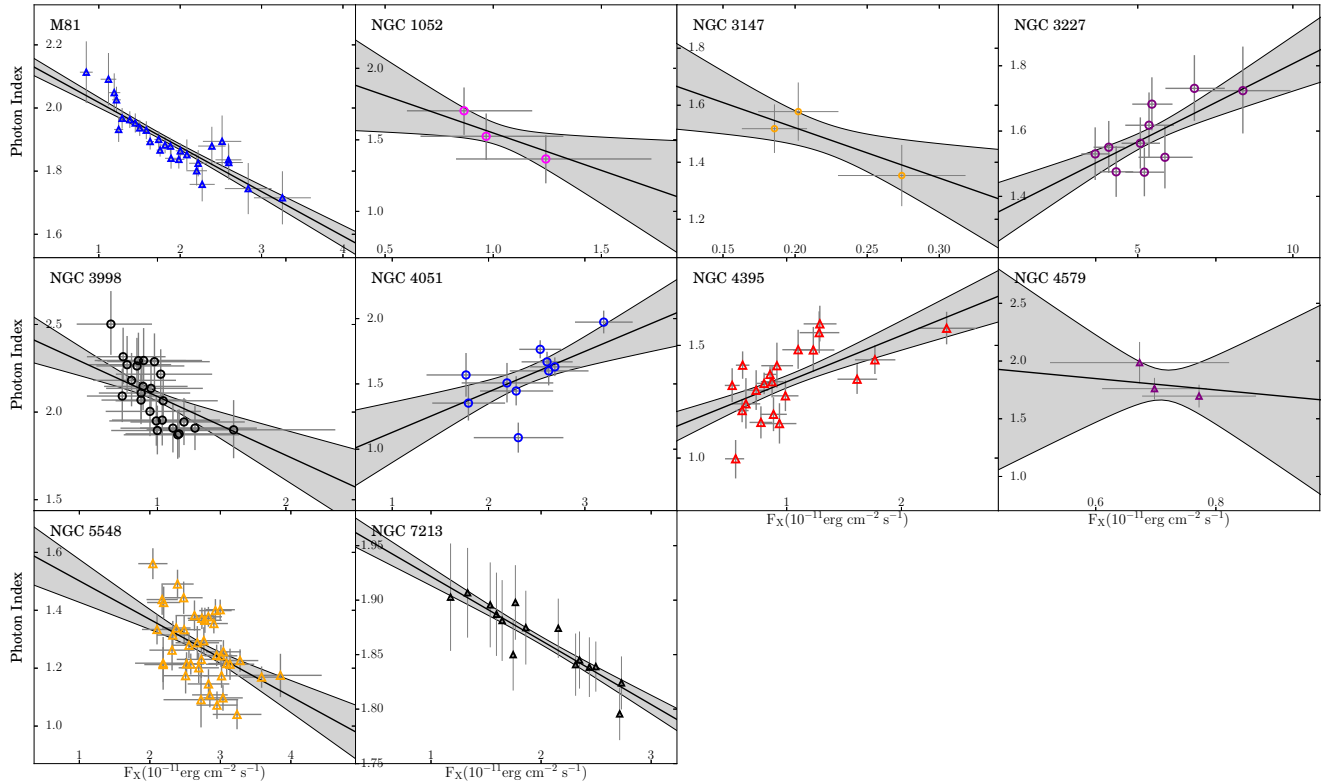}

	\caption{Plots of the photon index $\Gamma$ against the X-ray flux of each of the sources for which flux-binned spectral fitting
	was carried out, using the best-fitting model in each case. The $\mathrm{0.5-10 keV}$ flux is used in all cases except for
	NGC 3998, for which the $\mathrm{3-20 keV}$ {\it RXTE} flux is used.
	 Error bars show the one-sigma confidence levels. The data are fitted with linear models; the best-fitting model is shown in black
	 and the one-sigma error regions of each fit are shown in grey.}
	\label{allGamma}
\end{figure*}

\begin{table*}
\footnotesize
\begin{tabular}{l c  c c  c c}
\hline
Object & $\mathrm{\chi^2 / DoF}$ & Gradient & Intercept  & Pearson coefficient & Behaviour when brighter \\ 
\hline
M81       &  0.95	 & $-0.14^{+0.02}_{-0.024}$	& $2.16^{+0.04}_{-0.05}$ &	-0.89	& harder \\
NGC 1052  &  1.70	 & $-0.52^{+0.36}_{-0.53}$	& $2.07^{+0.52}_{-0.37}$ &	-0.96	& harder \\
NGC 3147  &  0.58	 & $-1.77^{+1.46}_{-1.94}$	& $1.87^{+0.44}_{-0.31}$ &	-0.91	& harder \\
NGC 3227  &  2.81	 & $0.048^{+0.02}_{-0.01}$	& $1.33^{+0.07}_{-0.08}$ &	0.73	& softer \\
NGC 3998  &  3.11	 & $-0.34^{+0.10}_{-0.11}$	& $2.42^{+0.10}_{-0.09}$ &	-0.74	& harder \\
NGC 4051  &  13.22	 & $0.31^{+0.09}_{-0.09}$	& $0.82^{+0.21}_{-0.22}$ &	0.65	& softer \\
NGC 4395  &  10.21	 & $0.21^{+0.02}_{-0.02}$	& $1.13^{+0.02}_{-0.02}$ &	0.63	& softer \\
NGC 4579  &  1.59	 & $-0.49^{+1.52}_{-1.48}$	& $2.14^{+1.07}_{-1.16}$ &	-0.83	& undetermined \\
NGC 5548  &  19.46	 & $-0.13^{+0.02}_{-0.02}$	& $1.63^{+0.06}_{-0.05}$ &	-0.50	& harder \\
NGC 7213  &  0.23	 & $-0.06^{+0.02}_{-0.02}$	& $1.98^{+0.04}_{-0.04}$ &	-0.91	& harder \\
\hline 
\end{tabular}

\caption{ The reduced $\mathrm{\chi^2}$ and best-fit parameters (with one-sigma errors) for
linear fits to the photon index vs. flux plots of each object (see Fig. \ref{allGamma}). The Pearson correlation
coefficient for each object is also shown.
}

\label{gammaFit}

\end{table*}

\section{Summary \& Discussion}

\begin{figure*}
	\includegraphics[width = 18cm,clip = true, trim = 0 0 0 0]{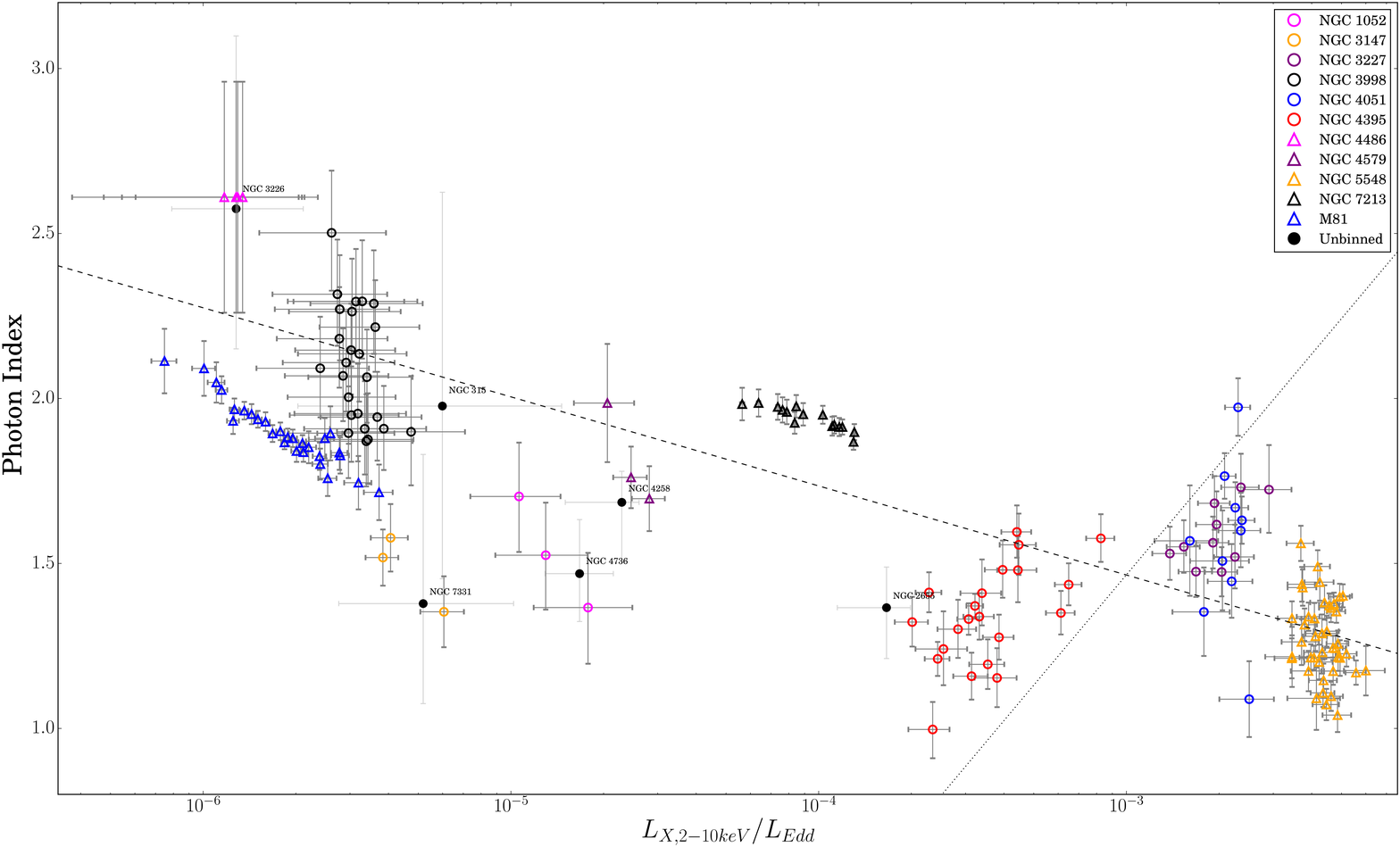}
	\caption{Plot of $\Gamma$ against $\mathrm{L_{X,2-10 keV} / L_{Edd}}$ (as a proxy for \mdote), for all of the Palomar {\it Swift} AGN  for which values of the
	photon index could be constrained, and NGC 7213 (data from \citet{Emmanoulopoulos2012}). 
		  In the cases of NGC 3998 and NGC 7213, the flux was scaled from the 3.3-20 keV flux according to the best-fitting model. Objects which did not have
		  enough data to be flux-binned are plotted individually in grey and labelled.
	The dashed line is the fit to a sample of LLAGN from \citep{Constantin2009}, the dotted line is the fit to a sample of higher Eddington rate radio-quiet AGN
	from \citep{Shemmer2006}. Error bars show the one-sigma confidence levels. The black hole masses and distances used to calculate the Eddington luminosities
	and the X-ray luminosities are shown in Table \ref{radio}.
		  }
	\label{gammaPlot}
\end{figure*}

\begin{table*}
	\begin{tabular}{l c c c c c c c c}

	\hline 
	 Object 	&  $\mathrm{Log (L_{R}, 5GHz}$) &  $\mathrm{Log (L_{X,2-10 keV}}$)& $\mathrm{Log (L_x/L_{R})}$  & $\mathrm{L_{bol}}$		& Eddington Ratio 		& Log(Mass) & Dist  \\
			&  ($\mathrm{ergs\ s^{-1}}$) 	&  ($\mathrm{ergs\ s^{-1}}$)  	 & 				&($\mathrm{ergs\ s^{-s}}$)	& ($\mathrm{L_{bol}/L_{Edd}}$)   &  (\Msol) & (Mpc)	\\
	\hline 
	NGC 1052 	& 	35.4$^a$		& 39.8				& 4.40				& 42.2$^p$			& $\mathrm{7.3 \times 10^{-5}}$	& 8.3$^r$ & 19.0 \\
	NGC 1068 	& 	36.9$^b$		& 39.8 				& 2.90				& 45.0$^r$			& $\mathrm{5.0 \times 10^{-1}}$		& 7.2$^r$ & 12.25  \\	
	NGC 2655 	& 	36.3$^d$		& 41.2 				& 4.91				& 42.2$^*$			& $\mathrm{3.6 \times 10^{-4}}$	& 7.8$^{ee}$& 24.4   \\	
	NGC 3031 	& 	36.8$^c$		& 41.9	 			& 5.10				& 41.5$^m$			& $\mathrm{3.5 \times 10^{-5}}$	& 7.9$^t$ & 3.63   \\
	NGC 3147 	& 	38.0$^d$		& 40.4				& 2.40				& 41.4$^*$			& $\mathrm{5.3 \times 10^{-6}}$	& 8.8$^bb$& 43.7   \\
	NGC 315	 	& 	39.7$^d$		& 41.6	 			& 1.87				& 42.6$^*$			& $\mathrm{6.1 \times 10^{-5}}$	& 8.9$^r$ & 51.3   \\	
	NGC 3226 	& 	37.3$^d$		& 40.4	 			& 3.10				& 41.4$^*$			& $\mathrm{2.0 \times 10^{-5}}$	& 8.2$^{bb}$& 26.35   \\	
	NGC 3227 	& 	36.3$^d$		& 41.0	 			& 4.70				& 43.1$^s$			& $\mathrm{1.3 \times 10^{-2}}$	& 7.6$^x$ & 20.85   \\
	NGC 3628 	& 	39.2$^e$		& 40.2	 			& 1.00				& 41.2$^*$			& -				& -	  & 11.25   \\	
	NGC 3998 	& 	38.2$^d$		&  41.4				& 3.20				& 43.5$^r$			& $\mathrm{3.2 \times 10^{-4}}$	& 9.0$^r$ & 18.65   \\	
	NGC 4051 	& 	37.0$^b$		&  41.8				& 4.80				& 42.6$^s$			& $\mathrm{1.9 \times 10^{-2}}$	& 6.2$^x$ & 15.5   \\	
	NGC 4151 	& 	39.8$^f$		&  44.1				& 4.30				& 44.0$^s$			& $\mathrm{1.8 \times 10^{-2}}$	& 7.7$^y$ & 16.2   \\
	NGC 4258 	& 	35.8$^d$		&  40.4				& 4.57				& 43.5$^r$			& $\mathrm{6.1 \times 10^{-3}}$	& 7.6$^z$ & 7.56   \\	
	NGC 4321 	& 	35.7$^g$		&  39.7				& 4.01				& 43.5$^n$			& $\mathrm{3.6 \times 10^{-2}}$	& 6.8$^{bb}$& 16.0   \\
	NGC 4388 	& 	39.0$^b$		&  43.1				& 4.10				& 43.7$^s$			& $\mathrm{6.5 \times 10^{-2}}$	& 6.8$^{bb}$& 18.55  \\
	NGC 4395 	& 	34.8$^d$		&  42.0				& 7.20				& 41.4$^s$			& $\mathrm{3.9 \times 10^{-2}}$	& 5.6$^{aa}$& 4.3   \\
	NGC 4472	& 	36.5$^d$		& 40.3	 			& 3.80				& 41.5$^*$			& $\mathrm{1.5 \times 10^{-6}}$	& 8.8$^{ff}$& 15.9  \\ 	
	NGC 4486 	& 	38.9$^d$		&  42.6				& 3.70				& 43.8$^*$			& $\mathrm{1.6 \times 10^{-4}}$	& 9.5$^{bb}$& 16.5   \\
	NGC 4579 	& 	37.7$^d$		& 42.2				& 4.50				& 42.0$^o$			& $\mathrm{1.1 \times 10^{-4}}$	& 7.9$^{bb}$& 20.5   \\
	NGC 4736 	& 	37.6$^f$		& 39.9				& 2.3				& 41.1$^*$			& $\mathrm{5.6 \times 10^{-5}}$	& 7.0$^{cc}$& 4.86   \\	
	NGC 5194 	& 	36.7$^f$		& 39.2	 			& 2.46				& 43.8$^r$			& $\mathrm{5.7 \times 10^{-2}}$	& 7.0$^{bb}$& 7.94  \\
	NGC 5548 	& 	36.1$^b$		& 40.8				& 4.70				& 44.1$^s$			& $\mathrm{1.6 \times 10^{-2}}$	& 7.8$^y$ & 92.50  \\
	NGC 5806 	& 	-			& 39.9				& -				& 40.9$^*$			& -				& -	  & 26.1  \\
	NGC 7331 	& 	37.2$^e$		& 39.8	 			& 2.61				& 40.8$^*$			& $\mathrm{2.3 \times 10^{-5}}$	& 7.6$^{dd}$& 14.55  \\ 			
	\hline 
	 NGC 7213 	&	38.7$^i$		& 42.1$^k$			& 3.40				&43.0$^n$			&  $\mathrm{7.4 \times 10^{-4}}$& 8.0$^r$  & 22.00	\\
	 Sag A*		&	32.6$^j$		& 33.4$^l$ 			& 0.80				&41.0$^o$			&  $\mathrm{1.0 \times 10^{-5}}$& 6.6$^v$  & $\mathrm{8.33 \times 10^{-3}}$	 \\
	\hline
	\end{tabular}
		
 	\caption{Radio luminosities at $\mathrm{5 Ghz}$ and 2-10 keV X-ray luminosities, Eddington ratios, masses and distances for all 24 AGN, plus NGC 7213 and Saggitarius A* for comparison. 
 	The ratios between the luminosities of each object are also given. Data taken from: 
 		 (a) \citet{Horiuchi2004}, (b) \citet{Gallimore2004}, (c) \citet{Perez-Olea1996}, (d) \citet{Nagar2005}, 
 		 (e) \citet{Cossio1975}, (f) \citet{Reich2000}, (g) \citet{Filho2006}, (h) \citet{Irwin2015}, 		 
 		 (i) \citet{Bell2011},(j) \citet{Zhao2001}, (k) \citet{Emmanoulopoulos2012}, (l) \citet{Baganoff2003},
 		 (m) \citet{Ho1996}, (n) \citet{Ptak2004}, (o) \citet{Lewis2006},
 		 (p) \citet{SkipperThesis},  		 
 		 (q) \citet{Starling2005}, 		 
 		 (r) \citet{Woo2002}, (s) \citet{Vasudevan2009},
 		 (t) \citet{Devereux2003}, (u) \citet{Walsh2012}, (v) \citet{Gillessen2008}, (w) \citet{Mchardy2014}, (x) \citet{Denney2010}, 
 		 (y) \citet{Bentz2006}, (z) \citet{Herrnstein2005}, (aa) \citet{Peterson2005}, (bb) \citet{Merloni2003} (cc) \citet{Hernandez-Garcia2014}
 		 (dd) \citet{Filho2004}, (ee) \citet{Gu2009}, (ee) \citet{Panessa2013}, (ff) \citet{Ho1999}, $^*$ - using bolometric corretion, x16.
 		 Dashes indicate that no measurement known to the authors exists.}
 		 
	\label{radio}
\end{table*}

We have analysed the long-term observations with {\it Swift} of a sample of 24 nearby AGN from the Palomar sample in order to determine the reason for their variability. 
This study concentrates on trying to determine long-term systematic trends rather than investigating short-term fluctuations which may not be representative of the overall underlying behaviour.
The main results of this study are:

\begin{itemize}
\item From examination of simple hardness ratios, we find that 18 of the AGN show hardening of their X-ray spectra with increasing luminosity, 2 
showed softening of their X-ray spectra with increasing luminosity, 2 showed more complex behaviour although generally softening with increasing luminosity
and the remaining 2 did not show any correlation between their spectral hardness and luminosity.

\item  There are 13 AGN for which there were sufficient data for flux-binned spectral fits to be made. Of these, NGC 1068, NGC 4395, NGC 4151 and NGC 5548 
show complex absorption, i.e. variable partial-covering absorbers, in their spectra, both in our Swift observations and in previous observations (see notes on individual sources), which makes it 
difficult to determine the true cause of the hardness variability. The other 8 are fit by relatively simple power law plus absorption models, making 
interpretation of the hardness variations simpler. Of these 8, NGC 3227 and NGC 4051 show increasing $\Gamma$ with increasing luminosity. Of the other 6,  
NGC 3031 (M81) and NGC 3998, for which there are a large number of observations, all show a decrease of $\Gamma$ with increasing luminosity.  
NGC 1052, NGC 3147 and NGC 4579, for which there are fewer observations, show the same behaviour. All of the AGN which show decreasing $\Gamma$ 
with increasing luminosity have values of $\mathrm{L_x/L_{Edd}}$ of $\mathrm{< 2 \times 10^{-4}}$. 
The 2 AGN which show increasing $\Gamma$ with increasing luminosity have values of $\mathrm{L_x/L_{Edd}}$ of $\mathrm{> 10^{-3}}$.

\item For the 11 objects for which spectral analysis could only be carried out on a summed spectrum of all their data, 3 were found to have spectra which could be described
      by a simple absorbed power law - NGC 3226, NGC 3628 and NGC 5806. The spectral modelling therefore implies that the harder-when-brighter behaviour of these 3 objects can 
      only be due either to changes in their photon indices which are negatively correlated with luminosity, or due to changes in absorption
      which are negatively correlated with luminosity. As with the harder-when-brighter objects whose spectra were flux-binned, the value of 
      $\mathrm{L_x/L_{Edd}}$ is $\mathrm{< 2 \times 10^{-4}}$ for NGC 3226.
      The masses of NGC 3628 and NGC 5806 have not been determined, so their Eddington luminosities could not be calculated, 
      but their low X-ray luminosities imply that they are very likely to have values of $\mathrm{L_x/L_{Edd}}$ which are $\mathrm{< 2 \times 10^{-4}}$;
      They would have to have very low masses of $\mathrm{< 6.3 \times 10^{5}}$ \Msol and $\mathrm{< 3.2 \times 10^{5}}$ \Msol respectively in order 
      to have values of $\mathrm{L_x/L_{Edd}}$ which are $\mathrm{> 2 \times 10^{-4}}$.

\end{itemize}

\subsection{Correlations between Photon Index and Luminosity}

We find that below $\mathrm{L_x/L_{Edd}}$ of $\mathrm{2 \times 10^{-4}}$, all AGN with well measured spectral variability show a decrease in $\mathrm{\Gamma}$ with increasing luminosity
(i.e. hardening of the underlying spectral slope). 
Above that value of $\mathrm{L_x/L_{Edd}}$, $\mathrm{\Gamma}$ increases with $\mathrm{L_X}$.  For an assumed bolometric correction factor of $\mathrm{L_{bol} / L_{x} = 16}$ \citep{Ho2008,Constantin2009}, 
this value would correspond to $\mathrm{L_Bol/L_{Edd}}$  (i.e. \mdote) of $\mathrm{3.2 \times 10^{-3}}$.
NGC 5548 and NGC 4395 both show complex behaviour in their hardness ratios (see Fig. \ref{swbhardness}), with the
largest scatter in our sample and are both known to possess complex variable absorption \citep{Kaastra2014, Nardini2011} which is likely to complicate the interpretation of their observed behaviour. 
Previous studies \citep[e.g.][]{Sobolewska2009} have shown an increase in $\mathrm{\Gamma}$ with increasing Luminosity within individual AGN at high accretion rates,
with the exception of NGC 5548.

In Fig. \ref{gammaPlot} we show all of the measurements of $\mathrm{\Gamma}$ for our sample, excluding those for which we had reason to distrust the measured values of $\mathrm{\Gamma}$ 
(see Appendix \ref{individual}). We also plot the fits to $\mathrm{\Gamma}$ from previous samples of LLAGN 
by \citet{Constantin2009} and, at higher accretion rates, from \citet{Shemmer2006}.  

The variations of $\mathrm{\Gamma}$ in M81 and NGC 7213, which are relatively bright sources 
with well measured spectra, follow the trend of \citet{Constantin2009} fairly well. However, the individual observations of some objects show a steeper 
variation with luminosity. This difference might be real or it might be due to an unknown systematic effect, e.g. over-subtraction of the hard background spectrum.
If the steeper gradients are not a systematic effect, they could be explained in a number of ways. A weak constant soft component which
is not resolved in the spectrum would artificially enhance the flattening of the photon index with increasing luminosity, as its relative contribution to the soft
end of the spectrum decreases. Absorption variability could also enhance the change in the photon index, but the amount of absorption would be required to increase
with increasing luminosity, which is hard to explain physically; it is perhaps possible if a higher luminosity leads to an increase in the wind-driving ability
of the system, due to e.g. a hotter disk or more ionised matter available for line-driving, but the timescales for this mechanism to lead to an increase in absorption are likely to be too long 
to explain the observed spectral changes \citep[see e.g.][]{Proga2007}. 
Finally, the difference in the rate of change of $\Gamma$ with luminosity between individual AGN and 
sample averages could be real.  We note that the spread of data points around the best fit relationship for both \citet{Constantin2009} and \citet{Shemmer2006} is very large. 
Thus although the change in $\Gamma$ within any individual AGN is determined only by the change in accretion rate and any subsequent related changes, e.g. change in absorption, 
the spread of $\Gamma$  within samples can also be affected by changes related to the black hole mass. For example mass-related changes in disc temperature at a given Eddington 
ratio could lead to differences in the relative importance of disc seed photons and coronal synchrotron seed photons \citep{Shakura1973, Mchardy2014}.

All of the AGN with $\mathrm{L_x/L_{Edd}}$ $\mathrm{< 2 \times 10^{-4}}$ are seen to broadly lie on a similar track, 
indicating that a similar physical process is responsible for the behaviour of all of the objects in the sample. 
The rate of change in $\Gamma$ with luminosity in these lower accretion rate objects is relatively similar in all harder-when-brighter sources, 
implying similarities between all of the objects in the physical causes of this anticorrelation.
Some of the sources show similar gradients 
to that of the \citet{Constantin2009} relationship, but several show slightly steeper gradients. 
The accuracy of the measurement of the distances and black hole masses of the AGN 
is limited in some cases and consequently the values of $\mathrm{L_{X}/L_{Edd}}$ may be offset for some objects. It should be noted that, in
addition to the aforementioned caveats, the bolometric correction could also vary between objects, meaning that their position on this plot is not
precisely correct.

For the softer-when-brighter sources, NGC 4051 and NGC 3227 are broadly consistent with the trend shown by \citet{Shemmer2006} and can be explained by the standard model of 
cooling of the corona with increasing accretion rate and disc seed photon flux. As discussed above, the behaviour of NGC 4395 and NGC 5548 is more complex.

The shift from a negative to a positive correlation between $\Gamma$ and \mdote, and therefore from harder-when-brighter to softer-when-brighter behaviour, 
is probably due to a transition of the dominant seed photon population for X-ray production, from synchrotron or 
cyclo-synchrotron emission from the X-ray emitting corona itself, or from a hot inner flow/ADAF, to black body photons from a surrounding accretion disc 
\citep[e.g.][]{Skipper2013, Sobolewska2011}. 

The same synchrotron self-Compton emission process which dominates the X-ray emission from coronae at low accretion rates could also come from a jet;
harder-when-brighter behaviour due to a $\Gamma$-\mdotes anticorrelation is common in 
blazars \citep[e.g.][]{Krawczynski2004,Gliozzi2006,Zhang2006},
e.g. due to shock acceleration of electrons in a jet producing both higher energy synchrotron seed photons and higher energy (inverse) Compton scattering.
This mechanism has been proposed to explain e.g. the spectral evolution of the BHXRB XTEJ1550-564 as it transitions from the `soft state' towards
the `hard state' \citep{Russell2010,Emmanoulopoulos2012}. There is no evidence yet for highly relativistic outflows in LLAGN, such as are seen in blazars, 
but weak jets are detected \citep[e.g][]{Marti-Vidal2011} and models for such jets \citep[e.g.][]{Markoff2008} can explain the observed X-ray spectral variability.

One further explanation for harder-when-brighter behaviour is a hot outflowing corona above an X-ray illuminated untruncated disc \citep{Sobolewska2011}, 
in which the seed photons are from the disc. In this model, the seed photon flux is limited
by relativistic beaming of the outflowing corona. At higher luminosities, a higher outflow velocity decreases the illumination of
the disc and therefore the seed photon population, softening the spectrum. However, this model predicts that at the lowest luminosities there
should be a strong reflection component from an untruncated disk, whilst many of the low-luminosity sources in this study, including M81, show little evidence
for reflection, but good evidence for an ADAF flow and a truncated disc from the x-ray spectrum and the presence of radio emission \citep{Ptak2004, 
Brenneman2009, Markoff2008, Young2007, Nowak2010}.

\subsection{AGN and BHXRB State Analogues}

Both the harder-when-brighter relation and the softer-when-brighter relation are also well established in black hole X-ray binaries (BHXRBs) 
and have been observed both in samples of single-epoch observations and in multi-epoch observations of individual objects \citep[e.g.][]{Kalemci2004,Yuan2007,Yamaoka2005}. 
As in samples of AGN, they are seen to switch from a positive correlation to a negative correlation between $\Gamma$ and \mdot,
and therefore from harder-when-brighter to softer-when-brighter behaviour, when \mdotes increases
above a critical value, \mcrit. This value is identified as $\mathrm{\sim 10^{-2}}$ \mdotes from samples of AGN  \citep{Wu2008}, and directly
observed in the BHXRB Cyg X-1, \citep{Axelsson2008,Skipper2013,Skipper2016}. 

Previous studies have looked for changes in the photon index with luminosity of individual LLAGN before \citep[e.g.]{Younes2011,Hernandez-Garcia2013,Hernandez-Garcia2014},
but been unsuccessful; our ability to detect changes in the  photon index more easily is likely due to the larger flux-ranges covered by our data. As the amplitude of
the change in $\Gamma$ is proportional to the change in the luminosity of the system, a large flux range is essential. A large amplitude of variability is therefore
also important, as changes in $\Gamma$ will be hard to find in systems with very low variability regardless of the timescales covered. Methodological differences,
such as the use of flux-binning, may also contribute, but we believe these effects to be less significant.

For BHXRBs, the transition from harder-when-brighter to softer-when-brighter takes place entirely within the hard state \citet{Wu2008,Skipper2016} 
and softer-when-brighter behaviour continues throughout the hard-intermediate state, at accretion rates below that of the change to the soft state. 
Thus all of the AGN in our sample are probably analogues to either hard state or hard-intermediate state BHXRBs. Thus LINERS, with accretion rates 
$\mathrm{<\sim 10^{-2}}$ \mdotes are analogues to hard state BHXRBs whilst Seyferts may be analogues to either higher accretion rate hard state, or to hard-intermediate state BHXRBs.

A so-called `fundamental plane' relationship between X-ray luminosity, radio luminosity and black hole mass has previously been shown for samples of AGN and BHXRBs 
\citep[e.g.][]{Merloni2003, Kording2006}.  In Table \ref{radio} we give the X-ray luminosities for our sample from our observations together with radio luminosities and 
black hole masses from the literature. We do not show the 'fundamental plane' derived from the data in Table \ref{radio}, but note that there is no obvious distinction between 
Seyferts and LINERs, indicating again that they are parts of a continuous distribution. The X-ray luminosities are from our present data and the radio are from the literature.  

The discovery that the individual variations in $\Gamma$ in the {\it Swift} AGN sample are consistent with previous fits to samples of AGN 
as well as similar studies showing the variability in $\Gamma$ of individual BHXRBs therefore adds strong evidence the idea that AGN are scaled-up analogues of the
BHXRBs. Based on the present data, which are the best data in which a harder-when-brighter correlation may be found in individual AGN, we suggest that this behaviour
is very common amongst LLAGN.

\section*{Acknowledgments}

SDC thanks the STFC for support under a studentship and IMcH thanks the STFC for support via grant ST/G003084/1. We note our appreciation to Tom Dwelly, 
who wrote our original Swift data analysis system. We thank the anonymous referee for constructive comments which greatly strengthened the paper.



\bibliography{library}{}
\bibliographystyle{mn2e}

\appendix

\section{Notes on Spectral Fitting of Individual Palomar AGN}
\label{individual}

The following section contains details on the results of spectral fitting for each of the 24 AGN in our sample.

\begin{itemize}
 
\item{\bf NGC 315 }

Only a single summed spectrum of NGC 315 could be produced from the available {\it Swift} data. 
Our best-fitting model consists of an absorbed power law, plus a hot gas component ($\mathrm{\chi^2_R = 1.02}$).
Previous spectral modelling of NGC 315 has employed very similar models \citep[e.g][]{Matsumoto2001,Terashima2002, Younes2011}.
The photon index we measure is very similar to that found by \citet{Matsumoto2001}  and \citet{Younes2011} 
($\mathrm{\Gamma = 1.94}$; slightly higher than that found by \citet{Terashima2002}).

\item{\bf NGC 1052 }

Three flux-binned spectra could be produced from the available {\it Swift} data of NGC 1052.
We found that a single partial coverer provided a very good fit to the spectra. 
A fairly good fit is obtained when all the parameters are tied ($\mathrm{\chi^2_R = 1.11}$). However, allowing $\Gamma$, the absorbing column or the covering fraction
to produces a fit which is statistically better in each case ($\mathrm{f = 9.5 ~and ~p=1.3 \times 10^{-4}}$,$\mathrm{~f = 3.5 ~and ~p=3.3 \times 10^{-2}}$, 
$\mathrm{f~ = 8.0 ~and ~p=4.7 \times 10^{-4}}$ respectively).  The best fit is obtained when only $\Gamma$ is left free ($\mathrm{\chi^2_R = 1.01}$).
Models in which both the absorbing and column and the covering fraction are free to vary but $\Gamma$ is tied, or in which all three are free to vary, provided
lower $\mathrm{\chi^2}$ values, but the added complexity is not required statistically when compared to the model in which only $\Gamma$ is free to vary between flux-binned spectra 
($\mathrm{f = 1.7 ~and~ p=0.19, f = 1.6 ~and ~p=0.186}$ respectively).  
Previous studies by \citet{Weaver1999} and \citet{Hernandez-Garcia2013}, using {\it ASCA} ,  {\it Chandra} and  {\it XMM-Newton}  data,
also concluded that the spectrum was heavily absorbed by one or more partial-coverers.  

For the best-fitting model, in which the absorber is constant and $\Gamma$ is free to vary between flux-binned spectra,  $\Gamma$ is anticorrelated with the flux. 
Furthermore, the same anticorrelation is seen in our spectral fits when the absorbing column is allowed to vary as well as the photon index. 
The values of $\Gamma$ found are between those found by \citet{Weaver1999} and by \citet{Hernandez-Garcia2013} $\mathrm{\Gamma = 1.37-1.70}$ for similar models,
though \citet{Hernandez-Garcia2013} attribute the primary cause of spectral changes in NGC 1052 to variations in the absorbing column, not the photon index.
As the source displays harder-when-brighter behaviour, if the spectral variations are caused by
absorption variation alone this absorption must be negatively correlated with the luminosity of the system, as in e.g. NGC 1365 \citep{Connolly2014}.

\item{\bf NGC 1068 }
\label{ngc1068}

Three flux-binned spectra could be produced from the available {\it Swift} data of NGC 1068.
The components of our best-fitting model to the spectra required an absorbed
power law, a hot gas component for contamination from the host. 
This is consistent with previous modelling of the source, which used similar models \citep[e.g.][]{Bauer2014,Brightman2015}. However,
the photon index we obtain is higher than that found with previous data from e.g. {\it NuStar} which included higher energies
than those detected by {\it Swift}, though indices of $\mathrm{>2}$ have been found. 
NGC 1068 is widely thought to be a Compton-thick AGN, making the intrinsic emission difficult to model \citep{Antonucci1985,Ghisellini1994,Brightman2015},
as the degree of absorption could not be constrained with the energy range of {\it Swift}, the flux extracted from this spectrum was 
much lower than the true value. For these reasons, NGC 1068 was not included in the subsequent discussion of spectral variability behaviour.

\item{\bf NGC 2655 }

Insufficient {\it Swift} data was available to produce multiple flux-binned spectra for NGC 2655, so a single-summed spectrum was produced.
We also find a partial covering model to be the best-fitting to the {\it Swift} data ($\mathrm{\chi^2_R = 1.15}$), 
when combined with an additional hot gas component.
Previous work on NGC 2655 has also required a partial covering model, and a hot gas component in some cases \citep[e.g.][]{Terashima2002,Gonzalez-Martin2009}. 
The best-fitting model has a very hard photon index ($\mathrm{\Gamma = 1.35}$), but this is consistent with what others have found previously \citep{Terashima2002}.

\item{\bf NGC 3147 }

Three flux-binned spectra of NGC 3147 were produced from the {\it Swift} data. 
We find the best-fitting model to the {\it Swift} data to be a simple power law for which $\mathrm{\Gamma}$ was allowed to vary ($\mathrm{\chi^2_R = 0.73}$), 
Previous modelling of the spectrum of NGC 3147 has also shown a lack of any strong absorption \citep[e.g.][]{Ptak1996}.  
As the {\it Swift} data are already slightly over-fitted by this model ($\mathrm{\chi^2_R < 1}$),
we assume that more complex models are incorrect, despite producing lower $\mathrm{\chi^2}$ values.
We find similar values of the photon index to those found by \citep{Ptak1996} ($\mathrm{\Gamma = 1.35-1.58}$).
This model show a a reduction in $\mathrm{\Gamma}$ with increasing intrinsic luminosity of the system.

\item{\bf NGC 3226 }

Only a single summed spectrum of NGC 3226 could be produced from the {\it Swift} data. 
We find that the spectrum is very well-fitted by a power law and a neutral absorber ($\mathrm{\chi^2_R = 0.90}$).
Past studies disagree somewhat as to whether the best-fitting model for the spectrum of NGC 3226 is
bremsstrahlung or a simple power law  \citep[e.g.][]{Gonzalez-Martin2009,Gondoin2004, Younes2011}, but previous
work agrees on the presence of a neutral absorber.
We find a high value for the photon index, but with large uncertainties ($\mathrm{\Gamma = 2.57 \pm 0.42}$); \citet{Gondoin2004} and \citet{Younes2011}
find value which is lower, but which agree with ours within our errors.

\item{\bf NGC 3227 }

Sufficient {\it Swift} data of NGC 3227 was available to produce 10 flux-binned spectra.
The best-fitting model to the {\it Swift} data 
consisted of a power law and an  ionised absorber, with $\mathrm{\Gamma}$ free to vary but all parameters of the absorber tied ($\mathrm{\chi^2_R = 1.07}$),
which is statistically superior to a neutral absorber ($\mathrm{f = 16.3 ~and ~p=1.1 \times 10^{-7}}$). 
Previous spectral modelling of NGC 3227 has also shown its spectrum to be absorbed by an ionised absorber \citep{George1998,Ptak1994},
though some more recent studies have used more complex models involving multiple ionised absorbers and a reflection
component \citep[e.g.][]{Beuchert2015}. 
The parameters of our best-fitting model show a positive correlation between the photon index and the
intrinsic luminosity of the source, consistent with \citet{Sobolewska2009}, 
who also found the same correlation and range of photon indices ($\mathrm{\Gamma = 1.43-1.65}$) in the $\mathrm{2-10 keV}$ band.

\item{\bf NGC 3628 }

Only a single summed spectrum of NGC 3628 could be produced from the {\it Swift} data. 
We found a power law with a neutral absorber to be best-fitting to the {\it Swift} data.
Previous studies have found the same spectral model to be best-fitting \citep{Dahlem1995,Gonzalez-Martin2009},
with a similarly low photon index ($\mathrm{\chi^2_R = 1.36}$) to that found in our best-fitting model.

\item{\bf NGC 3998}

As there were few {\it Swift} data available for NGC 3998, we instead used the {\it RXTE} data from \citet{SkipperThesis},
for which a large number of flux-binned spectra could be produced.
The {\it RXTE} data are of quite low S/N, as NGC 3998 is a very faint source. 
The spectrum can be fitted reasonably well by a power law absorbed by the galactic absorbing column of $\mathrm{1.05 \times 10^{21} cm^{-2}}$,
and shows no excesses which imply that this is not a good model. 
The fit is not improved by the addition of a further absorber, possibly because the 
  {\it RXTE}  does not extend below 3 keV, so is only sensitive to high absorbing columns.  If it is assumed that the spectrum
is absorbed and that the source of the spectral variability is changes in absorption instead of a changing $\Gamma$, a significantly worse fit is obtained.
This is consistent with previous spectral studies of this source \citep{Ptak2004,Awaki1991, Younes2011}.
The values obtained for the photon index in previous studies are all within the range found in this model ($\mathrm{\Gamma \sim 1.87 - 2.50}$). 
In the best-fitting model, the photon index shows a clear anticorrelation with the intrinsic luminosity of the system. \citet{Younes2011} also
found variations in $\mathrm{\Gamma}$, but had only two observations so could not test for any correlation.

\item{\bf NGC 4051 }

There were sufficient {\it Swift} data to produce 10 flux-binned spectra of NGC 4051.
We find that absorbed power law plus a hot gas component also fits the {\it Swift} data well ($\mathrm{\chi^2_R = 1.10}$). 
Previous spectral modelling of NGC 4051 with {\it XMM-Newton} data has also used an absorbed power law plus a hot gas component \citep[e.g.][]{Ponti2008}.
We find that the photon index is positively correlated with the flux, as found previously by \citet{Ponti2008,Sobolewska2009} with {\it XMM-Newton}  and 
{\it RXTE} in the $\mathrm{0.5-10 keV}$ and $\mathrm{2-10 keV}$ bands respectively, with the same range of values for the photon index ($\mathrm{\Gamma = 1.35-1.97}$). 
This correlation is less clear in the $\mathrm{2-10 keV}$ band, perhaps due
to the absorption variations found by e.g. \citet{Lamer2003,Ponti2008} making our parameter measurements less accurate over the narrower band.

\item{\bf NGC 4151 }

There were sufficient {\it Swift} data to produce 7 flux-binned spectra of NGC 4151.
We find a power law plus a neutral partially-covering absorber to be the best-fitting model to the {\it Swift} data ($\mathrm{\chi^2_R = 1.02}$). 
Previous work has also shown NGC 4151 to be best modelled with these components \citep[e.g.][with {\it ROSAT}, {\it NuSTAR} and {\it Sukaku} respectively]{Puccetti2007,Keck2015}.
The values of the photon index we find are, however, very low. The large absorbing column and complex, variable absorption are likely to have limited
accurate measurement of the underlying spectral index. We therefore exclude the results of the spectral fitting to NGC 4151 from the subsequent discussion.

\item{\bf NGC 4258 }

Only a single summed spectrum of NGC 4258 could be produced from the available {\it Swift} data. 
We find that a partial covering absorber and a hot gas component is required to obtain a good-fit ($\mathrm{\chi^2_R = 1.03}$), significantly better
than a fully-covering absorber ($\mathrm{\chi^2_R = 1.78}$). 
Previous studies of NGC 4258 have, however, found the spectrum to be best modelled by a simple absorbed power law and a hot gas component \citep{Yamada2009, Terashima2002}.
The value we find for the photon index ($\mathrm{\Gamma = 1.66}$) is consistent with that of \citet{Terashima2002} and \citep{Yamada2009}.

\item{\bf NGC 4321 }

Only a single spectrum of NGC 4321 could be produced from the {\it Swift} data.
We find that an absorbed power law plus a hot gas component is best-fitting ($\mathrm{\chi^2_R = 1.62}$),
consistent with previous spectral studies of NGC 4321 \citep{Gonzalez-Martin2009, Roberts2001}. 
In our data, however, the hot gas component from the host is dominant, meaning the 
underlying spectral index is not well constrained. The results of this spectral fitting are therefore not included in subsequent analysis.

\item{\bf NGC 4388 }

Sufficient {\it Swift} data were available to produce 4 flux-binned spectra of NGC 4388.
We also obtained a reasonable fit to the {\it Swift} data with a power law and a neutral absorber ($\mathrm{\chi^2_R = 1.16}$).
This model did not require the photon index (or absorption) to vary between flux-binned spectra, though the spectra only cover a small range of fluxes and therefore may not 
be expected to change significantly.
Previous studies have also fitted the spectrum with a model consisting of as absorbed power law 
\citep[e.g.][]{Elvis2004,Risaliti2002}. However, we obtain a very low value for the photon index ($\mathrm{\Gamma = 1.10}$), which does not agree
with the value found in previous studies - together with the high level of absorption ($\mathrm{N_H \sim 3 \times 10^{23} cm^{-2}}$), which is great enough to affect the majority of the {\it Swift XRT} 
energy range, were therefore conclude that the underlying spectrum, and therefore the intrinsic luminosity, of NGC 4388 are not well constrained.
The results of the spectral fitting to NGC 4388 are therefore not included in subsequent analysis.

\item{\bf NGC 4395 }

The large quantity of {\it Swift} data of NGC 4395 allowed us to produce 22 flux-binned spectra.
We find that a model consisting of a power law and an ionised absorber gives a good fit to the {\it Swift} data.
The best-fitting model of the flux-binned spectra requires only the photon index to vary ($\mathrm{\chi^2_R = 1.03}$); although the model in which the ionisation state 
and absorbing column are free to vary does give a lower $\mathrm{\chi^2}$, the extra complexity is not statically required ($\mathrm{f = 0.31, p = 1.00}$).
The spectrum has previously been modelled with multiple absorbers with varying ionisation states and absorbing columns 
 \citep[e.g.][]{Iwasawa2010,Nardini2011}. 
The lack of a need for variability in the absorption, despite previous work showing absorption variability to be present, is perhaps due to the flux-binning 
averaging the absorption over the period covered by the data, but this is likely to add scatter to the parameter values. Some of the spectral indices obtained from this fit
are very hard ($\mathrm{\Gamma = 0.91-1.60}$), so absorption variability is not ruled out as the source of spectral variability,
however the values are consistent with the values found by \citet{Iwasawa2010}, who also found evidence of variations in $\mathrm{\Gamma}$. 
Our spectral modelling therefore also implies that there is not any flux dependence in the degree of absorption.

\item{\bf NGC 4472 }

Spectral modelling of NGC 4472 has been limited due to its very low-luminosity.
We find that an absorbed power law fits the spectrum of NGC 4472 very well in conjunction with a hot gas component  ($\mathrm{\chi^2_R = 1.03}$)
consistent with previous work, which has used a simple power law model \citep{Loewenstein2001}.
However, the hot gas component dominates the spectrum in our data due to the low luminosity of the x-ray source, making
constraints on the intrinsic spectrum less reliable; we find an unphysically high photon index of $\mathrm{\Gamma = 3.38}$.
We therefore conclude that the spectral fitting could not reliably find the parameters of the intrinsic spectrum
and exclude the results from subsequent analysis.

\item{\bf NGC 4486 }

Four flux-binned spectra were produced from the available {\it Swift} data of NGC 4486.
We find that a single power law plus a hot gas model gives a good fit to the spectra ($\mathrm{\chi^2_R = 1.03}$). 
The model in which $\mathrm{\Gamma}$ does not vary is best-fitting, as the model in which $\mathrm{\Gamma}$ varies
is not statistically better ($\mathrm{f = 0.1, p = 0.96}$), meaning we obtain only a single value for the whole range of luminosities.
Previous spectral modelling of NGC 4486 has also mostly used a single, absorbed power law \citep[e.g.][]{Gonzalez-Martin2006, Dudik2005},
though some have used more complex, multi-component models \citep[e.g.][]{Gonzalez-Martin2009}.
 As the flux range is relatively small, it may be that in reality there are small variations which cannot be successfully resolved in the {\it Swift} data. The value for the photon index obtained
 ($\mathrm{\Gamma = 2.68}$) is relatively high, but consistent with the value found by \citep{Gonzalez-Martin2009} within our errors.
 
\item{\bf NGC 4579 }

Three flux-binned spectra were produced from the available {\it Swift} data of NGC 4486.
We find an absorbed powerlaw and a hot gas component to be a good fit ($\mathrm{\chi^2_R = 1.00}$).
Previous studies also required an absorbed powerlaw and a hot gas component \citep{Gonzalez-Martin2009,Dewangan2004}.
We find a range of photon indices very similar to those found by previous studies ($\mathrm{\Gamma = 1.7 - 2.0}$). 
The photon index decreases with increasing luminosity in this model.

\item{\bf NGC 4736 }

Only a single summed spectrum of NGC 315 could be produced from the available {\it Swift} data. 
We find an absorbed power law plus a hot gas component to be the best-fitting to the {\it Swift} data. 
Previous studies successfully also model the spectrum of NGC 4736 with an absorbed power law and a hot gas component \citep[e.g.][]{Roberts2001, Terashima2002}. 
We find a very similar value for the photon index to previous studies ($\mathrm{\Gamma = 1.45}$).

\item{\bf NGC 5194 }

We find the spectrum of NGC 5194 a heavily absorbed power law with and a Gaussian to be the best fit to the {\it Swift} data. 
Previous spectral studies of NGC 5194 have also found it to be well fit by a heavily absorbed power law, with a strong Fe emission line \citep{Terashima2002, Ptak1999}.
As the source is Compton thick, however, we cannot constrain the absorption well with the {\it Swift} data; we obtain an extremely low photon index ($\mathrm{\Gamma = 0.7}$), significantly lower
than that found in previous studies. We therefore conclude that both the value of $\Gamma$ and the x-ray luminosity we obtain are unreliable and consequently do
not include the results of the spectral fits to NGC 5194 in subsequent analysis.

\item{\bf NGC 5548 }

The large quantity of {\it Swift} data of NGC 5548 allowed us to produce 42 flux-binned spectra.
We find that the flux-binned {\it Swift} data are well-fitted by a single, ionised absorber for which $\Gamma$, the absorbing column and the ionisation state
are all free to vary ($\mathrm{\chi^2_R = 0.94}$). The data are also well fitted by changes in each of these parameters alone, but statistically the model
in which they are all free to vary is superior ($\mathrm{f = 3.3 ~and ~p=6.9 \times 10^{-21}}$, $\mathrm{f = 2.2 ~and ~p=5.4 \times 10^{-9}}$, $\mathrm{f = 662.7 ~and ~p=7.7 \times 10^{-141}}$
compared to the models in which only $\Gamma$, the absorbing column or the ionisation state were free to vary, respectively).
This is consistent with previous studies which hav found NGC 5548 possesses complex variable absorption \citep[e.g.][]{Steenbrugge2005,Kaastra2014}.

In the best-fitting model, there does not appear to be any strong correlation between $\Gamma$ and the intrinsic luminosity of the x-ray source,
though there are hints of a negative correlation. The lack of any correlation being found may be due to the complex, variable absorption adding a large amount of scatter to the measure values, 
however a previous study by \citep{Sobolewska2009} with {\it RXTE} (whose data is less affected by absorption at $\mathrm{3-20 keV}$) also did not find any 
significant correlation between $\Gamma$ and luminosity in the $\mathrm{2-10 keV}$ band, with the same scatter in  $\Gamma$ that we find.
 They also find a higher range of photon indices, further suggesting that the variable absorption in NGC 5548 is affecting the accuracy of our measurements
 of the photon index.

\item{\bf NGC 5806 }

Only a single summed spectrum of NGC 5806 could be produced from the available {\it Swift} data. 
We find that the {\it Swift} data for NGC 5806 are best-fitted by an absorbed power law ($\mathrm{\chi^2_R = 1.16}$),
with a fairly typical photon index of $\mathrm{\Gamma = 1.65}$.
Although NGC 5806 has been classified as an AGN optically \citep{Seth2008}, there have been no X-ray spectral
studies of this object to date; the X-ray data used in this study are taken from observations intended to look
at the iPTF13bvn supernova which occurred in NGC 5806 in 2013 (the supernova occured near the edge of the galaxy
and could therefore not have contaminated the spectrum of the nucleus) \citep{Cao2013}.

\item{\bf NGC 7331 }

Only a single summed spectrum of NGC 7331 could be produced from the available {\it Swift} data. 
We find an absorbed power law and a hot gas component to be best-fitting to the {\it Swift} data. 
Previous X-ray studies of NGC 7331 employed similar models \citep{Gonzalez-Martin2006, Gallo2006}.
We find a slightly lower photon index than previous studies, but the value obtained is the compatible within our errors ($\mathrm{\Gamma = 1.38 \pm 0.3}$).

\end{itemize}

\begin{figure*}

	\includegraphics[width = 7cm,clip = true, trim = 0 0 0 0]{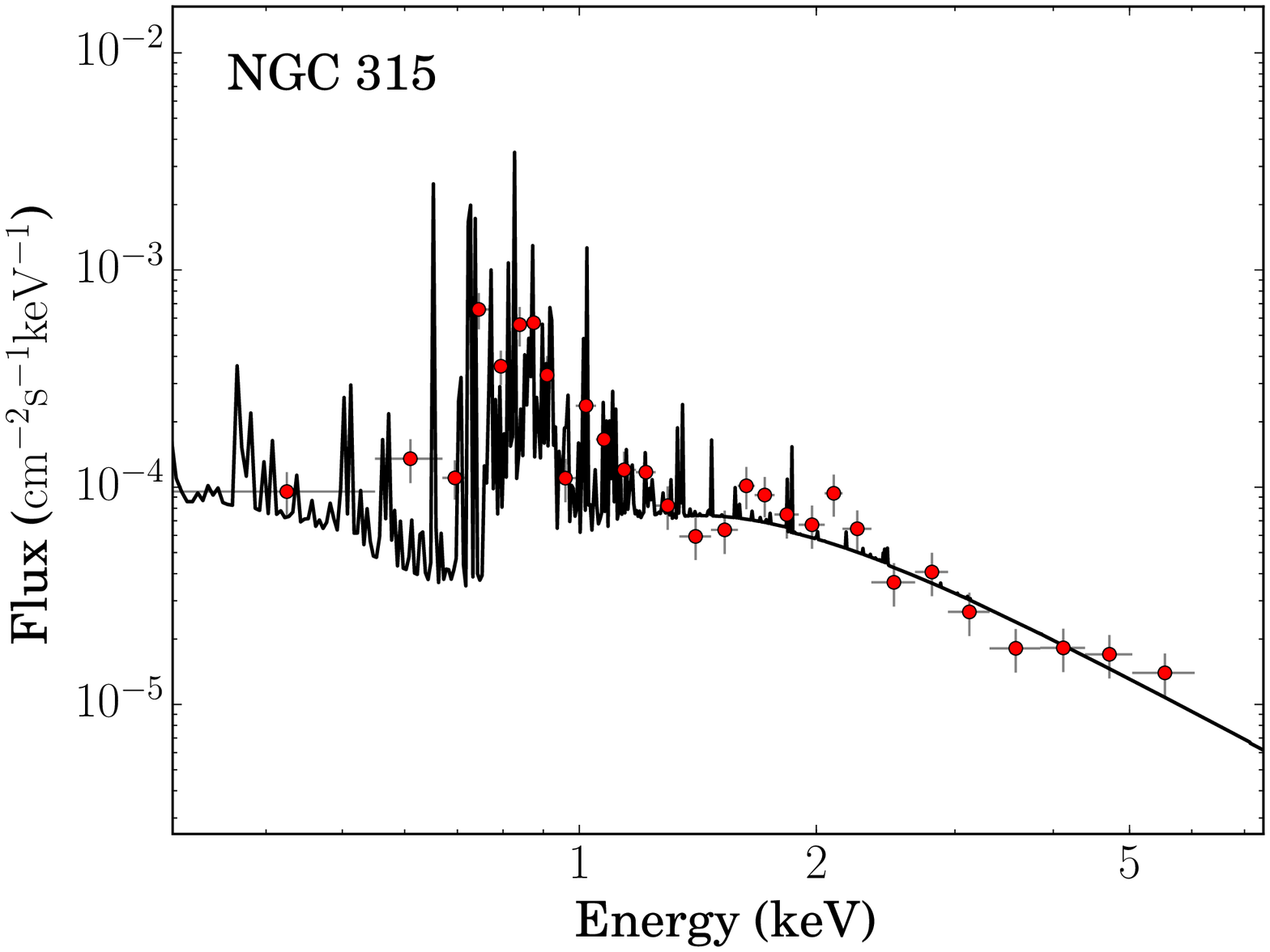}
	\includegraphics[width = 7cm,clip = true, trim = 0 0 0 0]{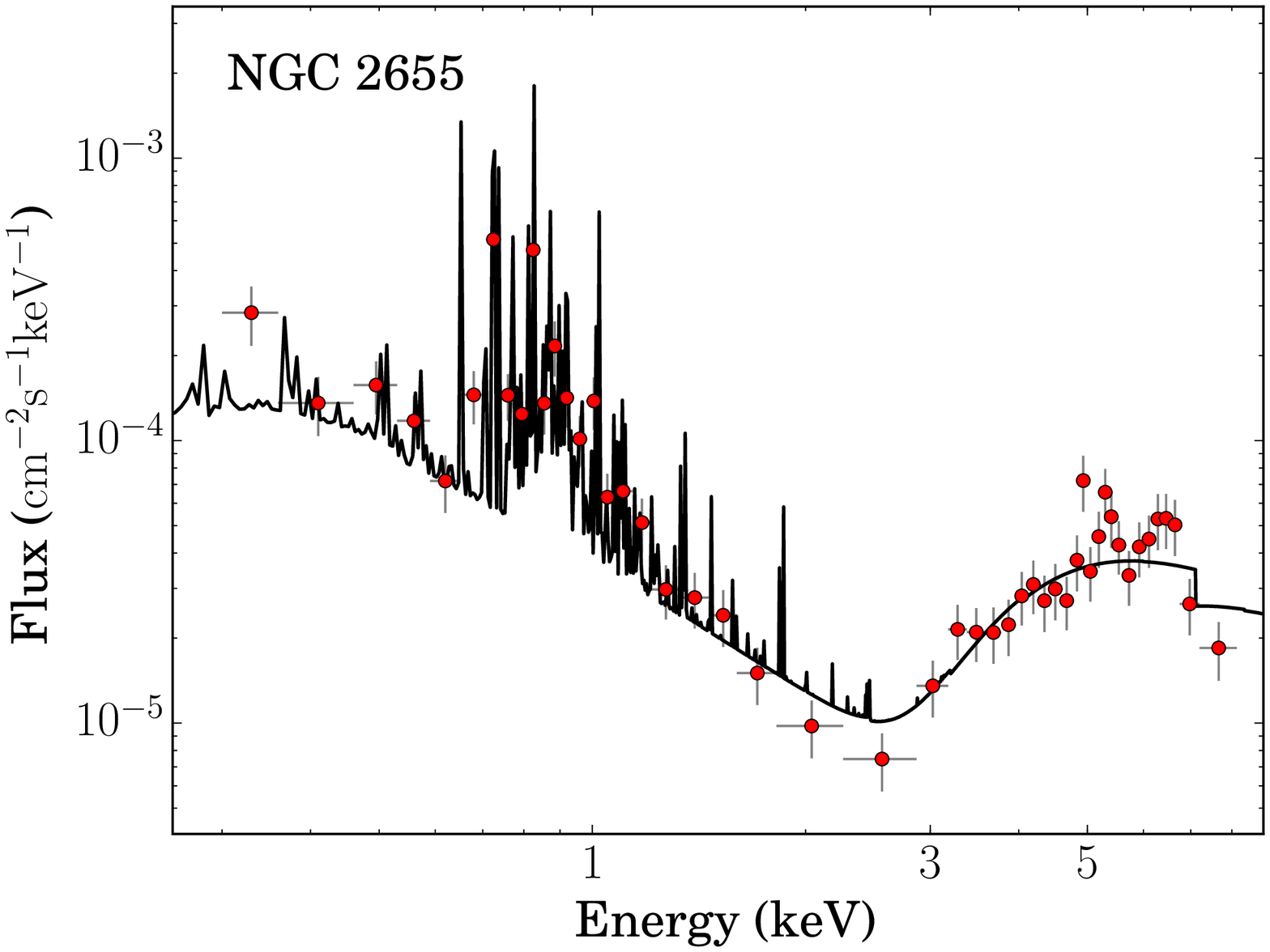}
	\includegraphics[width = 7cm,clip = true, trim = 0 0 0 0]{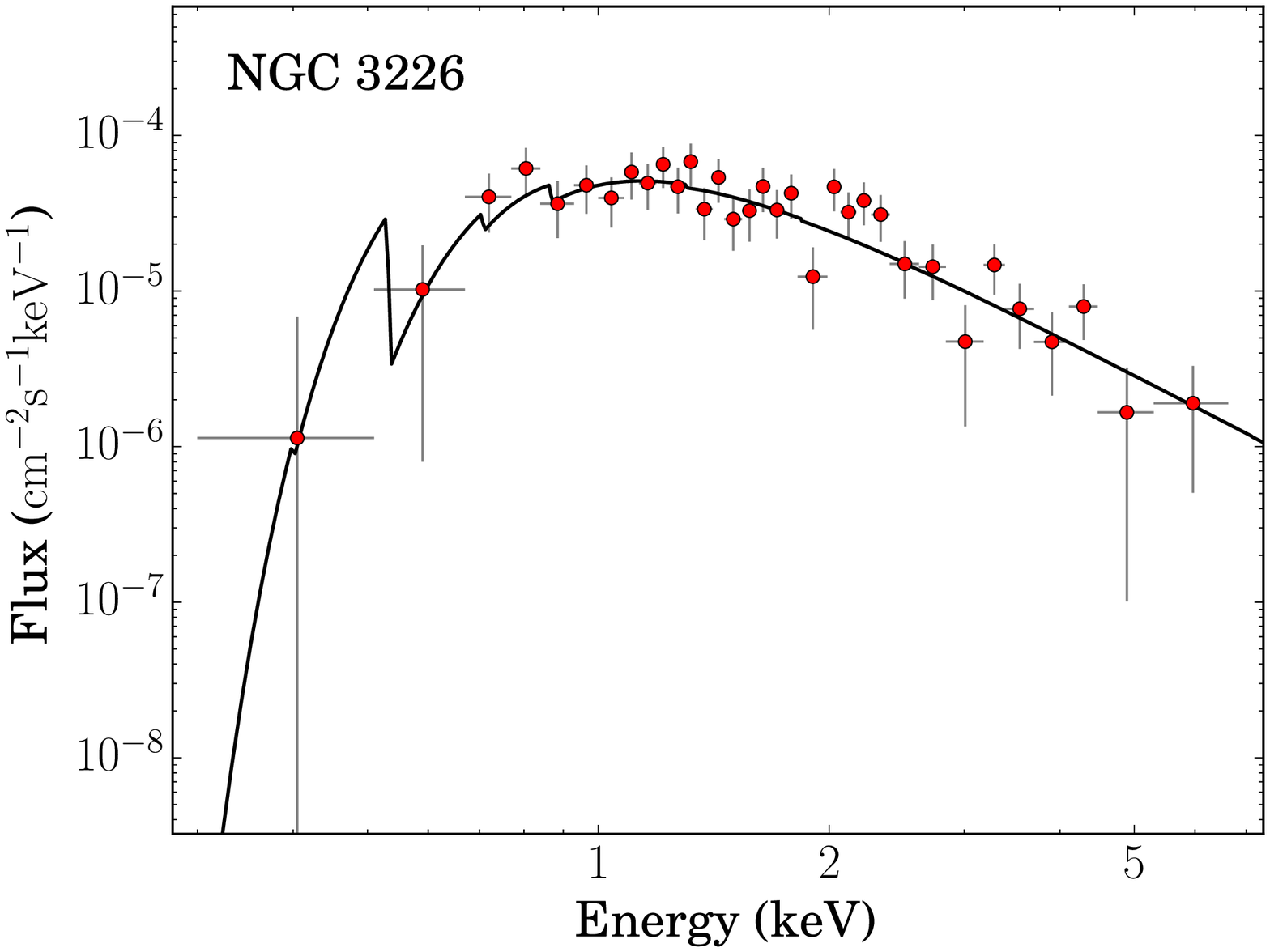}
	\includegraphics[width = 7cm,clip = true, trim = 0 0 0 0]{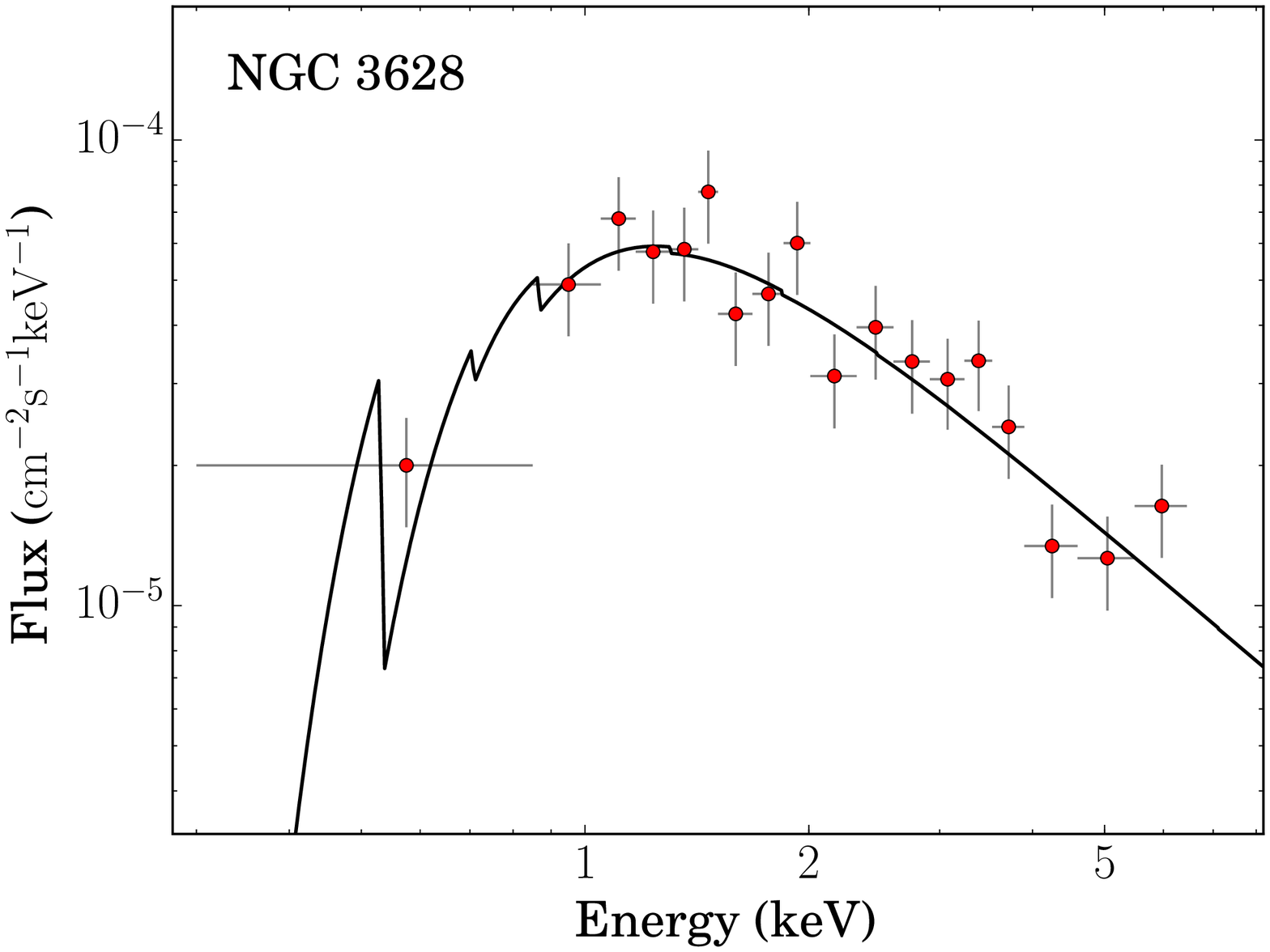}
	\includegraphics[width = 7cm,clip = true, trim = 0 0 0 0]{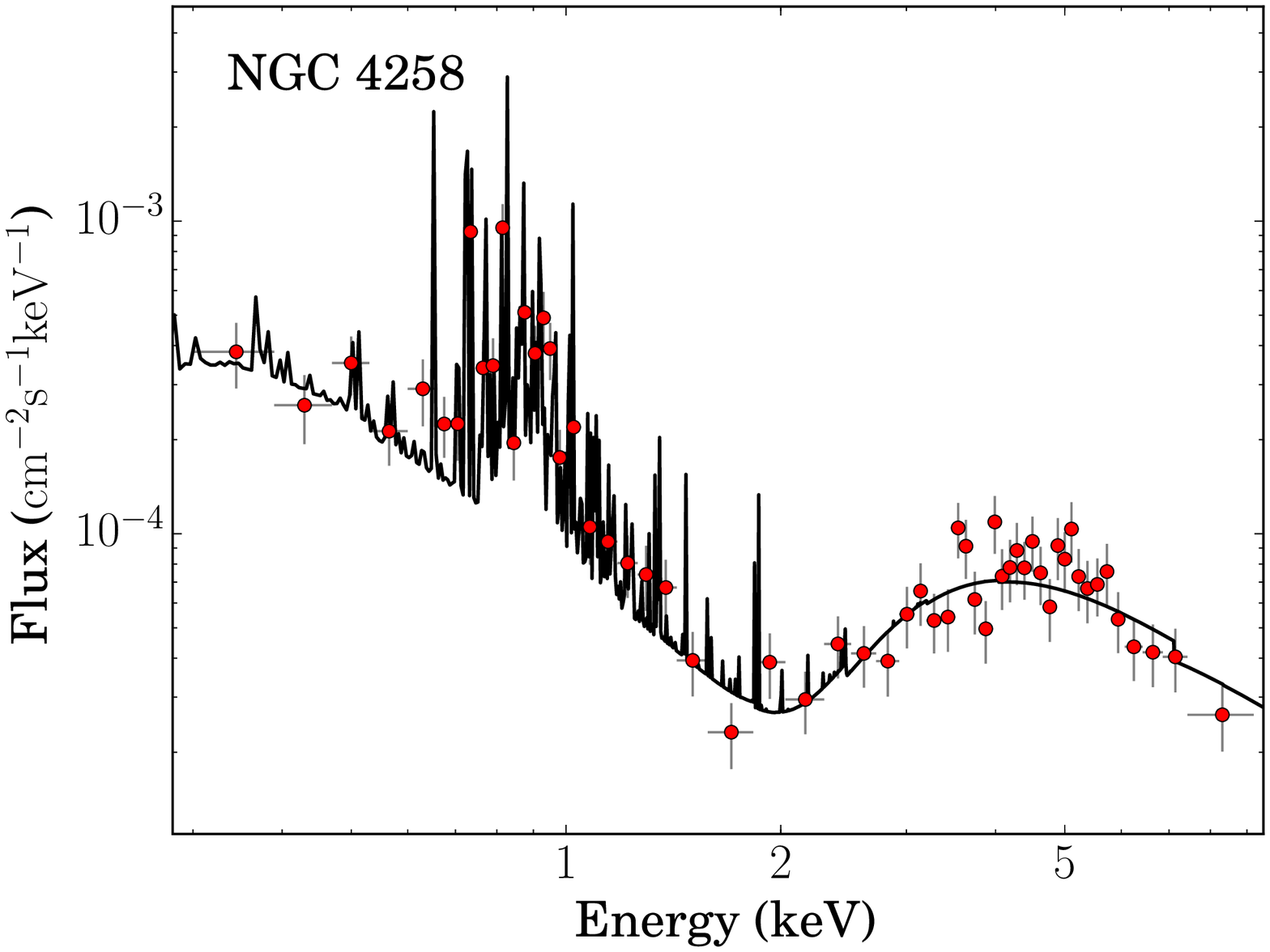}
	\includegraphics[width = 7cm,clip = true, trim = 0 0 0 0]{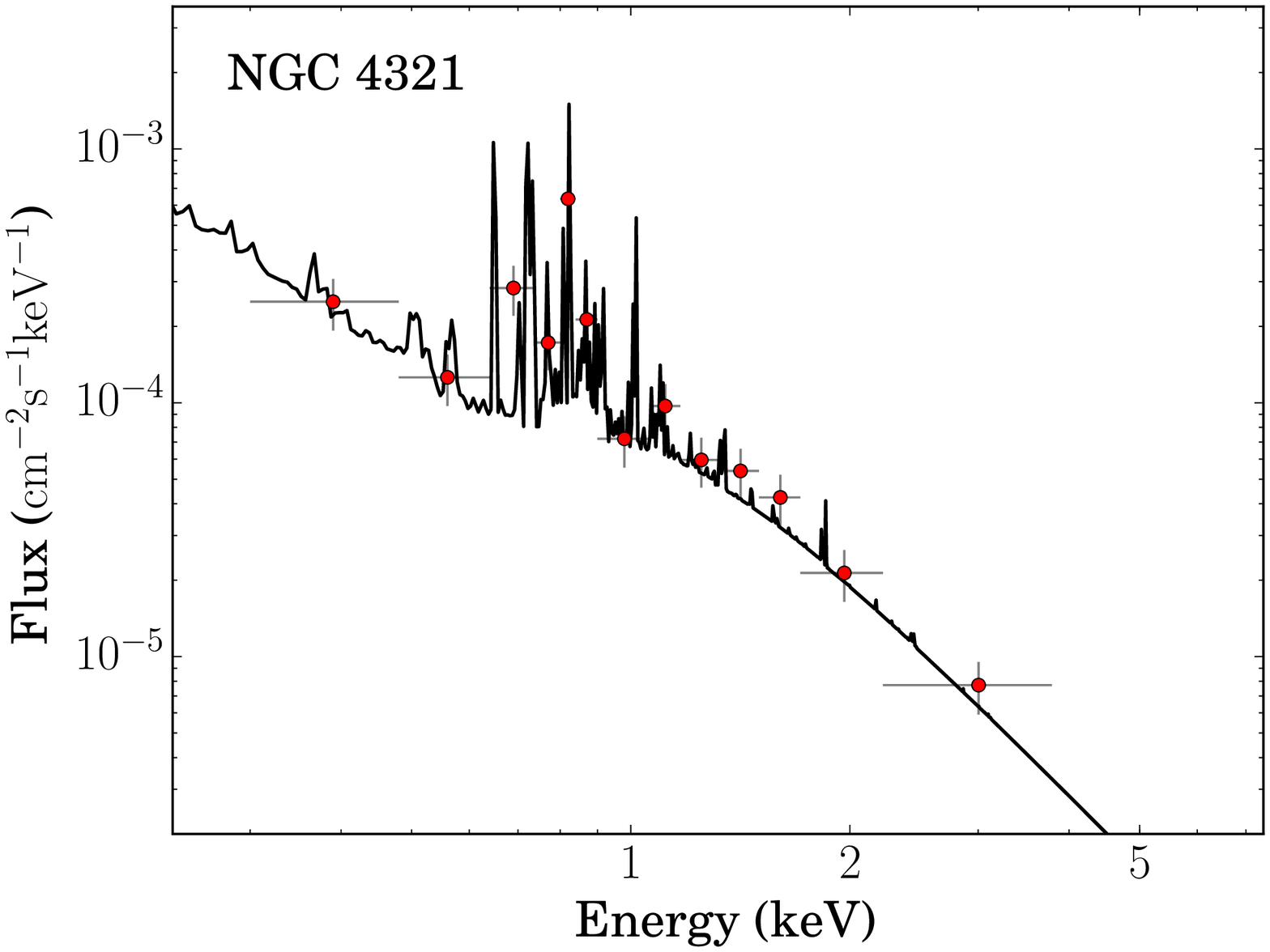}
	\includegraphics[width = 7cm,clip = true, trim = 0 0 0 0]{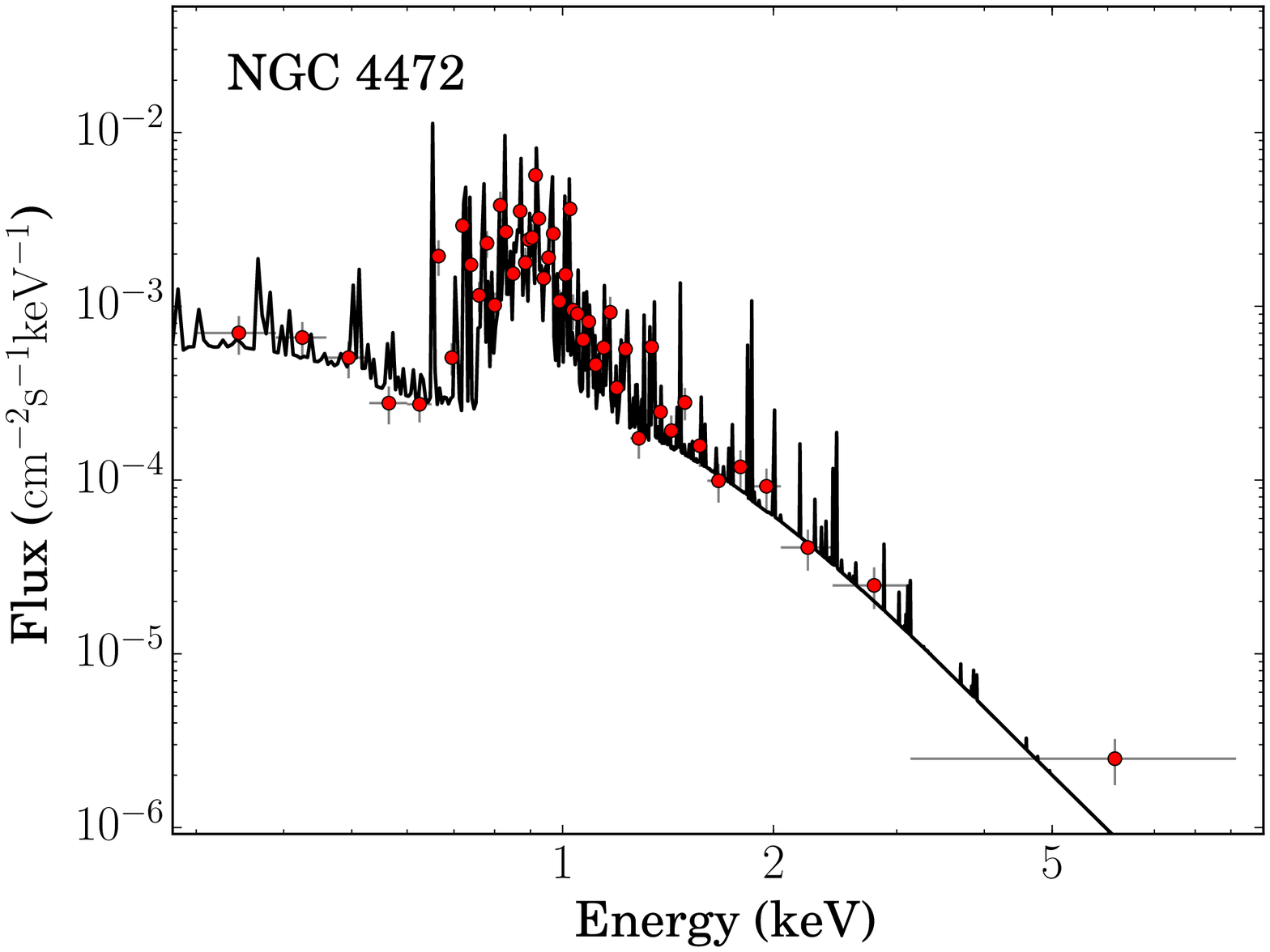}
	\includegraphics[width = 7cm,clip = true, trim = 0 0 0 0]{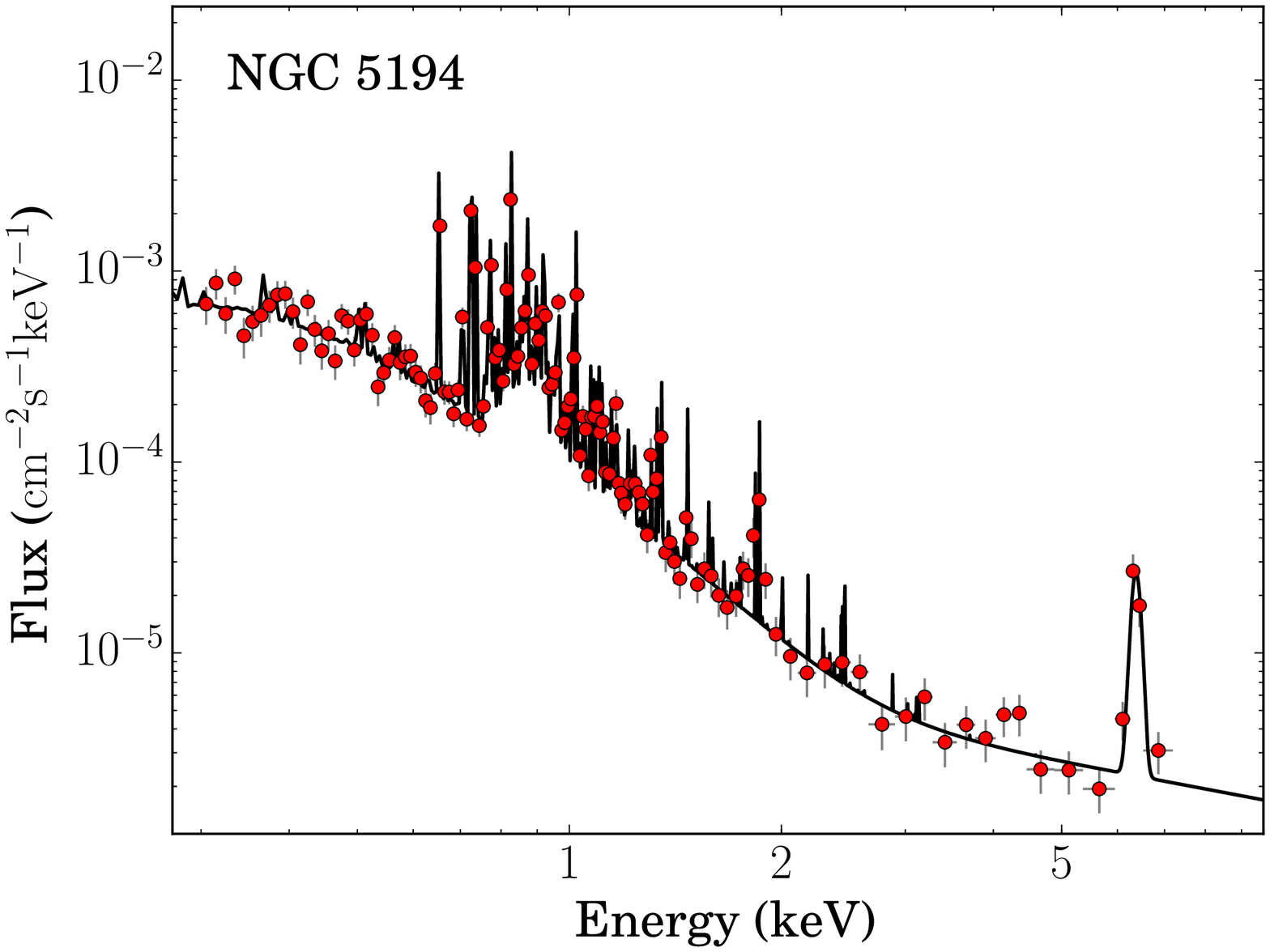}

	\label{totalSpectra}
	\caption{The middle flux-binned spectrum (unfolded) of each of the AGN in the sample for which flux-binning was carried out 
		(excluding M81 - see Fig. \ref{m81spectra}). The best-fitting spectral model for each object is also shown in black.}
\end{figure*}

\begin{figure*}

	\ContinuedFloat
	\includegraphics[width = 7cm,clip = true, trim = 0 0 0 0]{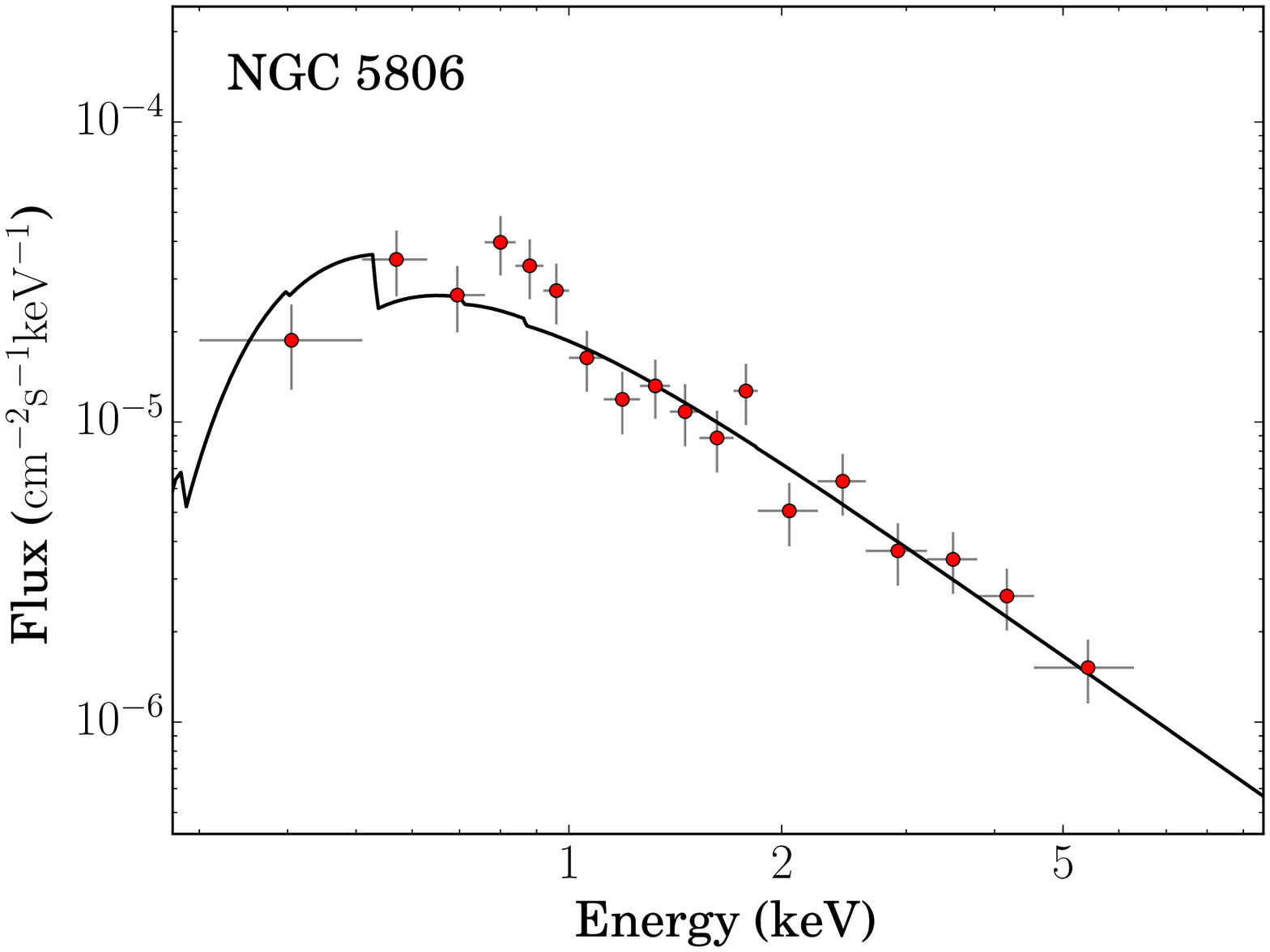}
	\includegraphics[width = 7cm,clip = true, trim = 0 0 0 0]{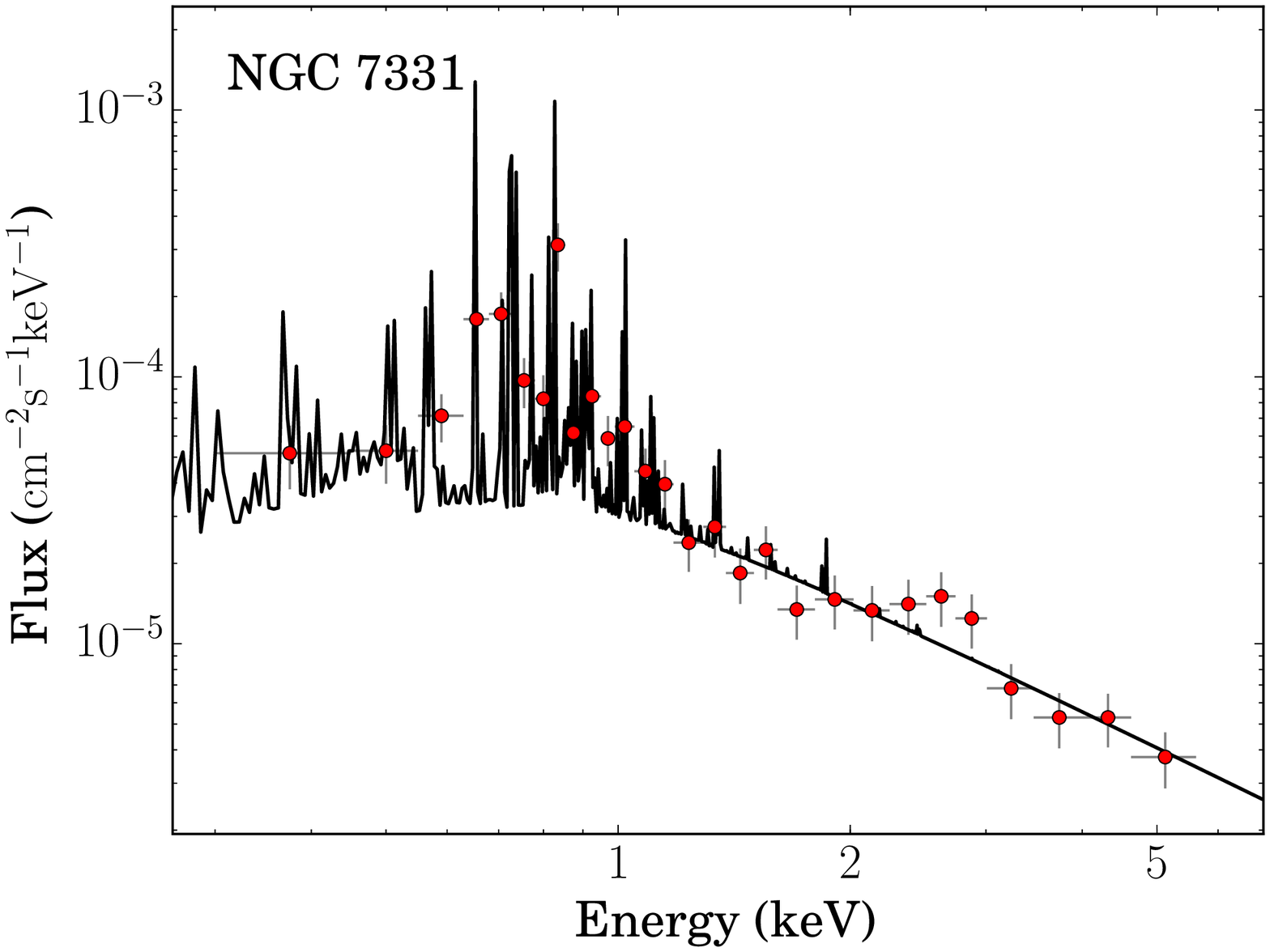}

	\caption{ (Continued)}

\end{figure*}

\begin{figure*}

	\includegraphics[width = 7cm,clip = true, trim = 0 0 0 0]{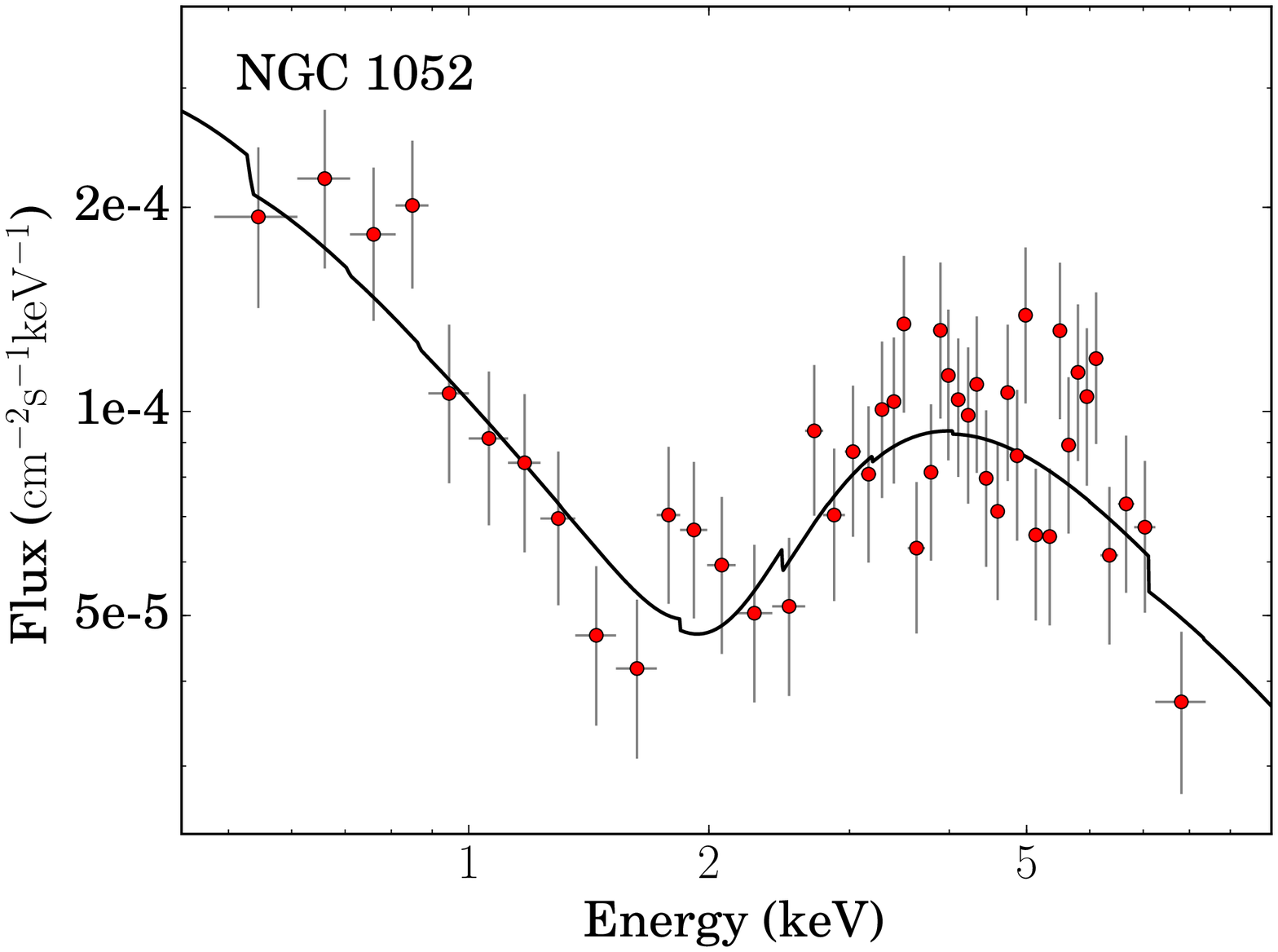}
	\includegraphics[width = 7cm,clip = true, trim = 0 0 0 0]{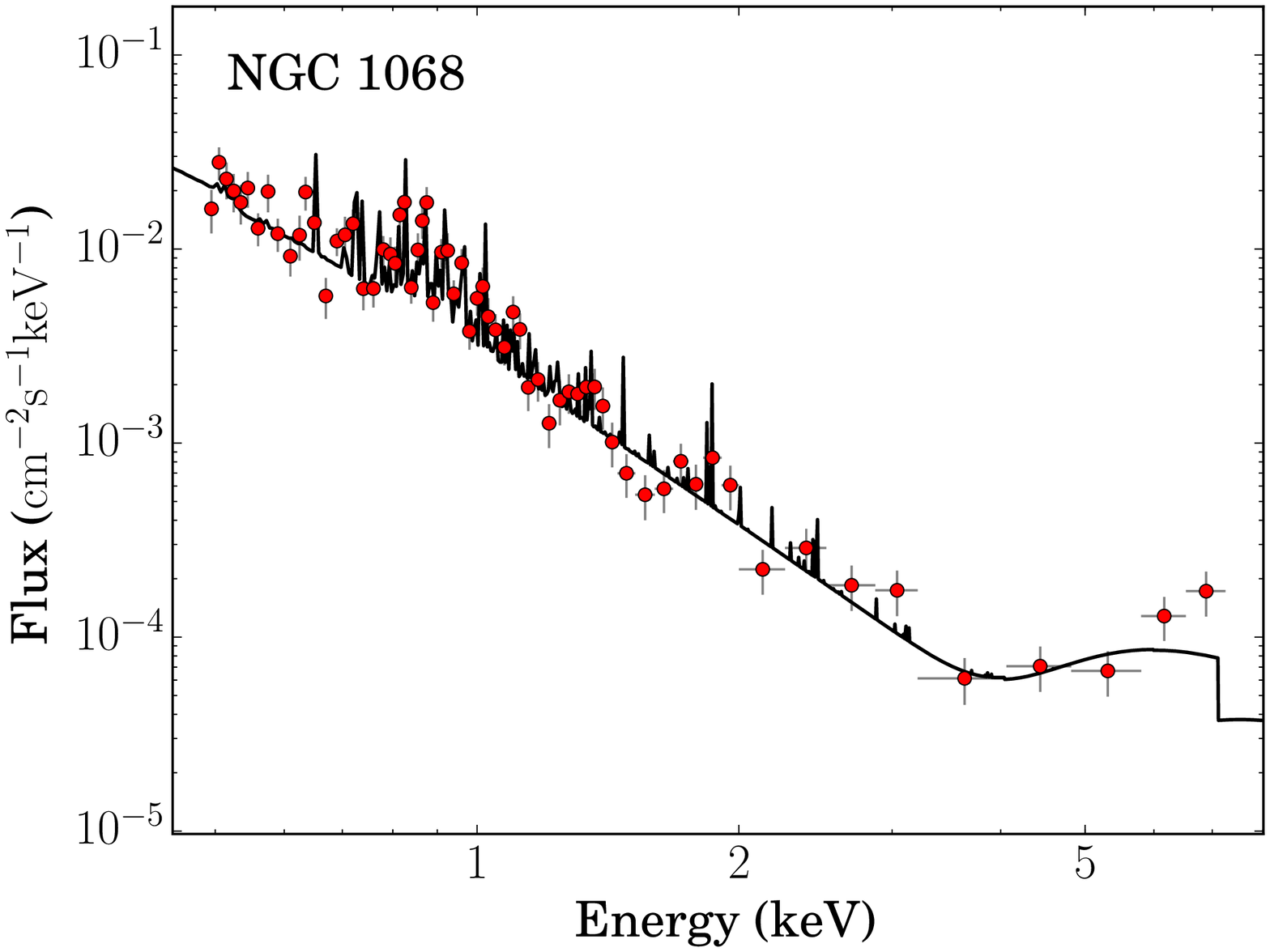}
	\includegraphics[width = 7cm,clip = true, trim = 0 0 0 0]{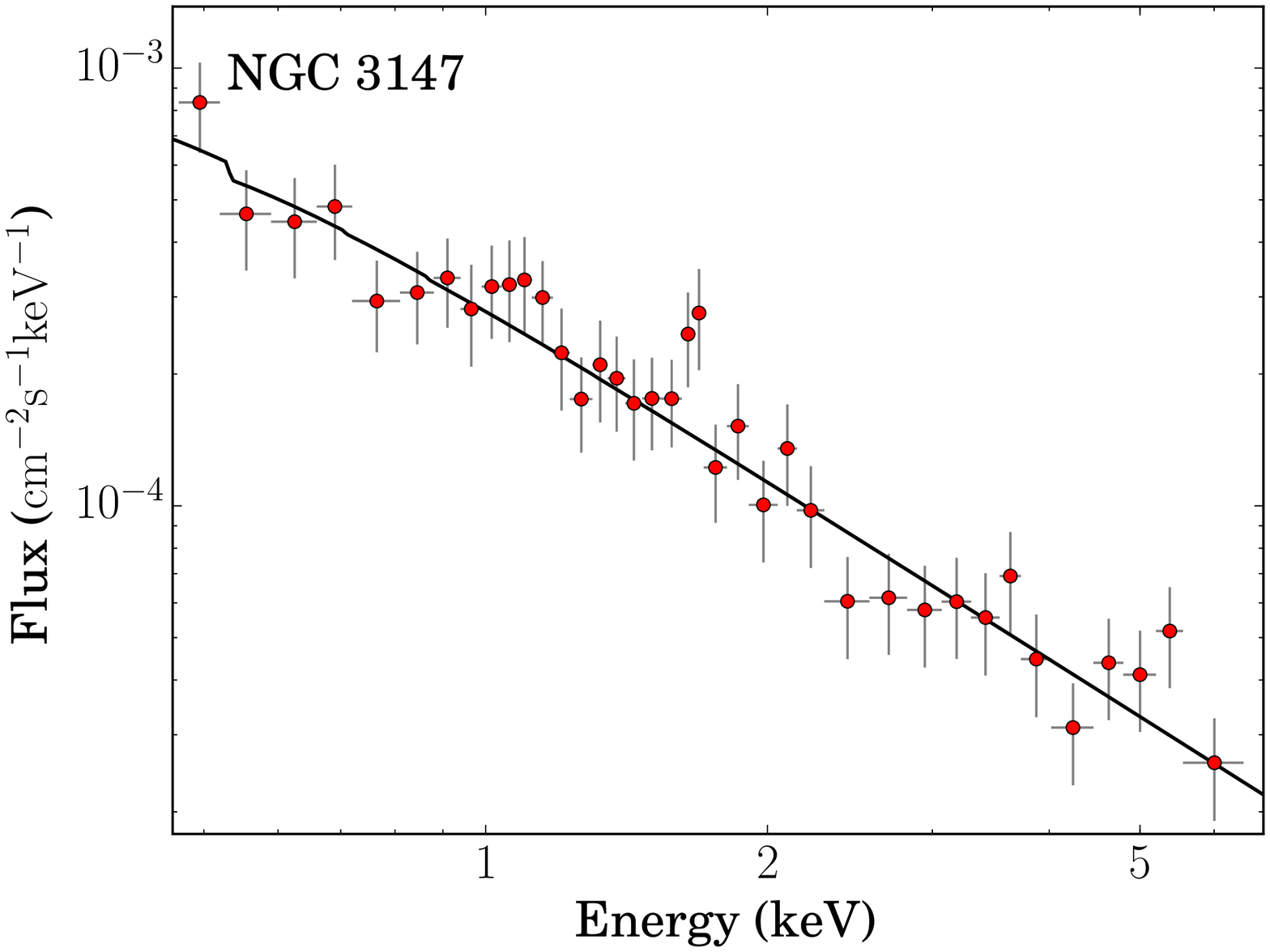}
	\includegraphics[width = 7cm,clip = true, trim = 0 0 0 0]{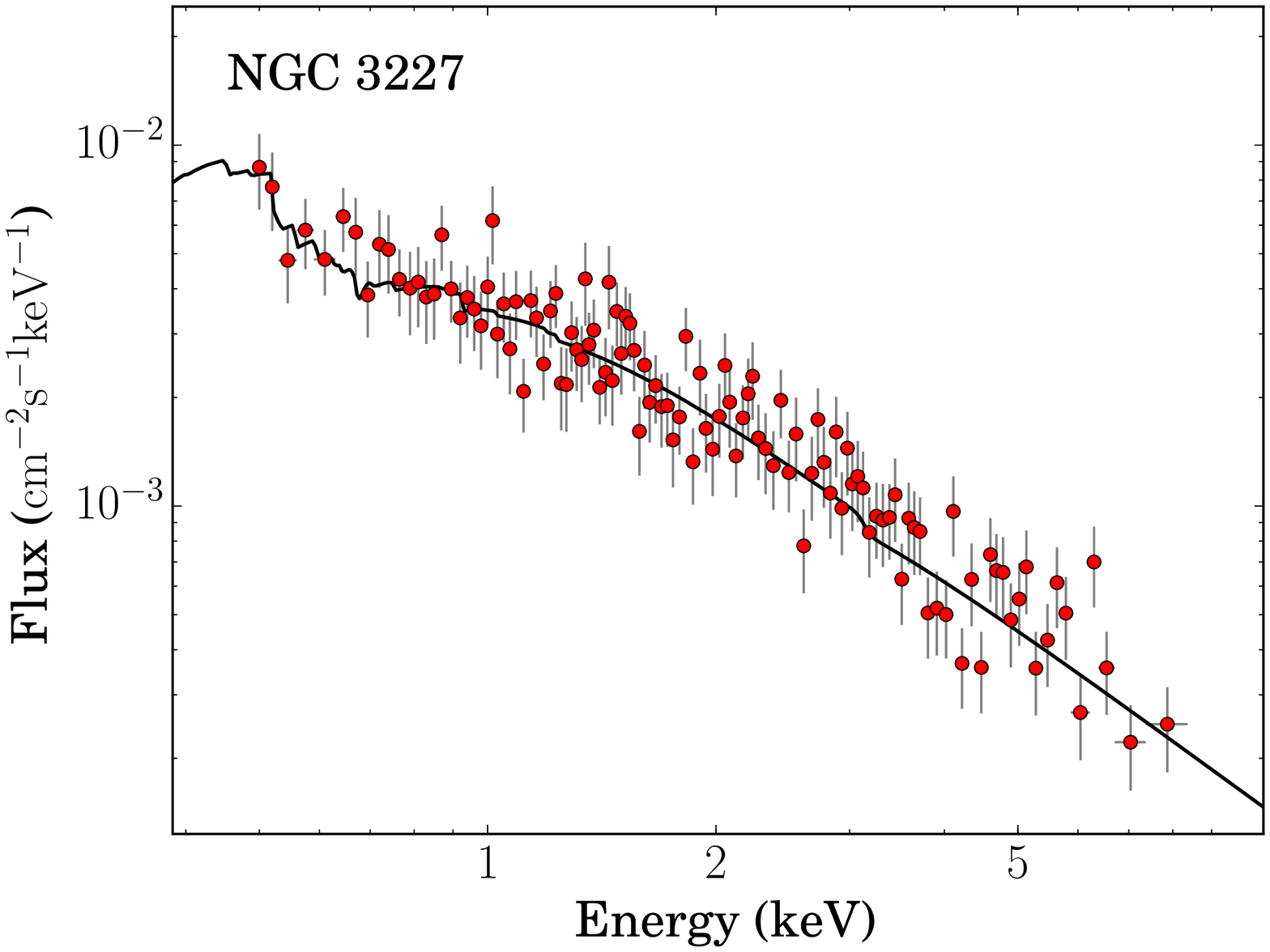}
	\includegraphics[width = 7cm,clip = true, trim = 0 0 0 0]{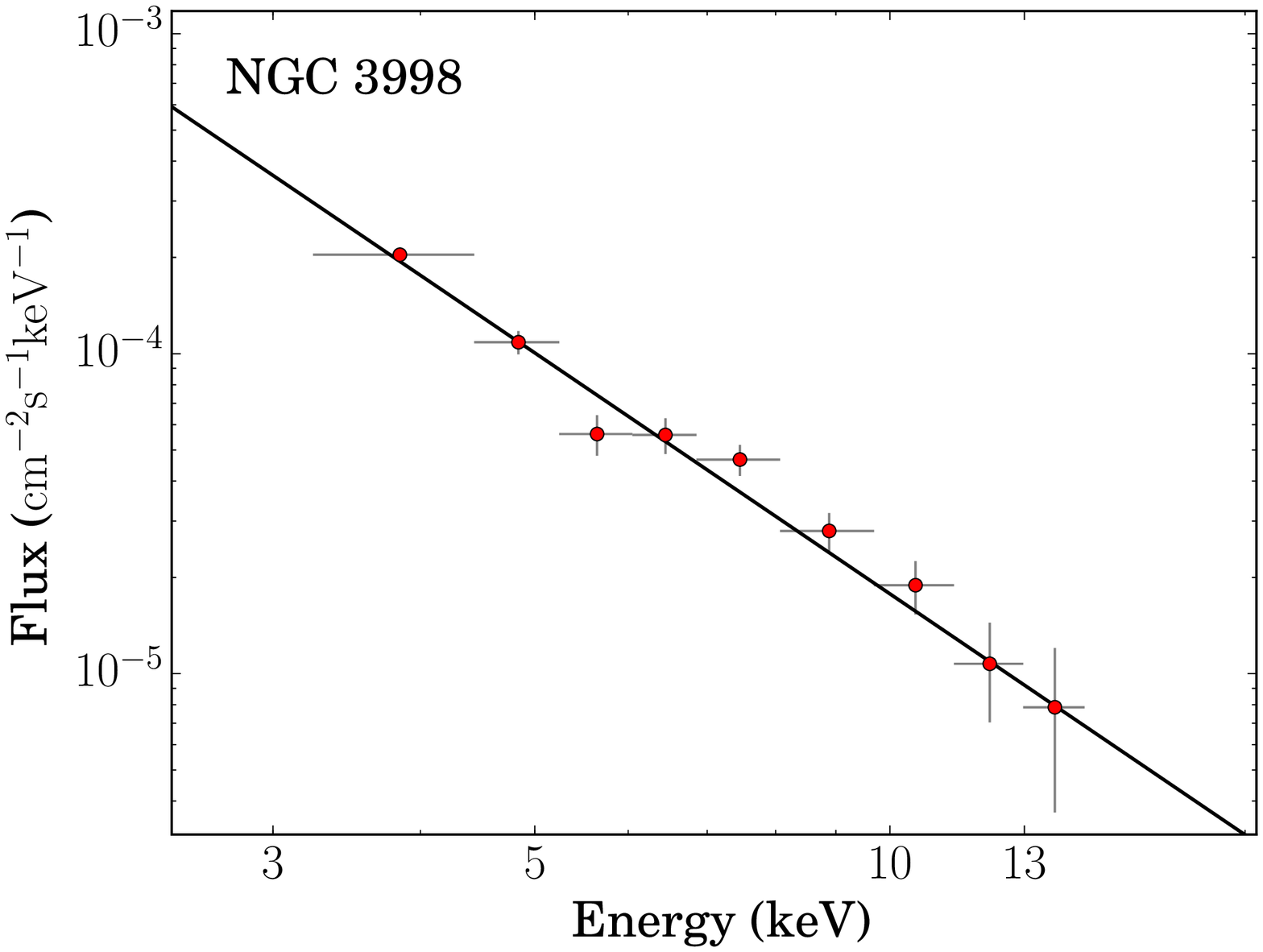}
	\includegraphics[width = 7cm,clip = true, trim = 0 0 0 0]{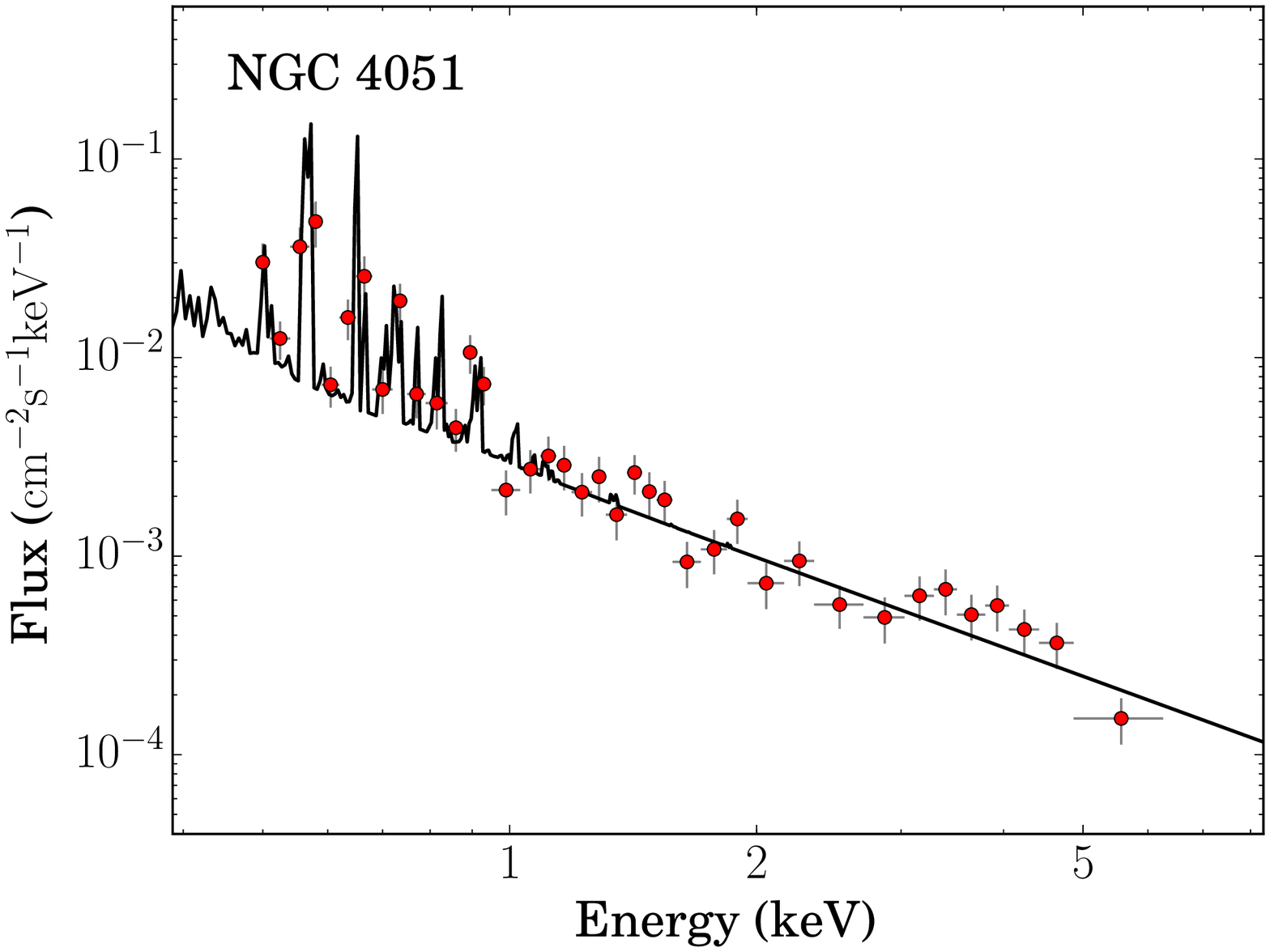}
	\caption{The total spectrum (unfolded) for each of the AGN in the sample for which flux-binning could not be carried out.
		 The best-fitting spectral model for each object is also shown in black.}
	\label{fluxSpectra}
\end{figure*}

\begin{figure*}

	\ContinuedFloat
	\includegraphics[width = 7cm,clip = true, trim = 0 0 0 0]{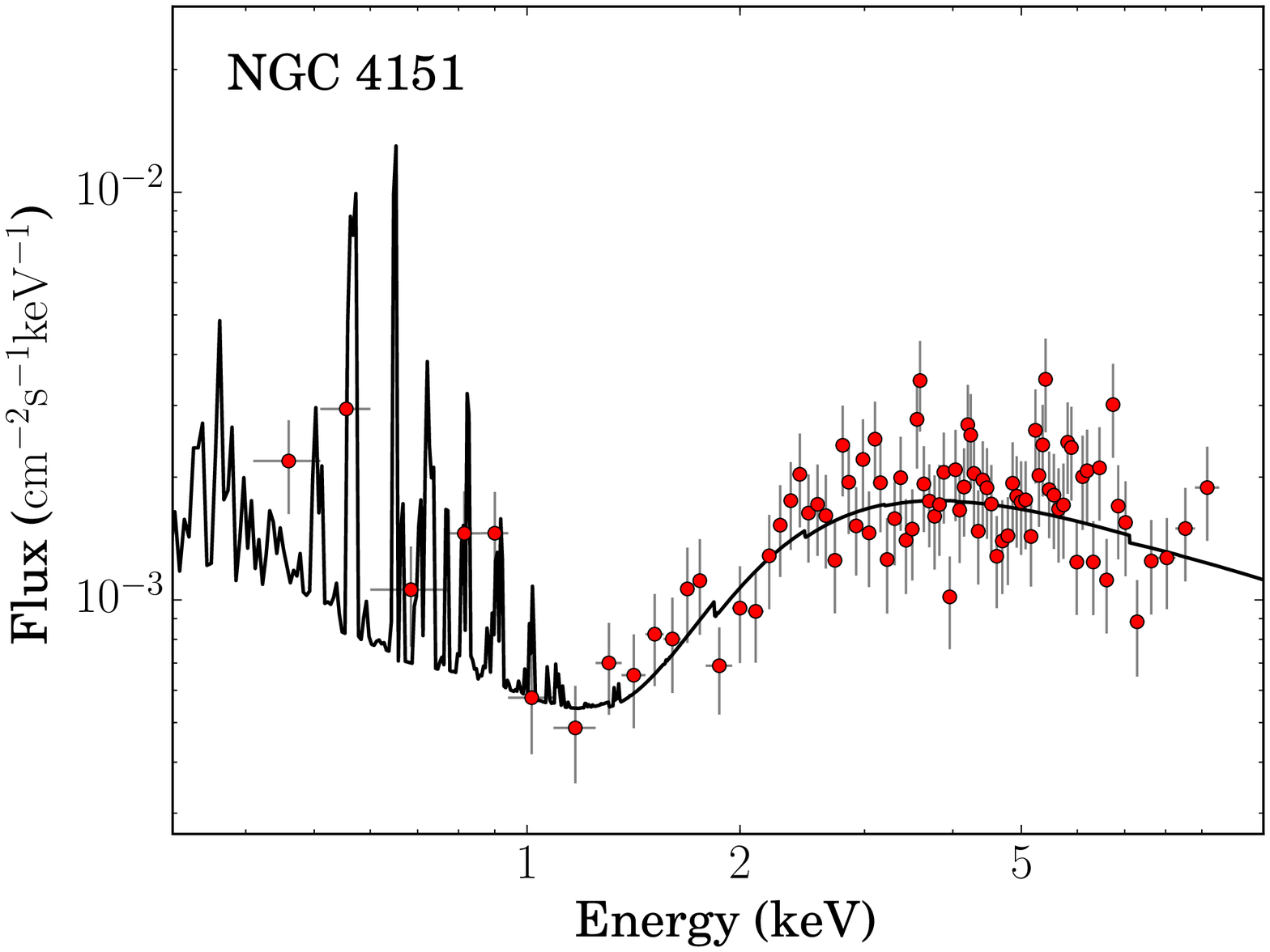}
	\includegraphics[width = 7cm,clip = true, trim = 0 0 0 0]{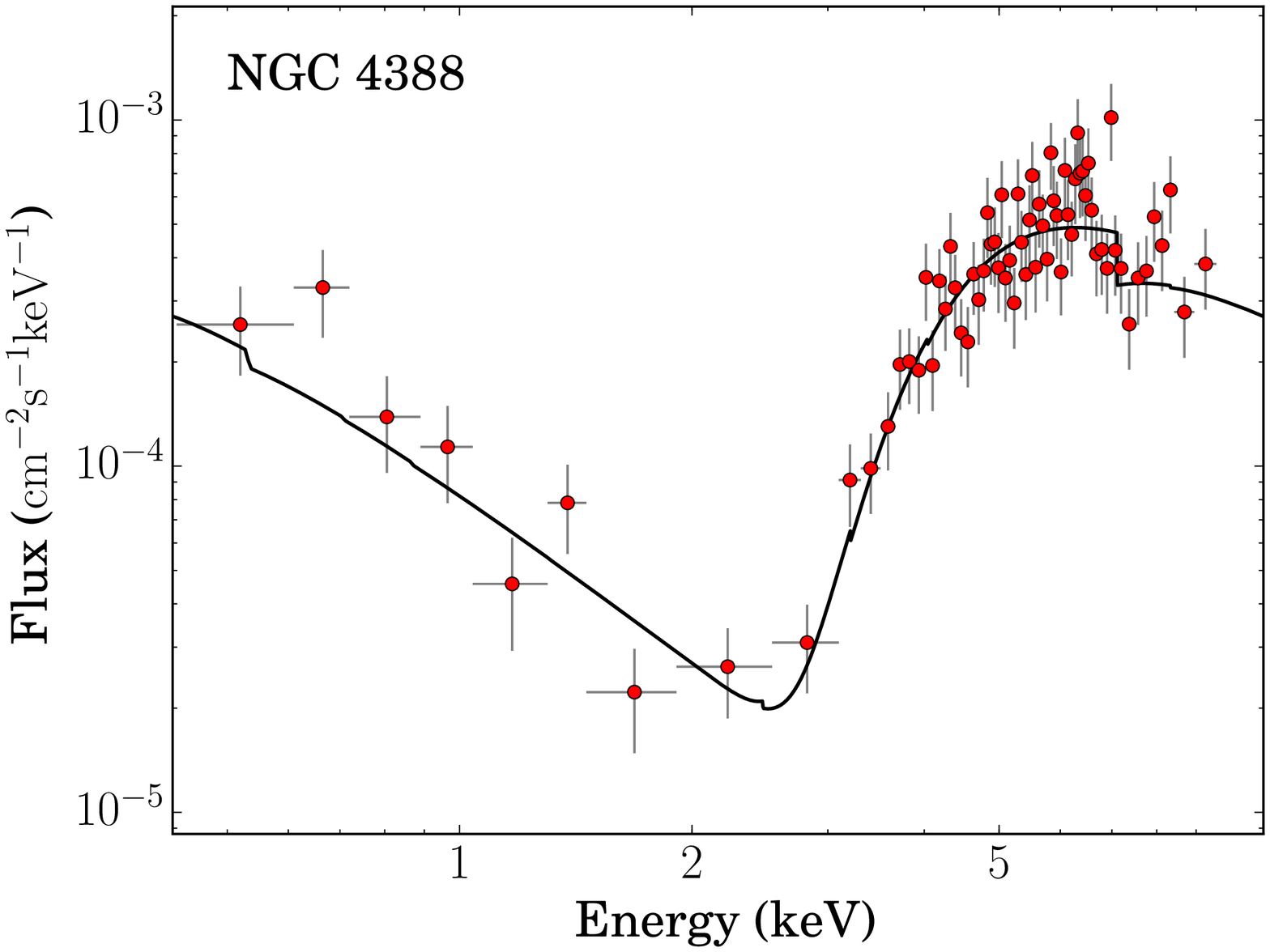}
	\includegraphics[width = 7cm,clip = true, trim = 0 0 0 0]{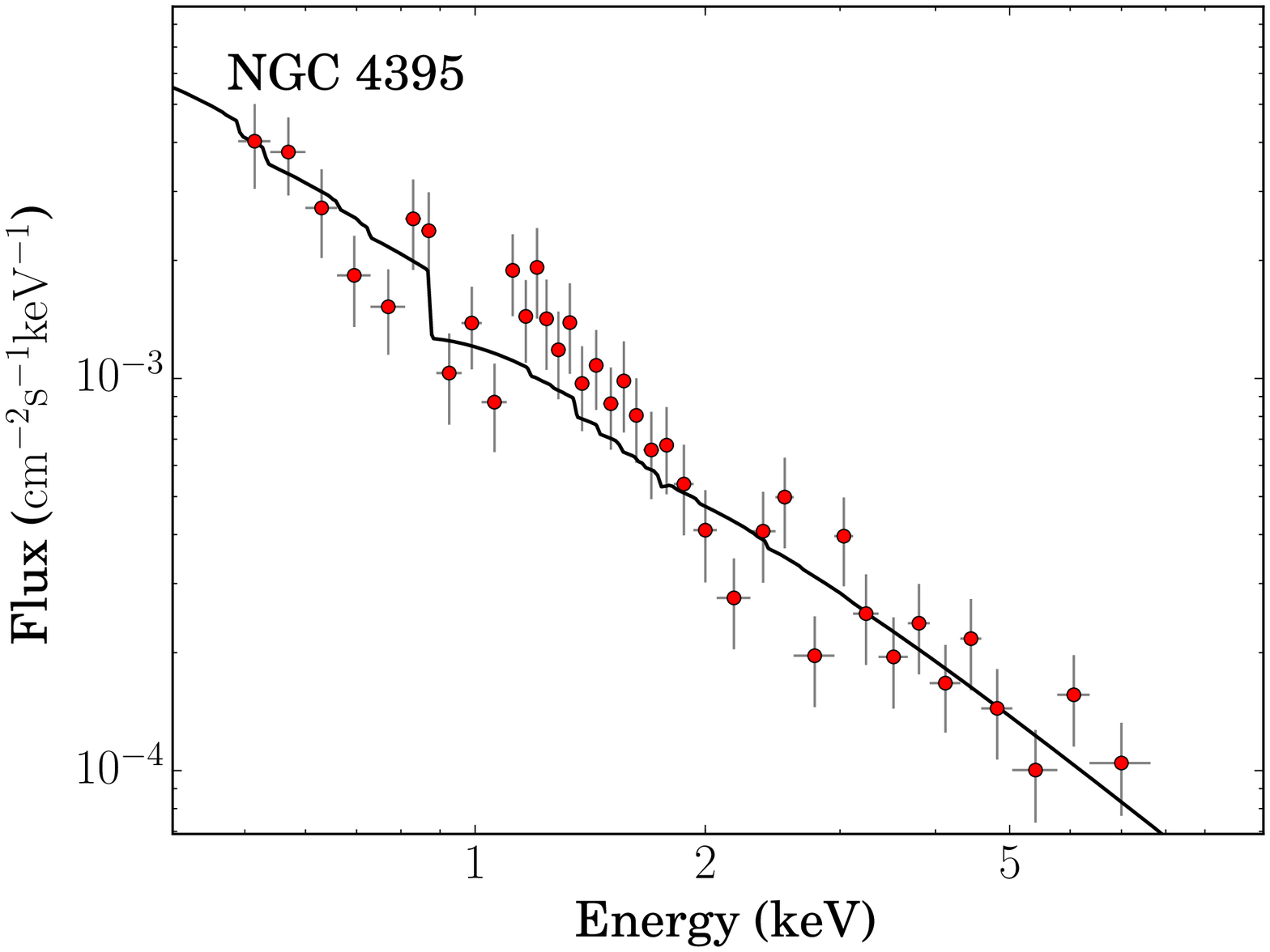}
	\includegraphics[width = 7cm,clip = true, trim = 0 0 0 0]{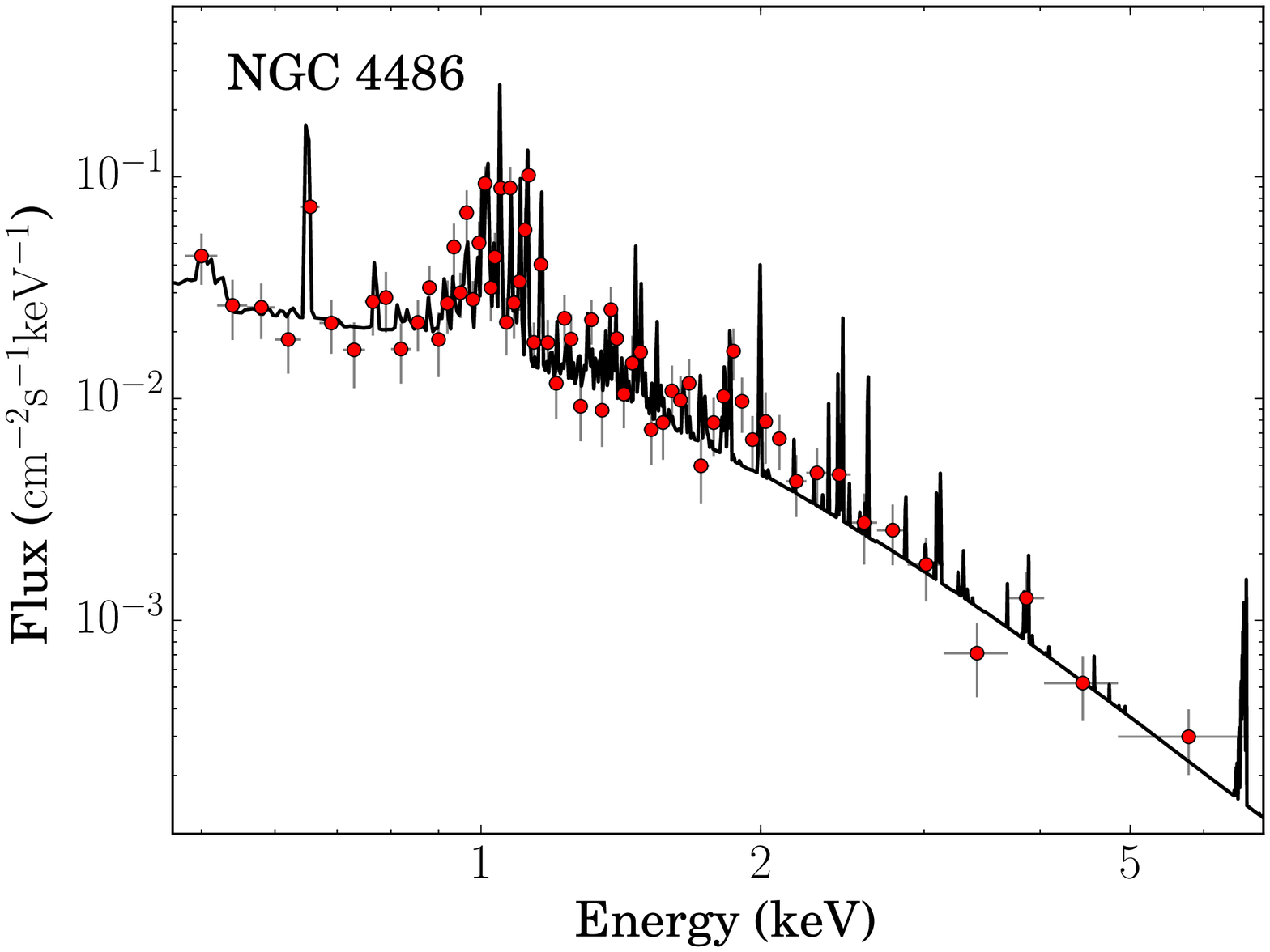}
	\includegraphics[width = 7cm,clip = true, trim = 0 0 0 0]{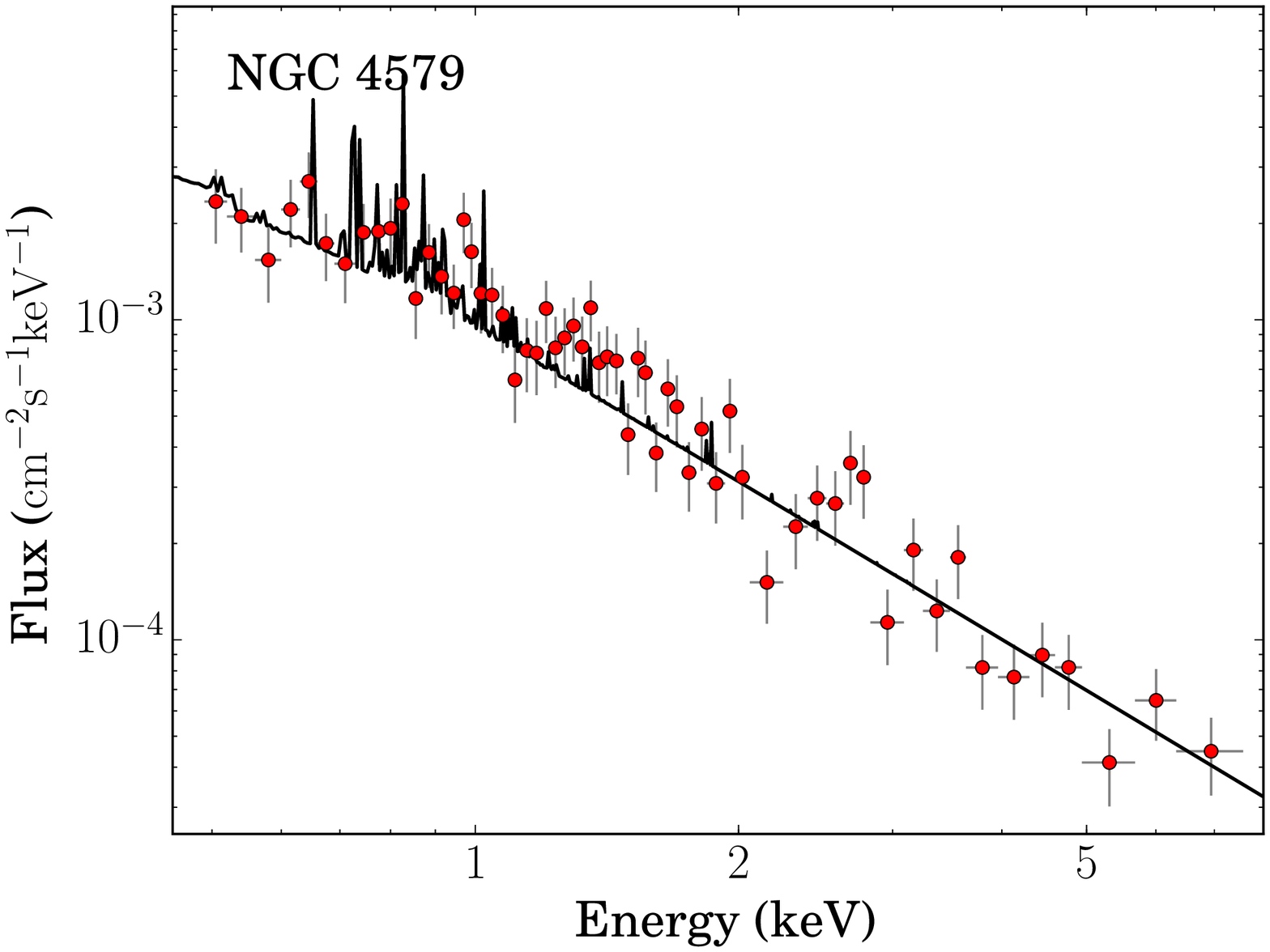}
	\includegraphics[width = 7cm,clip = true, trim = 0 0 0 0]{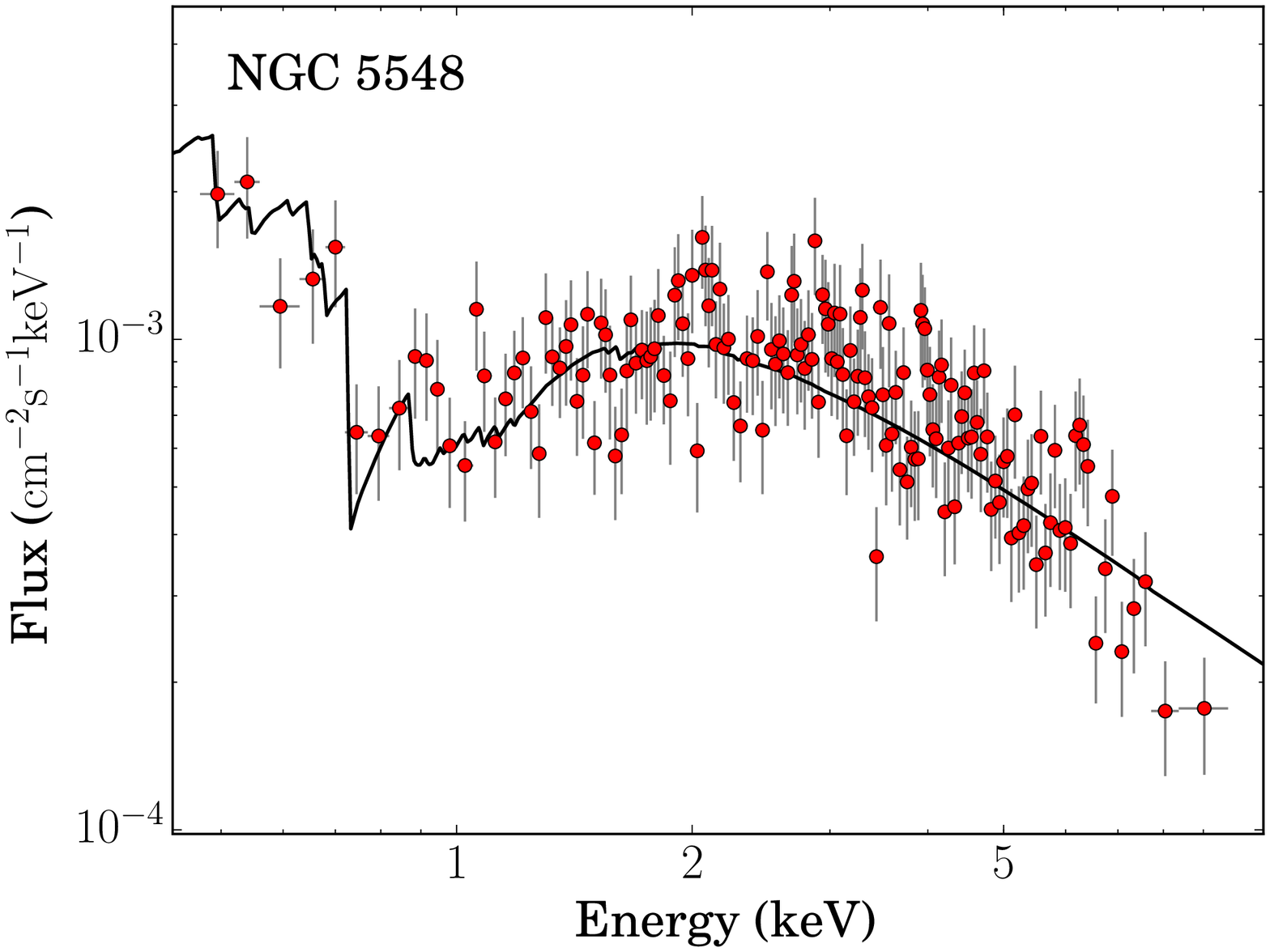}
	
	\caption{(Continued)}

\end{figure*}

\label{lastpage}

\end{document}